\PassOptionsToPackage{dvipsnames}{xcolor}

\documentclass[a4paper,11pt]{article}
\usepackage{jheppub}

\usepackage{lineno}
\usepackage{graphicx}
\usepackage{lipsum}
\usepackage{slashed}
\usepackage{amsmath}
\usepackage[normalem]{ulem}
\usepackage{enumitem}
\usepackage{mathtools}
\usepackage{bm}
\usepackage[all]{hypcap}

\usepackage{tikz}
\usepackage[compat=1.1.0]{tikz-feynman}
\tikzset{/tikzfeynman/warn luatex=false}

\usepackage{bbm}

\usepackage{mathabx}
\usepackage{pifont}
\usepackage{longtable}
\usepackage{multirow}
\usepackage{tablefootnote}
\usepackage{array}
\usepackage{makecell}
\usepackage{url}
\usepackage{duckuments}
\usepackage{cleveref}

\newcolumntype{R}[1]{>{\raggedleft\arraybackslash}p{#1}}

\newcommand{\Lagr}{\mathcal{L}}
\newcommand{\Order}{\mathcal{O}}

\newcommand{\cC}{\mathcal{C}}
\newcommand{\cL}{\mathcal{L}}
\newcommand{\cB}{\mathcal{B}}

\newcommand{\cO}{\mathcal{O}}

\newcommand{\SM}{{\rm SM}}

\title{
    New measurements in leptonic kaon and pion decays: experimental and theoretical perspectives
}

\author[a]{Lukas Allwicher,} 
\emailAdd{lukas.allwicher@desy.de}

\author[b]{Stefan Alexandru Ghinescu,} 
\emailAdd{stefan.alexandru.ghinescu@cern.ch}

\author[c]{Federico Mescia,} 
\emailAdd{federico.mescia@lnf.infn.it}

\author[c]{Tommaso Spadaro,} 
\emailAdd{tommaso.spadaro@lnf.infn.it}

\author[d,a]{and Lucine Tabatt} 
\emailAdd{lucine.marie.anais.tabatt@hu-berlin.de}

\affiliation[a]{Deutsches Elektronen-Synchrotron DESY, Notkestr. 85, 22607 Hamburg, Germany}
\affiliation[b]{Horia Hulubei National Institute for R\&D in Physics and Nuclear Engineering, Bucharest-Magurele, Romania}
\affiliation[c]{INFN, Laboratori Nazionali di Frascati, C.P. 13, 00044 Frascati, Italy}
\affiliation[d]{Institut f\"ur Physik, Humboldt-Universit\"at zu Berlin, 12489 Berlin, Germany}

\abstract{
We study the future prospects for precision measurements of purely leptonic charged-current kaon and pion decays. Together with the ratio of leptonic widths $R_K^\nu=\Gamma(K\to e\nu)/\Gamma(K\to\mu\nu)$, we consider the single ratios $R_{K\pi}^{\ell=\mu,e} = \Gamma(K \to \ell\nu) / \Gamma(\pi \to \ell\nu)$, for which many experimental uncertainties cancel, as well as the double ratio  $R_{K\pi}^e / R_{K\pi}^\mu$,  which provides a particularly sensitive probe of new physics. This study is timely in light of two converging developments. On the experimental side, a new conceptual design for a dedicated setup based on a slow-extracted proton beam and a multi-year data-taking program, is expected to achieve a precision at the few $10^{-4}$ level. On the theory side, recent lattice-QCD calculations have reached a comparable level of accuracy. Finally, we assess the impact of these next-generation measurements on searches for physics beyond the Standard Model and show that they can provide constraints that are either competitive with or complementary to existing bounds.
}

\begin{document}

\preprint{
DESY-26-117
HU-EP-26/29-RTG
}

\maketitle
\flushbottom

\section{Introduction}

In the search for physics beyond the Standard Model, precision tests of SM-allowed transitions play a crucial role.
Such tests, pushing the so-called \textit{intensity frontier}, have been historically of great importance, for example through electroweak precision tests at LEP and SLC, and continue to play a crucial role in constraining possible new physics scenarios. 
Such precision SM tests are however not confined to the electroweak sector, but can be performed in complementary channels at lower energies through flavour observables. 

The common argument in favour of flavour observables usually relies on accidental suppressions occurring in the SM, such as the GIM mechanism, or the conservation of lepton number, which is why flavour-changing neutral currents and lepton-flavour violating transitions are cornerstones of our exploration of the BSM landscape.
In this work, instead, following the seminal proposal of~\cite{Masiero:2005wr} we want to turn our attention to charged-current decays, specifically to purely leptonic decays of light mesons, i.e. pions and kaons.
These decays occur at tree-level in the SM~\cite{FlaviaNetWorkingGrouponkaonDecays:2010lot,FlaviaNetWorkingGrouponkaonDecays:2008hpm, FlavourLatticeAveragingGroupFLAG:2024oxs,Aebischer:2025mwl}, through the exchange of a virtual $W$ boson. The only significant suppression playing a role is the helicity suppression due to the left-handed weak interaction, leading to $\mathcal{B}(M\to e \nu)/\mathcal{B}(M\to \mu\nu) \sim 10^{-4}$ ($10^{-5}$) for $M=\pi^+ (K^+)$.

These modes (and combinations thereof) are relevant both in our understanding of the SM, giving a clean probe of the first row of the CKM matrix, and in BSM explorations~\cite{Masiero:2005wr,Masiero:2008cb,Ellis:2008st,Girrbach:2012km,Fonseca:2012kr,Abada:2012mc,Abada:2013aba,Jung:2010ik,Gonzalez-Alonso:2016etj,Aebischer:2025mwl}, as they can provide important constraints on scenarios with scalar currents, or a test of lepton flavour universality complementary to those performed in $B$ and $D$ physics. This direction has also been explicitly emphasized in the recent update of the European Strategy for Particle Physics~\cite{deBlas:2025gyz}.

Current experimental measurements are around the per-mil level, while the SM predictions have been recently updated 
by lattice QCD simulations~\cite{Boyle:2026lrz,DiPalma:2025iud,DiCarlo:2019thl} to be well below that, at around $10^{-4}$.
This new determination motivates even further the need to assess what precision could be achieved with possible future measurements of the leptonic decay rates, and what we would learn from them.
We shall present a detailed analysis of both aspects, in the form of a conceptual study for an experimental setup aiming at measuring charged kaon and pion decays, and a theoretical study about the implications. The proposed programme is highly synergistic with the PIONEER experiment~\cite{PIONEER:2022alm,PIONEER:2022yag}, which focuses on the lepton-universality observables in pion decays. The simultaneous study of kaon and pion leptonic decays provides a unique opportunity to perform complementary and highly precise tests of lepton flavor universality, probe different effective operator structures, and enhance the sensitivity to a broad class of new-physics scenarios.

The remainder of the paper is structured as follows.
In Section \ref{sec:observables} we introduce the observables which are the focus of our study, highlighting their main features as new physics probes, alongside their current SM predictions.
Section \ref{sec:exp} contains the sketch of a new experiment, with details on the beam and detector setup, and estimates about running times and projected sensitivities for the final measurements.
Following up, Section \ref{sec:th} is a study of our observables as BSM probes, done both within an Effective Field Theory approach and in an explicit example model. The objective is to highlight their potential in comparison to other observables, within a coherent theoretical framework.
Finally, we conclude in Section \ref{sec:conclusions}.

\section{Observables of interest}\label{sec:observables}
A precision flavour universality test can be performed comparing expected and measured values for the ratio
\begin{align}
    R_K^\nu = \frac{\mathcal{B}(K^+ \to e^+ \nu)}{\mathcal{B}(K^+ \to \mu^+ \nu)}.
\end{align}
The experimental precision for $R_K^\nu$ is usually limited by systematic effects, which are primarily driven by the evaluation of absolute detection efficiencies for the positron and the muon in the final state. An attempt to partially circumvented such limitation through the exploitation of muon in flight decays to positrons is in progress at NA62, with an overall uncertainty expected to be at the level of a few per mil~\cite{lubos:LatticeLab2026}. 
The present experimental state of the art is~\cite{ParticleDataGroup:2026mpi,PIONEER:2022alm}:
\begin{align}
    R_K^{\nu,\,{\rm exp}} = 2.488(9)\times10^{-5}
    \qquad     R_\pi^{\nu,\,{\rm exp}} = 1.2327(23)\times 10^{-4}\,,
\end{align}
where the $R_\pi^\nu$ is the analogous ratio for the charged pion decays, which we report here for later convenience. 
On the theoretical side, recent lattice QCD determinations \cite{Boyle:2026lrz} have established
\begin{align}
    R_K^{\nu,\,{\rm SM}} = 2.47653(34) \times 10^{-5} \qquad R_\pi^{\nu\,{\rm SM}} = 1.23501(10) \times 10^{-4} \,.
\end{align}

In a variety of new-physics scenarios, the following ratios would be affected by NP contributions as well:
\begin{align}
R_{K\pi}^{\ell=\mu,e} = \frac{\Gamma(K^+\to\ell^+\nu)}{\Gamma(\pi^+\to\ell^+\nu)}.
\end{align}
The above ratios allow for improved cancellation of experimental systematic uncertainties. Their SM expectations derive from estimates of the ratio of decay constants $f_K/f_\pi$ and of the parameters of the CKM matrix $V_{us}/V_{ud}$, presently known to within a few $10^{-3}$.
The current experimental values and SM predictions are
\begin{align}
    \begin{aligned}
    R_{K\pi}^{\mu,\,{\rm exp}} &= 1.3377(40)\qquad R_{K\pi}^{\mu,\,{\rm SM}} =& &1.316(6) \, ,\\
    R_{K\pi}^{e,\,{\rm exp}} &= 0.2698(14)\qquad R_{K\pi}^{e,\,{\rm SM}} =& &0.2639(12) \, .
    \end{aligned}
\end{align}

More details on the SM values can be found in Appendix \ref{app:RKpiSM}. The experimental values are computed accounting for uncertainties of the $K$ and $\pi$ lifetimes and of the branching ratios of the channels involved. The PDG averages~\cite{ParticleDataGroup:2026mpi} are used, as explained in Section~\ref{sec:sensitivity_1_year}. No precise Standard Model prediction from lattice calculations currently exists for electrons~\cite{Boyle:2026lrz,DiPalma:2025iud,DiCarlo:2019thl}, though such a calculation would be highly desirable in the future. Therefore, we evaluated the experimental and SM values for $R_{K\pi}^{e}$ using the $R_{K\pi}^{\mu}$ together with the ratio of kaon and pion leptonic widths, $R_{K}^\nu$ and $R_\pi^\nu$. The  uncertainty on $R_{K\pi}^{e,\,{\rm exp}}$ is 50\% correlated to that of $R_{K\pi}^{\mu,\,{\rm exp}}$, because the latter is dominated by the measurement of the absolute branching ratio for $K\to\mu\nu$. On the contrary, the  uncertainty on $R_{K\pi}^{e,\,{\rm SM}}$ is fully correlated to that of $R_{K\pi}^{\mu,\,{\rm SM}}$, through the $V_{us}/V_{ud}$ and $f_K/f_\pi$ input.

The double ratio 
\begin{align}
R_{K\pi}^{e/\mu}=R_{K\pi}^{e}/R_{K\pi}^{\mu}
\end{align}
evades from the above limitations and, while allowing a thorough cancellation of the experimental systematic effects, is also theoretically very clean. 
The helicity suppression is stronger for kaons than for pions by more than one order of magnitude. Consequently, NP effects mediated by scalar currents can be expected to give comparatively larger contributions in kaon decays than in pion decays. For this reason we will typically assume that only kaons are affected by NP, such that effectively, from the theoretical perspective, the substitution $R_{K\pi}^{e/\mu} \to R_K^\nu$ can be made. This will simplify the expressions for the observables significantly. The double ratio serves as useful overall normalisation for the NP search in helicity suppressed leptonic modes.

\section{Experimental considerations}\label{sec:exp}
We consider proposing a new experimental setup to pursue precision measurements of $R_K^\nu$, $R_{K\pi}^\mu$, $R_{K\pi}^{e}$ and the related double ratio. 
The idea presented here is to exploit high-intensity, slowly-extracted 400~GeV beams such as those produced for some of the experiments of the CERN north areas. For a summary of the beam parameters considered, please refer to Tab.~\ref{tab:beamparameters}.
\begin{table}[]
    \centering
    \begin{tabular}{l|l}\hline
    {\bf Parameter} &  {\bf Nominal value}\\ \hline
    Protons per pulse per second on spill on target & $1.6\times10^{10}$  \\ 
    Acceptance of secondary beam definition system & 10\,$\mu$sr \\
    Momentum of the secondary beam & 25~GeV \\
    Number of $\pi^+$,~$K^+$,~$p$ transmitted per proton [$10^{-4}$]& 8,~0.8,~0.8\\ 
    Longitudinal space allocated for beam definition & 60\,m \\ 
    $\pi^+$,\,$K^+$,\,$p$ fluxes at the detector in~4~s-long spill [$10^{6}$] & $46$,~$3.7$,~$4.8$\\
    $\pi^+$\,:\,$K^+$\,:\,$p$ relative flux at the detector & 84\%\,:\,7\%\,:\,9\% \\
    Spills per year & $6\times10^5$ \\
    Number $\pi^+$,\,$K^+$ tagged per year [$10^{12}$] & $28$,\,$2.2$\\ \hline
    \end{tabular}
    \caption{Summary of the most relevant beam parameters}
    \label{tab:beamparameters}
\end{table}

\subsection{Beam concepts}

After impact of the beam protons on a light-material target, a momentum-selected, positively-charged secondary hadron beam can be produced to be directed towards a sensitive volume located further downstream. A system of magnetic elements (one or more large-acceptance quadrupole triplets) immediately downstream of the target focuses the secondaries emitted at a given angle (not necessarily different from zero) with respect to the incoming proton beam, with an angular acceptance of $\mathcal{O}(10~\mu\mbox{sterad})$. A momentum selection is performed ensuring a $\pm1\%$ accepted range around a nominal value.

The value of the nominal momentum is critical for the proposed measurement. To detect in-flight decays $\pi\to\mu\nu$ in a downstream decay volume of reasonable length, a minimum angle must be ensured for the muon emission in the laboratory. After considering a number of possibilities, we propose a 25~GeV momentum.  The beam momentum corresponds to decay lengths of 188~m and 1396~m for $K^+$ and $\pi^+$, respectively. With a 25\% (70\%) probability, a muon emitted from a pion (kaon) leptonic decay moves at least 8~cm and less than 100~cm away from the beam line of flight at 25 m, so that it can be tracked. 

The estimated number of $\pi^+$, $K^+$, protons transmitted by the momentum-selection system slightly depends on the angle around which the secondary beam is selected.\footnote{Establishing the angle value depends on a number of infrastructure considerations which are beyond the scope of the present document.} For angles of 6~mrad or less we expect $8\times10^{-5}$  $K^+$ and $8\times10^{-4}$ $\pi^+$ transmitted per incident beam proton~\cite{VanDijk:2641500}. The abundance of secondary protons is of the same order of that of the $K^+$.


We consider an overall length of 60~m allocated for beam preparation. This figure can be considered reasonable, given past beam line designs~\cite{NA62:2017rwk}. 
The fraction of $K^+$ and $\pi^+$ reaching the decay volume 60~m downstream of the target are 73\% and 96\%, respectively. The fractional beam composition at the entrance of the detector area is $K^+$~:~$\pi^+$~:~$p=0.07$~:~$0.84$~:~$0.9$.

\subsection{Detector scheme}
\label{sec:detector}
\begin{figure}
    \centering
    \includegraphics[width=\linewidth]{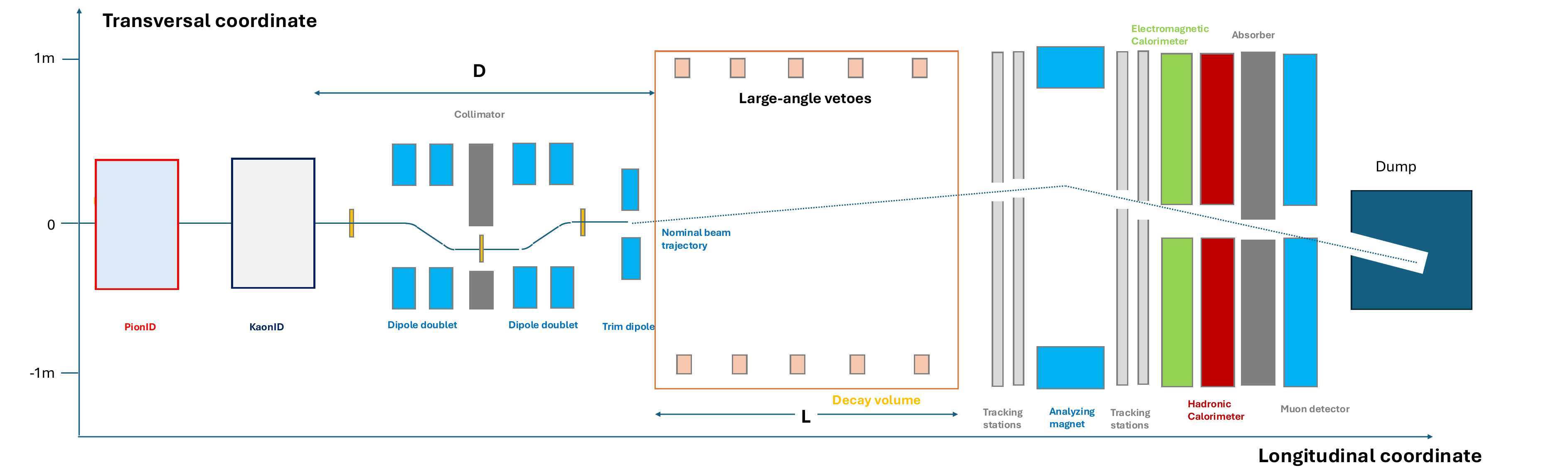}
    \caption{Conceptual transverse view of the detector region.}
    \label{fig:detector}
\end{figure}

A conceptual sketch of the detector region is shown in Figure~\ref{fig:detector}.
Proceeding from upstream to downstream, the detector elements are:
\begin{itemize}    
    \item Two Cherenkov detectors determine a positive and non-destructive identification of pions in the beam and, downstream of that, of kaons. The model for such detectors is the CEDAR developed for the CERN North Area experiments~\cite{Bovet:1975bx}. They are indicated as PionID and KaonID, respectively, and provide a precise determination of the beam particle times. Alternative schemes might be pursued, for example including a single ring-imaging Cherenkov detector instead of two separated CEDARs. The distance $D$ from the KaonID to the decay volume must be kept as small as possible to prevent limitations due to the knowledge of the kaon lifetime. A time resolution at the level of 200~ps is required.
    \item A fast beam tracker to determine the position, time, and three-momentum of each incoming beam particle. A possible implementation consists of three stations of silicon detectors: the first determines position and time of beam particles and is followed by a pair of dipoles with opposite polarities inducing a spatial offset which is a linear function of the momentum. The position measured by the second station provides a momentum measurement. A further dipole pair restores the beam particle to their original direction. Finally a third station provides the beam direction measurement. A time resolution at the level of 500~ps is required.
    \item A decay volume in vacuum including detectors to detect daughter particles emitted at large angles. The length $L$ of the decay volume is a crucial parameter, since it impacts on the number of observed events.
    \item A spectrometer system including gas-based tracking stations upstream and downstream of a large-bore dipole magnet, to determine position and direction of any charged decay daughter particle emitted. Centrally, it includes a non-instrumented region (``hole'') to allow passage of beam particles. In this study we considered the hole to be circular and 6~cm in radius. A relative momentum resolution at the level of 0.3\% is required.
    \item A calorimeter system with capability to detect and identify positrons, pions, and muons: it should include an electromagnetic calorimeter and a muon hodoscope following a passive material block thick enough to absorb any non minimum-ionizing particle. The electromagnetic calorimeter will host a non instrumented region to allow passage of the beam particles, here considered circular and with a radius of 6 cm. The outer radius of the calorimeter must be of the order of 1.1~m. The calorimeter system will participate in the online trigger system together with the KaonID/PionID systems, thus providing streams of data with a programmable number of daughter $\pi^+$, $\mu^+$, $e^+$. The trigger time jitter must be reduced to within 200~ps, to limit pile-up effects.
\end{itemize}

\subsection{Expected yield and sensitivity to leptonic width ratios}
\subsubsection{Intensity considerations}
We consider a beam intensity of $1.6\times10^{10}$ protons on target per second on spill, which is roughly a factor of 70 lower than that provided for a number of CERN North Area experiments~\cite{ANDRIEUX2022166069,NA62:2024pjp}. Considering spills with 4~s flat top, the flux of incoming $\pi^+$, $K^+$, and protons at the downstream detector per spill are: $4.6\times10^{7}$, $3.7\times10^{6}$, $4.8\times10^{6}$, respectively. 
Given such intensities, the setup proposed might not require to be installed in an underground area.

\subsubsection{Particle fluxes at the detector}
The flux of beam particles at the detector is 13~MHz, so that the probability of having an accidental beam particle to within 0.5~ns from the trigger time is 0.7\%, which should limit enough pileup effects. Without a sweeping system in the beam preparation section, the flux of muons from in-flight $K^+$ and $\pi^+$ decays upstream of the decay volume would be approximately $2.7\times10^{6}$ per spill (25\% of which from $K^+$ meson decays), corresponding to an accidental matching probability of 0.03\% in a 0.5~ns time window. If the secondary beam is taken at an angle larger than 6--7~mrad, the hard component of the muon flux from hadrons created in the most upstream collimator might be directed out of the acceptance of the detector, thus decreasing the needed sweeping power.

\subsubsection{Trigger considerations}
With a roughly 50\% acceptance, the expected number of decays within a 50~m long fiducial volume is $0.8\times10^6$ $\pi\to\mu\nu$ and $\approx0.3\times10^{6}$ $K\to\mu\nu$ per spill. As said, in addition to the in-flight decay in the decay volume, the muons from upstream decays amount to $2.7\times10^{6}$. The decays to $e^+\nu$ are obviously negligible in terms of trigger bandwidth. A two-level trigger system might be considered. The level zero would be based on logical conditions applied at the hardware level through field-programmable gate arrays (FPGAs) and would include the PionID, the KaonID and a cluster counting in the calorimeters and in the large-angle vetoes. The level-zero yield will flag around 4 million events per spill. The level one would be software based and should run a full event reconstruction, including the tracking algorithms both for the beam and the daughter particles. The computing request for the Level 1 operation is not negligible and its infrastructure and design will need to be carefully addressed in due time. Due to data-throughput and storage considerations, we will produce two different streams at the output level:
\begin{enumerate}
    \item Will store the data in electronic format (``raw'') for all the events with positron modes and a heavily scaled-down subset for the muon modes.
    \item Will store all the events with muon modes in a heavily-reduced, high-level format. The format will include the track parameters for the parent and daughter particles and only high-level information for the other detectors. 
\end{enumerate}
Additional trigger streams will select radiative decays or events for the search for heavy neutral leptons (HNLs), to complement the main physics goal of the experiment.
\subsubsection{Sensitivity expected for two years of data taking}
\label{sec:sensitivity_1_year}
\begin{figure}
    \centering
    \includegraphics[width=0.6\linewidth]{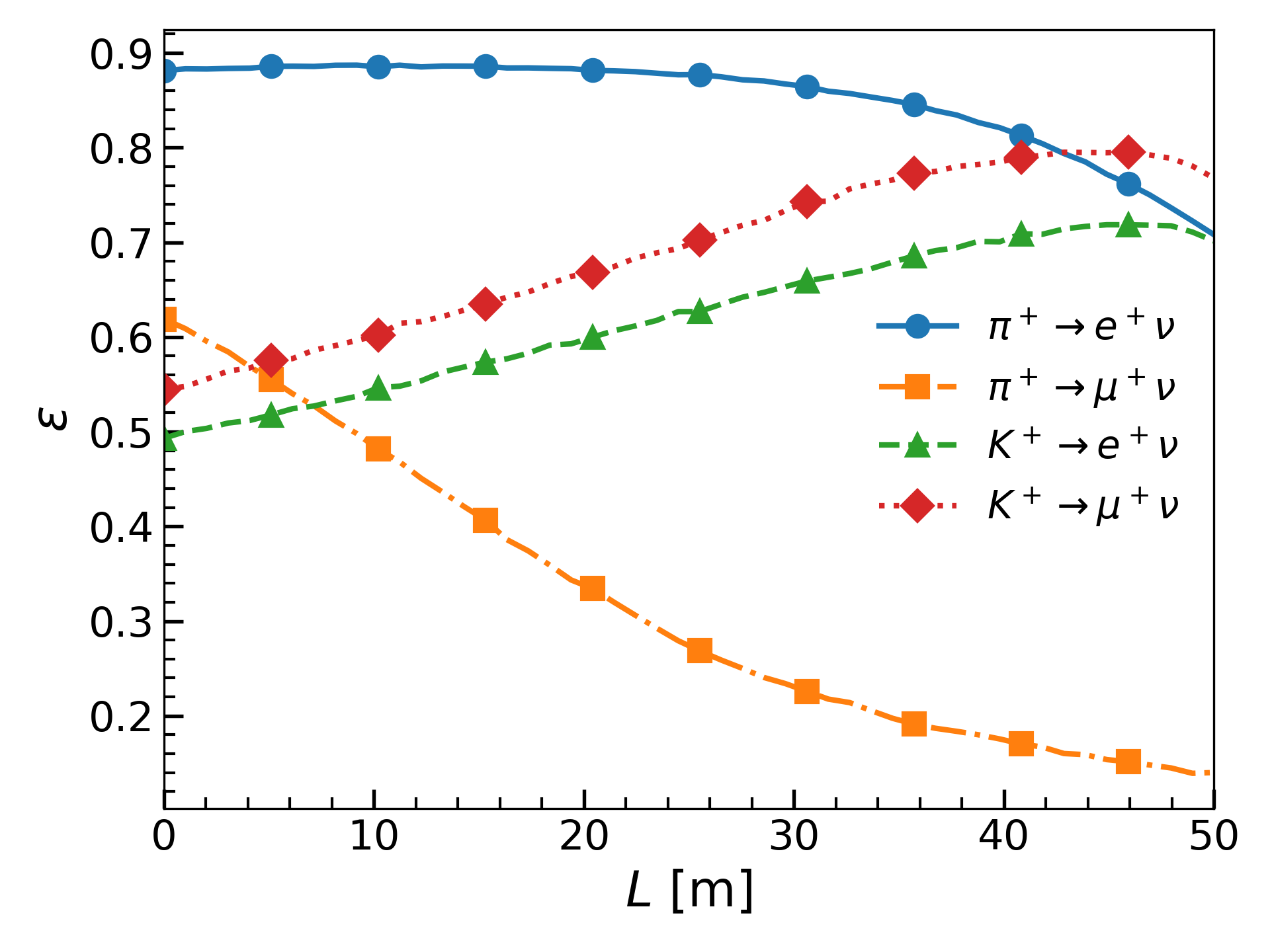}
    \caption{Geometrical acceptance of the leptonic decays of $K^+$ and $\pi^+$ as a function of the length of the fiducial volume $L$. For each of the configurations tested, the distance to the tracking spectrometer is the same: 51~m from the entrance of the fiducial volume to the first tracking station.}
    \label{fig:acceptances}
\end{figure}

In two years, we assume to collect 1.2 million spills corresponding to $9\times10^{13}$ tagged $\pi^+$ and $4.4\times10^{12}$ tagged $K^+$ mesons. The measurement of the decay widths for $K^+\to\ell^+\nu$ and $\pi^+\to\ell^+\nu$ and of the double ratio $R_{K\pi}^e/R_{K\pi}^\mu$ depend on the ratio of observed decays $N_{D,\mathrm{obs}}$ to those expected from the number of tagged mesons, $N_T$. The  number of observed decays is usually restricted to decay vertices within a ``fiducial'' volume used for analysis optimization:
\begin{equation}
    N_{D,\mathrm{avg}} = N_{T} \times p
\end{equation}
where $p$ is the probability for a beam particle to decay within the fiducial volume and its decay products to arrive in the sensitive volume of the tracker and the downstream forward detectors,
\begin{equation}
    p = e^{-D/\lambda}(1-e^{-L/\lambda})\epsilon\mathcal{B},
\end{equation}
where $D$ is the distance from the tagging detector (PionID, KaonID) to the entrance of the fiducial volume and $L$ is the length of the fiducial volume up to the decay, $\epsilon$ is the geometric acceptance (i.e. the probability that given a decay occurs in the fiducial volume, the products reach the sensitive volume of the tracker) and 
\begin{equation}
    \lambda = \gamma \beta c\tau = \frac{\gamma\beta\hbar c}{\Gamma}.
\end{equation}
In principle the length $L$ of the fiducial volume can be chosen differently for each of the four leptonic modes considered. Due to the different distributions of the lepton angle of emission, each of the acceptances for the four decay modes shows a different dependence on $L$, see Figure~\ref{fig:acceptances}.
A bayesian analysis has been performed, also accounting for the a-priori knowledge on the kaon and pion lifetime, $\Gamma_{K,\pi}$, as well as of the branching ratio $\mathcal{B}$ for each of the decay modes. The input parameters for $\mathcal{B}_\mathrm{obs}$ and $\Gamma_\mathrm{obs}$ are taken from the PDG {\it averages}~\cite{ParticleDataGroup:2026mpi}, with uncertainties already conservatively scaled up in case of internal inconsistency observed among the various measurements, see Section 5.2 of the Introduction from~\cite{ParticleDataGroup:2026mpi}. The input values used for the evaluation of the expected sensitivity are: $\Gamma_{K,\mathrm{obs}} = 5.317\times 10^{-17}$~GeV, $\Gamma_{\pi,\mathrm{obs}}=2.528\times 10^{-17}$~GeV, $\mathcal{B}_{\mathrm{obs}}(K^+\to\mu\nu)=0.636$, $\mathcal{B}_{\mathrm{obs}}(K^+\to e^+\nu)=1.582\times 10^{-5}$, $\mathcal{B}_{\mathrm{obs}}(\pi^+\to\mu\nu)=0.999877$, $\mathcal{B}_{\mathrm{obs}}(\pi^+\to e^+\nu)=1.23\times 10^{-4}$. 
The likelihood for the evaluation of the width for each decay mode is written as follows:
\begin{equation}
\label{eq:likelihood}
    \mathcal{L} = \mathcal{N}(N_{D,\mathrm{obs}}|N_{D,\mathrm{avg}},\sigma_{D}) \times \mathcal{N}(\mathcal{B}_\mathrm{obs}|\mathcal{B},\sigma_{B}) \times \mathcal{N}(\Gamma_\mathrm{obs}| \Gamma,\sigma_{\Gamma})
\end{equation}
where $\Gamma$ and $\mathcal{B}$ are the ``true'' total width of the parent particle and the ``true'' branching ratio for the channel considered and $\mathcal{N}$ stands for the normal distribution. They are nuisance parameters, to be estimated in the fit. The quantities $\sigma_{B}$ and $\sigma_{\Gamma}$ are considered fixed and taken to be equal to the PDG values for all considered particles and channels. The standard deviation of $N_{D}$ is then given by
\begin{equation}
    \sigma_{D} = \sqrt{N_{T}\times p\times(1-p)},
\end{equation}
following the normal approximation to the binomial distribution. When evaluating the double ratio $R$, ten terms enter the product of probabilities, related to the number of decay $N_{D}$ and branching ratio $\mathcal{B}$ for each mode, and the total widths $\Gamma_K,\pi$ for the parent.

We performed two evaluations of the expected uncertainties implied by the minimization of $-\log{\mathcal{L}}$. The default one is based on a numerical simulation of the decay paths and daughter particle emission and accounts for the geometry of the decay volume and of the tracker. An alternative analysis assumes that the acceptances $\epsilon$ are independent of the decay width and proceeds through the Fisher information matrix, defined as:
\begin{equation}
    I(\theta) = V^{-1}(\theta) = E\left[\left(-\frac{\partial^{2}\ln\mathcal{L}}{\partial \theta_{i}\partial\theta_j}\right)_{i,j}\right]
\end{equation}
where $\theta_{i}, i=1,..,N$ are the model parameters and $V$ is the covariance matrix. For the measurement of a single leptonic width $\Gamma_c$ for mode $c$, the information matrix is
\begin{equation}
    I(B,\Gamma) = V^{-1}(B,\Gamma) = \begin{pmatrix}
E\left[-\frac{\partial^2\ln L}{\partial B^2}\right] & E\left[-\frac{\partial^2\ln L}{\partial B\partial \Gamma}\right] \\
E\left[-\frac{\partial^2\ln L}{\partial B\partial\Gamma}\right] & E\left[-\frac{\partial^2\ln L}{\partial \Gamma^2}\right] 
\end{pmatrix}
\end{equation}

From the expression above the covariance matrix can be written as
\begin{equation}
\label{eq:covmat}
    V(B,\Gamma) = \frac{1}{\det I}\begin{pmatrix}
E\left[-\frac{\partial^2\ln L}{\partial \Gamma^2}\right] &
E\left[\frac{\partial^2\ln L}{\partial B\partial \Gamma}\right] \\
E\left[\frac{\partial^2\ln L}{\partial B\partial\Gamma}\right] & 
E\left[-\frac{\partial^2\ln L}{\partial B^2}\right]
\end{pmatrix} =
\begin{pmatrix}
\mathrm{var}(B) &
\mathrm{cov}(B,\Gamma) \\
\mathrm{cov}(B,\Gamma) &
\mathrm{var}(\Gamma)
\end{pmatrix}
\end{equation}
To first order, the variance of $\Gamma_c$ is given by
\begin{equation}
    \mathrm{var}(\Gamma_c) = \left(\frac{\partial \Gamma_c}{\partial B}\right)^2\mathrm{var}(B)+\left(\frac{\partial \Gamma_c}{\partial \Gamma}\right)^2\mathrm{var}(\Gamma) + 2\frac{\partial \Gamma_c}{\partial B \partial \Gamma} \mathrm{cov}(B,\Gamma).
\end{equation}

\begin{figure}
    \centering
     \includegraphics[width=0.45\linewidth]{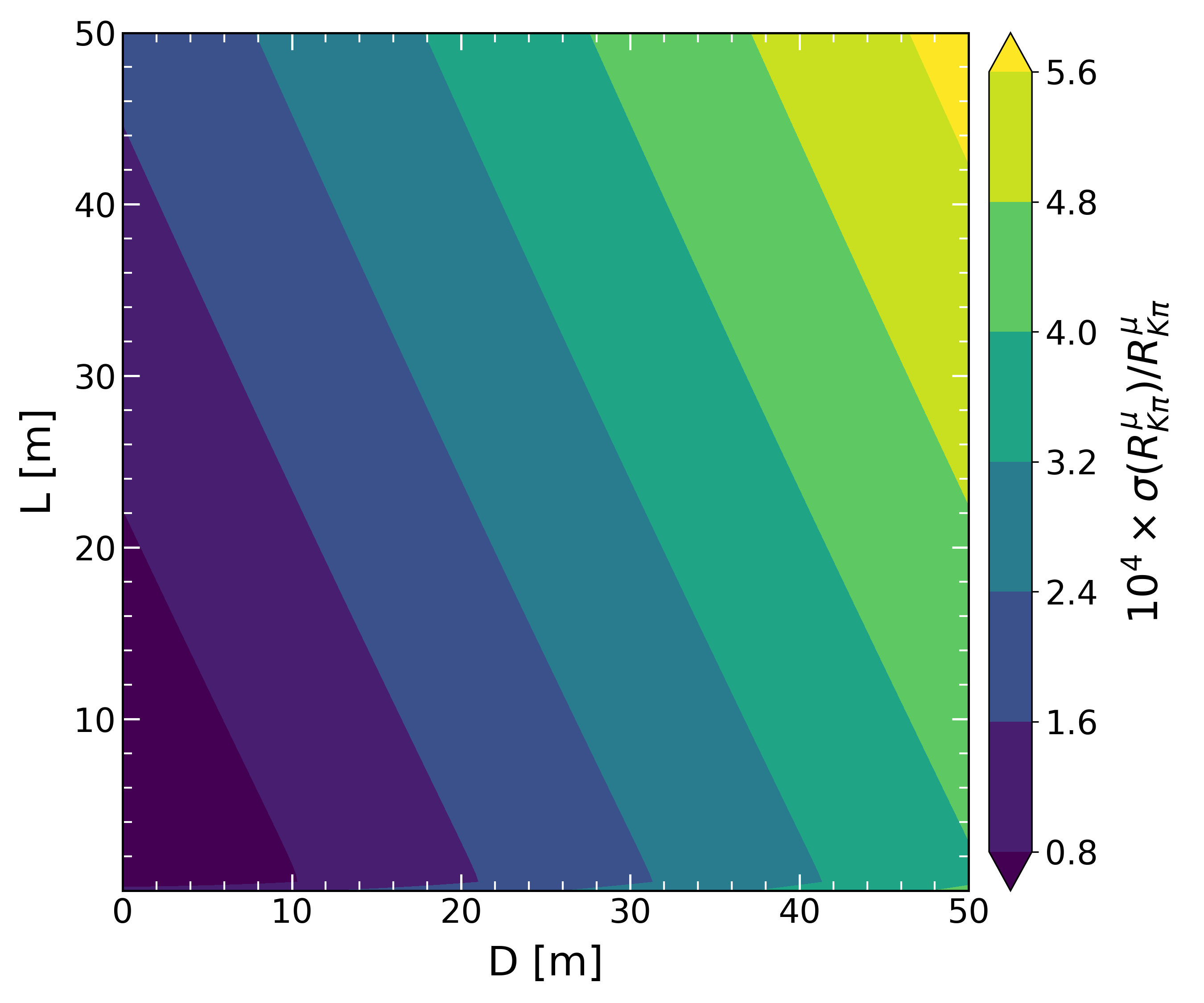}
    \includegraphics[width=0.45\linewidth]{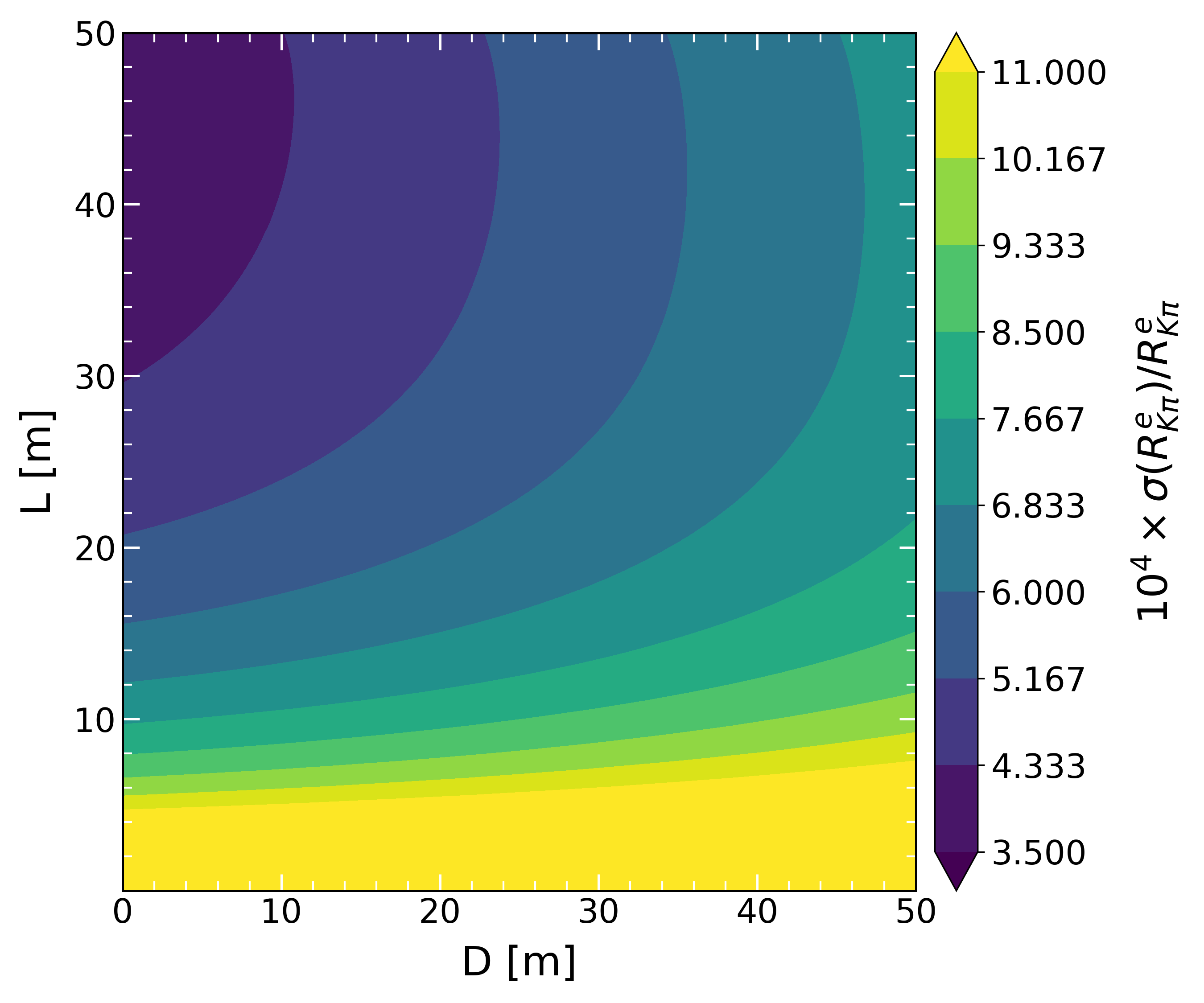}
    \caption{Expected relative uncertainty for $R_{K\pi}^\mu$ (left) and $R_{K\pi}^e$ (right) as a function of the distance $D$ from the tagging detector to the entrance of the decay volume and of the length $L$ of the fiducial volume. We assume to collect $9\times10^{13}$ tagged $\pi^+$ and $4.4\times10^{12}$ tagged $K^+$ mesons.}
    \label{fig:relerr_singleratio}
\end{figure}

\begin{figure}
    \centering
    \includegraphics[width=0.45\linewidth]{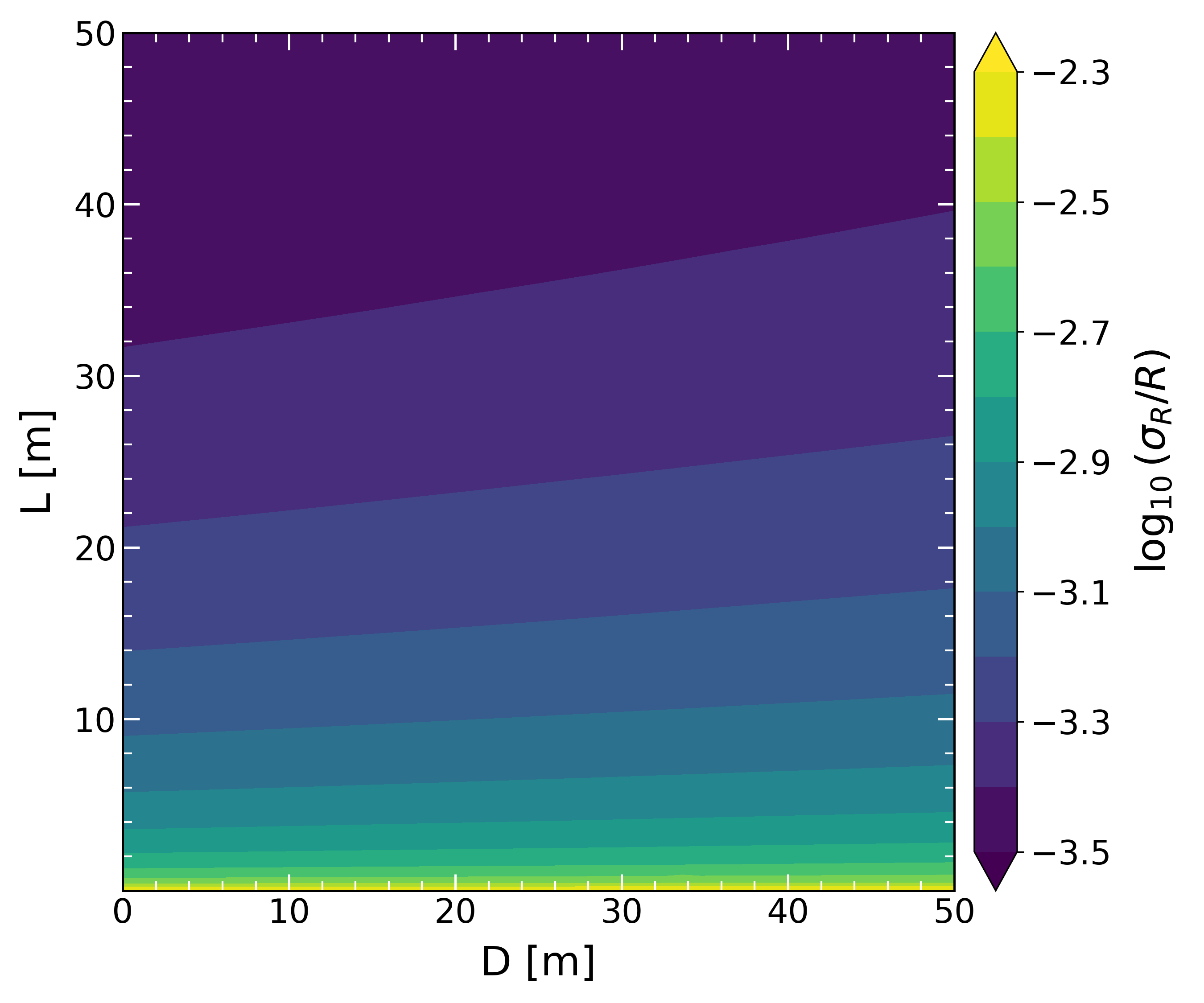}
    \includegraphics[width=0.45\linewidth]{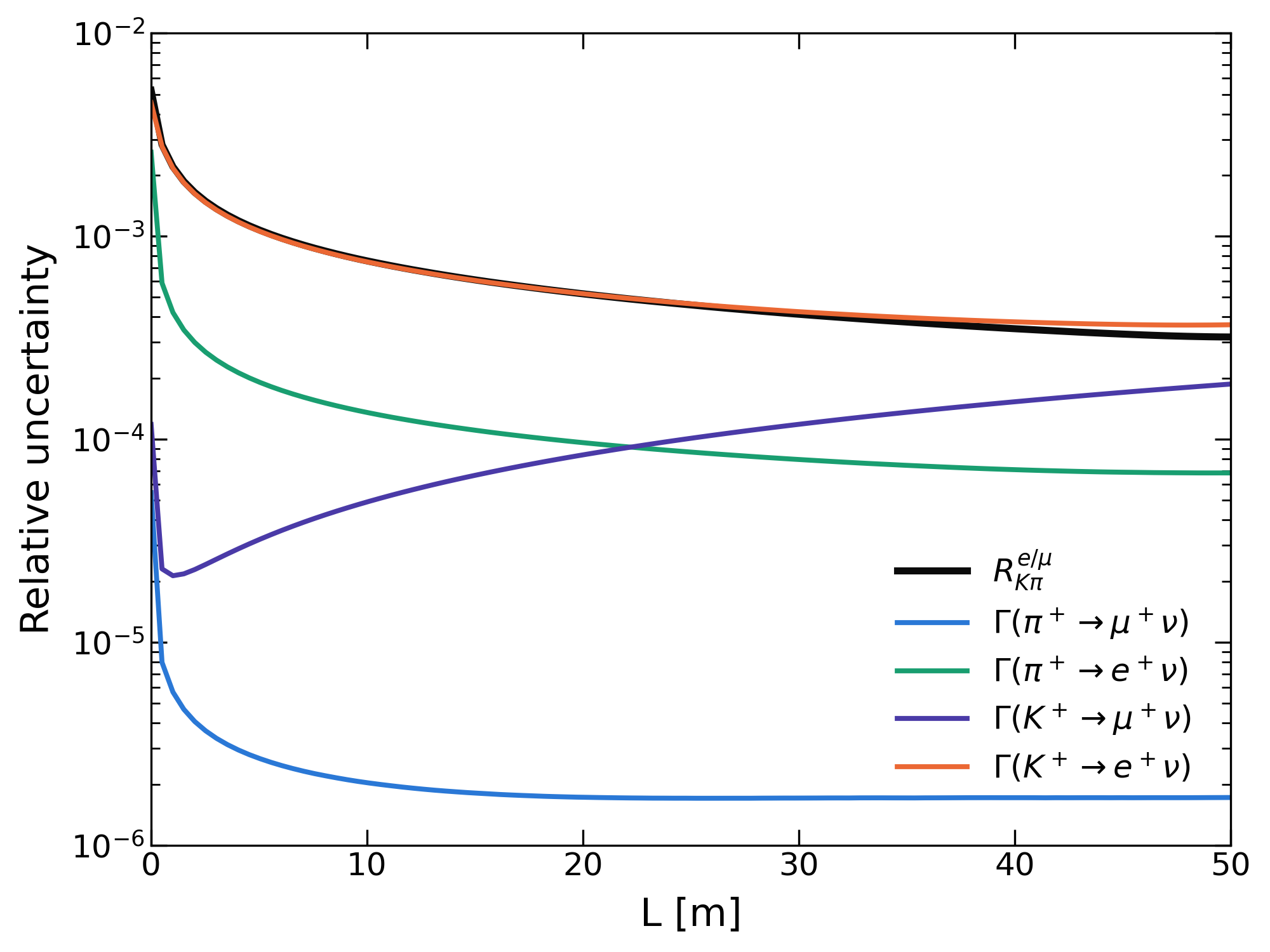}
    \caption{Left: expected relative uncertainty for $R_{K\pi}^e / R_{K\pi}^\mu$ as a function of the distance $D$ from the tagging detector to the entrance of the decay volume and of the length $L$ of the fiducial volume. We assume to collect $9\times10^{13}$ tagged $\pi^+$ and $4.4\times10^{12}$ tagged $K^+$ mesons (2 years of data taking). Right: the various contributors to the total relative uncertainty, as a function of $L$ (evaluated at $D=2$~m). In both plots, the minimum corresponds to a relative uncertainty of $3\times10^{-4}$. }
    \label{fig:relerr_doubleratio}
\end{figure}

The relative error of the partial width is 
\begin{equation}
    \frac{\sigma_{\Gamma_c}}{\Gamma_c} = \sqrt{\frac{\text{var}(B)}{B^2} +\frac{\text{var}(\Gamma)}{\Gamma^2}+2\frac{\mathrm{cov}(B,\Gamma)}{\Gamma_c^2}}.
\end{equation}
The results are:
\begin{itemize}
    \item At the statistical level, the ratio $R_{K\pi}^\mu$ can be measured below $10^{-4}$. The expected relative uncertainty is shown in the left panel of Figure~\ref{fig:relerr_singleratio} as a function of the distances $D$ and $L$. The optimal point is obtained with $D$ values as small as possible and $L$ values around 10~m. To achieve such a stringent level for the systematic uncertainty, we will modify the distributions of the daughter particles by weighting them so that the emitted muons from pions and kaons will overlap at the detector (tracking spectrometer, calorimeters). The available statistics is such that this optimization can be performed without limiting the measurement uncertainty to above the level of $10^{-4}$. 
    \item At the statistical level, the ratio $R_{K\pi}^e$ can be measured to the level of $3.5\times10^{-4}$ in two years. The expected relative uncertainty is shown in the right panel of Figure~\ref{fig:relerr_singleratio} as a function of the distances $D$ and $L$. The optimal point is obtained with $L$ values as large as possible, while limiting $D$ to below approximately 10~m. In this case, a cancellation of the systematic uncertainties is less expensive than for the muon measurements, given that the angle of emission of the daughter positrons is broadly distributed. 
    \item The uncertainty on the measurement of the double ratio $R_{K\pi}^e/R_{K\pi}^\mu$ is dominated by the statistics collected for the positron modes and in particular the $K^+\to e^+\nu$, see Figure~\ref{fig:relerr_doubleratio} (right). With $N_K=4.4\times10^{12}$, corresponding to 2 years of data taking, the minimum relative uncertainty is $3\times10^{-4}$. For a collected number of tagged kaons $N_K$ above $10^{11}$, the total relative error scales as $1/\sqrt{N_K}$, thus ensuring gains for extended data taking durations. It must be noted that, as expected, the dependence on $D$ is way weaker for the double ratio than for the single leptonic ratios.
    \item For the measurement of the kaon ratio $R_K^\nu$, the cancellation of experimental systematic uncertainties will exploit a weighting technique to maximise the overlap between the distributions of point of impact to the tracker between muons from $K\to\mu\nu$ and positrons from $K\to e\nu$. The particle identification efficiencies will remain to be evaluated separately for muons and positrons, which will be performed using control samples with the decay modes $K\to\pi^0\mu\nu$ and $K\to\pi^0 e\nu$, respectively. The precision goal achievable is at the level of $10^{-3}$.
\end{itemize}

\subsection{Other observables experimentally accessible}
Exploiting the full kinematic reconstruction of parent and daughter charged particles, the experiment will provide state-of-the-art sensitivities to the production of heavy neutral leptons (HNL, $N$) in a variety of flavour coupling patterns, via searches for the decay
\begin{equation}
    P\to \ell N,
\end{equation}
with $N$ undergoing invisible decays or escaping the detector without interaction and $P=K,\pi$. Dedicated trigger streams must be implemented for such searches with a careful optimization of the available bandwidth, both at Level 1 and for the trigger output.
The selection criteria for such channels are the same used for the $R_{K\pi}^{e,\mu}$ ratios and depend on the capability to efficiently reject the backgrounds. 

We expect to have large acceptances for the radiative modes as well, so that precision measurements connected to low-energy effective constants can be performed. 

\section{Theoretical considerations}\label{sec:th}

In this section we discuss theoretical aspects concerning the ratios $R_K^\nu$ and $R_{K\pi}^\ell$ as probes of heavy BSM physics.
We start by describing the low-energy parametrisation in terms of broken-phase effective operators, and how gauge symmetry restoration above the electroweak scale allows to connect to other decay channels or high-energy observables.
Working in the Standard Model Effective Field Theory, we want to study the impact of the proposed measurements in terms of the effectively probed scales, first considering one operator at a time, and then performing a combined fit to all kaon (and related) data.
An interesting aspect regarding the $R_{K\pi}$ ratios lies in tests of CKM unitarity in the first row, which currently shows a slight tension between experimental measurements of $R_{K\pi}^\mu$, $K_{\ell 3}$, and nuclear $\beta$ decays \cite{Seng:2021nar,CKMMatrixReview2026, 3161023}.
The tension can be lifted by a NP contribution in $R_{K\pi}$ \cite{Cirigliano:2023nol}, as we shall study in detail while providing future prospects from a $10^{-4}$-level measurement.

In addition to being very sensitive probes of new physics in kaon decays, $R_K^\nu$ can be used as a precise test of lepton flavour universality in the $\mu$-$e$ sector.
As such, it provides a complementary probe to the measurements in $b$ decays, among others.
Within a specific flavour assumption on the new physics couplings, $b$, $c$, and $s$ decays can all be related, which is what makes $R_K^\nu$ a very interesting measurement also in the context of heavy flavour.

Finally, the flavour of the neutrino in the final state of all the decays we consider cannot be determined, implying particularly that a new physics signal could involve a $\tau$ neutrino.
In this case, our ratios become also a probe of lepton flavour violating effects, to compare e.g. against $\tau\to K\ell$.

\subsection{Effective Field Theory approach}

Assuming that new physics lies well above the electroweak scale, an Effective Field Theory approach allows not only to parametrise all possible effects, but also to take into account the large scale separation between the new physics scale and the scale of the processes being considered.
The Standard Model Effective Field Theory (SMEFT) is well-suited to parametrise any effect above the electroweak scale, while below the mass of the $W$ and $Z$ bosons, effects are conveniently described in the Low-Energy Effective Field Theory (LEFT).

In the double ratio (or in $R_K^\nu$), we will always assume that new physics effects are present either in muons or electrons, separately, i.e. using it as a test of lepton-flavour universality.
The $R_{K\pi}$ ratios instead allow us to extract information about the CKM combination $|V_{us}|/|V_{ud}|$.

\subsubsection{LEFT}

The natural context to parametrise new physics effects in kaon decays is the Low-energy Effective Field Theory (LEFT).
This is the theory in which all heavy Standard Model fields ($Z$, $W$, $t$, and $h$) have been integrated out, electroweak symmetry is broken, and weak interactions are described by effective dimension-six four-fermion operators.
Normalising our Lagrangian as
\begin{align}
    \cL_{\rm LEFT} = \sum_i L_i O_i \,,
    \label{eq:LagrLEFT}
\end{align}
the LEFT operator structures relevant for our discussion are (using the notation and conventions from \cite{Jenkins:2017jig})
\begin{align}\label{eq:leftcoeffs}
    \begin{aligned}
    [O_{\nu edu}^{V,LL}]_{ij21} &= (\bar\nu_{L,i} \gamma_\mu e_{L,j})(\bar s_L \gamma^\mu u_L) \,, \qquad [O_{\nu edu}^{V,LR}]_{ij21} = (\bar\nu_{L,i} \gamma_\mu e_{L,j})(\bar s_R \gamma^\mu u_R) \,,  \\
    [O_{\nu edu}^{S,RR}]_{ij21} &= (\bar\nu_{L,i} e_{R,j})(\bar s_L u_R) \,, \qquad\qquad [O_{\nu edu}^{S,RL}]_{ij21} = (\bar\nu_{L,i} e_{R,j})(\bar s_R u_L) \,,
    \end{aligned}
\end{align}
where $i = 1,2,3$ and $j=1,2$, and we don't include the tensor operator, as it does not contribute to the purely leptonic decays under consideration.
As a function of the coefficients listed above, our observables can be expressed as
\begin{align}
    R_K^\nu &= \frac{\Gamma(K^+ \to \mu \nu)}{\Gamma(K^+ \to e \nu)} = \frac{m_{\mu}^2 \left(1 - \frac{m_{\mu}^2}{m_{K^+}^2}\right)^2}{m_{e}^2 \left(1 - \frac{m_{e}^2}{m_{K^+}^2}\right)^2}
    \left(1+\delta_{\rm ew}^{K\mu}-\delta_{\rm ew}^{Ke}\right)
    \times \dfrac{\displaystyle\sum_{i=1}^3 \left|C_{i221}\right|^2}{\displaystyle\sum_{i=1}^3 \left|C_{i121}\right|^2 } \,,
\end{align}
where 
\begin{align}
C_{ijmn}=
\left[L_{\nu edu}^{V,LL}-L_{\nu edu}^{V,LR} - \frac{m_{K^+}^2}{m_{\ell_j}(m_u+m_s)}\left(L_{\nu edu}^{S,RR} + 
L_{\nu edu}^{S,RL}\right)\right]_{ijmn},
\end{align}
from which one observes the customary chiral enhancement of the scalar contributions with respect to the SM (vector) one.
The $\delta_{\rm ew}^{K\ell}$ symbols indicate radiative and isospin-breaking corrections (see Appendix \ref{app:radcorr}).
The ratio $R_{K\pi}^\ell$, on the other hand, is given by
\begin{align}
    R_{K\pi}^\ell =& \frac{\Gamma(K^+ \to \ell^+_j \nu)}{\Gamma(\pi^+ \to \ell^+_j \nu)} =  \frac{f_{K^+}^2 m_{K^+} \left(1 - \frac{m_{\ell_j}^2}{m_{K^+}^2}\right)^2}{f_{\pi^+}^2 m_{\pi^+} \left(1 - \frac{m_{\ell_j}^2}{m_{\pi^+}^2}\right)^2} \left(1+\delta_{\rm ew}^{K\ell}-\delta_{\rm ew}^{\pi\ell}\right)\times  \dfrac{\displaystyle\sum_{i=1}^3 \left|C_{ij21}\right|^2}{\displaystyle \left|L_{\nu e du,\,{\rm SM}}^{V,LL}\right|^2 }
\end{align}
where $L_{\nu e du,\,{\rm SM}}^{V,LL}= V_{ud}\, G_F/\sqrt2$ stands for the SM contribution, assuming thereby no new physics contribution to the pion.

\subsubsection{SMEFT}
Above the electroweak scale, we define the SMEFT Lagrangian to be
\begin{align}
    \Lagr_{\rm SMEFT} = \sum_{i} \cC_i \Order_i \,,
\end{align}
where we use the Warsaw conventions for the effective operators \cite{Grzadkowski:2010es} (see Table \ref{tab:smeftops}).
The matching of the LEFT coefficients in Eq. \eqref{eq:leftcoeffs} is given by \cite{Jenkins2018}
\begin{align}
    [L_{\nu edu}^{V,LL}]_{prst} + h.c. &= 2 [\cC_{lq}^{(3)}]_{prst}- \frac{2}{v^2} (\delta_{pr}+v^2 [\cC_{Hl}^{(3)}]_{pr}) (\delta_{ts}+v^2 [\cC_{Hq}^{(3)}]_{ts}) \\
    [L_{\nu edu}^{V,LR}]_{prst} +h.c. &=  - \frac{2}{v^2} (\delta_{pr}+v^2 [\cC_{Hl}^{(3)}]_{pr}) (\frac{1}{2} v^2 [\cC_{Hud}]_{ts} )\\
    [L_{\nu edu}^{S,RL}]_{prst} &=  \,[\cC_{ledq}]_{prst} \\
    [L_{\nu edu}^{S,RR}]_{prst} &=  \,[\cC_{lequ}^{(1)}]_{prst},
\end{align}
in the weak eigenstate basis. 
In order to match to the mass basis below the electroweak scale, an assumption has to be made in SMEFT about the alignment of the left-handed quark fields inside the doublets $q_i$.
Two common choices are up- and down alignment, i.e. the limit cases where the up quarks or respectively the down quarks are in the mass eigenstates.
The other component is then given by a superposition of mass eigenstates, weighted by the CKM matrix elements as follows:
 \begin{align}
     q^{\text{up}}_i= \begin{pmatrix} u_i  \\ (V d)_i\end{pmatrix}\,, \qquad q^{\text{down}}_i= \begin{pmatrix}(V^\dagger u)_i  \\d_i\end{pmatrix} \,.
 \end{align}
 If, for example, we want to match the coefficient $[L_{\nu edu}^{S,RL}]_{pr21}$ onto the SMEFT in the down-aligned basis, we would get
 \begin{align}
     [L_{\nu edu}^{S,RL}]_{pr21}^{\text{mass}} = \sum_{i=1}^3 V^*_{1i} [\cC_{ledq}]_{pr2i}^{\text{down}} \,,
 \end{align}
 and analogous expressions for all other coefficients.

\subsection{SMEFT analysis}

\begin{table}[]
    \centering
    \renewcommand{\arraystretch}{1.3}
    \begin{tabular}{|cl|cl|}
        \hline
        $\cO_{\ell edq}$ & $=(\bar\ell e)(\bar d q)$ & $\cO_{H\ell}^{(3)}$ & $=(H^\dagger i \overleftrightarrow{D}_\mu^I H)(\bar \ell \gamma^\mu \sigma^I \ell)$ \\
        $\cO_{\ell equ}^{(1)}$ & $=\varepsilon_{ij}(\bar\ell^i e)(\bar q^j u)$ & $\cO_{Hq}^{(3)}$ & $=(H^\dagger i \overleftrightarrow{D}_\mu^I H)(\bar q \gamma^\mu \sigma^I q)$ \\
        $\cO_{\ell q}^{(3)}$ & $=(\bar \ell \gamma_\mu \sigma^I \ell)(\bar q \gamma^\mu \sigma^I q)$ & $\cO_{Hud}$ & $=(H^\dagger i \overleftrightarrow{D}_\mu \widetilde{H})(\bar u \gamma^\mu d)$ \\ \hline
    \end{tabular}
    \caption{SMEFT operators entering our analysis. Quark indices are always chosen to contain an $\bar s \Gamma u$ current, while lepton indices are varied. The covariant derivatives appearing are $\overleftrightarrow{D}_\mu = \overrightarrow{D}_\mu - \overleftarrow{D}_\mu$, and $\overleftrightarrow{D}_\mu^I = \overrightarrow{D}_\mu \sigma^I -\sigma^I \overleftarrow{D}_\mu$.}
    \label{tab:smeftops}
\end{table}

In this section we outline the connections of our observables with other relevant observables, in the flavour sector and at higher energies. 
Details about the individual flavour observables are given in Appendix \ref{app:obs}.
Taking SMEFT (at the electroweak scale) as a starting point, we consider the operators giving an effect in leptonic kaon decays, and match them to the LEFT.
Considering for the moment only the ratio $R_K^\nu$ (or equivalently the double ratio $R_{K\pi}^{\mu/e}$, as discussed in Section \ref{sec:observables}), there are a total of 5 relevant coefficients, barring flavour multiplicities on the lepton side and quark-flavour alignment:
\begin{align}
    [\cC_{\ell e d q},]_{ij21} \quad [\cC_{\ell equ}^{(1)}]_{ij21}, \quad [\cC_{\ell q}^{(3)}]_{ij12}, \quad
    [\cC_{H\ell}^{(3)}]_{ij}, \quad [\cC_{Hq}^{(3)}]_{12} \,.
\end{align}
Our procedure will be the following: for each of these operators, we identify the LEFT coefficients they match onto, and then proceed to identify the most relevant observable for each of them.
We do this assuming separately the up- and down-alignment of the left-handed quark doublet, as these two limiting cases lead to different and distinctive phenomenology.
As we consider SMEFT coefficients defined at the electroweak scale, we only account for RG evolution within the LEFT, between the electroweak scale and the relevant scale of the process, using {\tt DsixTools} \cite{Fuentes-Martin:2020zaz}.
The only numerically relevant running effects consist of QCD-induced self-renormalisation of scalar operators.
For each process, we extract a lower bound on the new physics scale as
\begin{align}
    \Lambda_{\rm NP\,,i} = \rm max \left\{|\cC_i^{\rm min}|^{-1/2},|\cC_i^{\rm max}|^{-1/2}\right\} \,,
\end{align}
where $[\cC_i^{\rm min}, \cC_i^{\rm max}]$ is the 95\% C.L. interval for the coefficient, assuming only one coefficient is switched on at a time.
The bounds are collected in the form of bar plots in Figures \cref{fig:ledq_bounds}, and the numerical results in tables in Appendix \ref{app:plots}.
This procedure will allow us to
\begin{enumerate}
    \item identify whether some observables may already be giving a stronger constraint than our future projection for $R_K^\nu$, effectively allowing us to exclude possible new physics scenarios;
    \item in the opposite case, determine which observables/scenarios can be indirectly constrained by measuring leptonic kaon decays precisely.
\end{enumerate}

\subsubsection*{$\cC_{\ell edq}$}

\begin{figure}
    \centering
    \includegraphics[width=\textwidth]{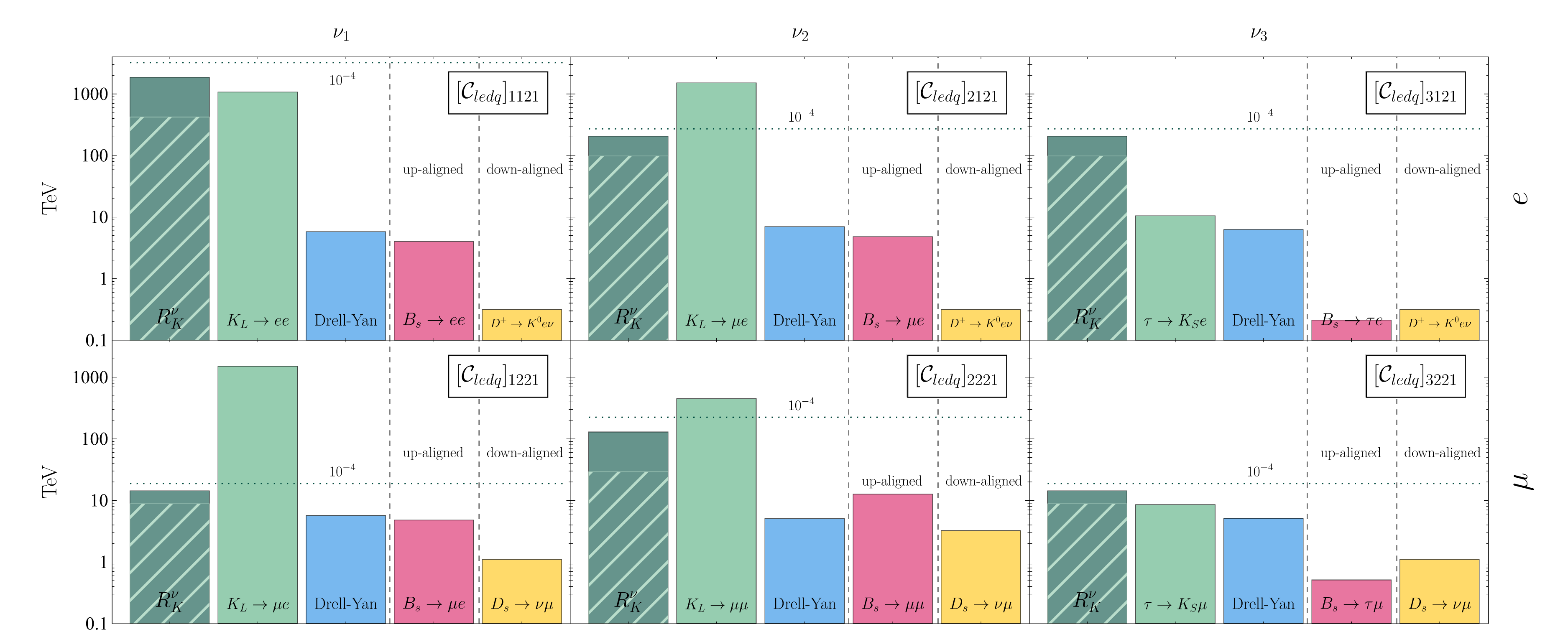}
    \caption{Bounds on the NP scale from the different flavour processes connected to $R_K^\nu$ under the assunmption of NP in $\cC_{\ell edq}$. For $R_K^\nu$ the hatched bars represent the bound from the current measurement, the solid bars the future projection with a precision of $3\times 10^{-4}$, and the dotted line the optimistic scenario of $1\times 10^{-4}$.}
    \label{fig:ledq_bounds}
\end{figure}

We start from $\cC_{\ell edq}$. At low energies, it gives a contribution to both charged and neutral currents, since
\begin{align}
    (\bar\ell e)(\bar d q) = (\bar \nu_L e_R)(\bar d_R u_L) + (\bar e_L \bar e_R)(\bar d_R \bar d_L) \,,
\end{align}
where we have suppressed flavour indices.
Looking in particular at $[\cC_{\ell edq}]_{1121}$, for example, we will have both a scalar $\nu e s u$ and a $e e s d$ operator, which would give a chirally-enhanced contribution to $K^+\to e^+ \nu_e$ and $K_L\to ee$ respectively.
The bounds on the new physics scale are shown in Figure \ref{fig:ledq_bounds}.
As can be seen, $K_L$ decays are today the strongest probe of $\cC_{\ell edq}$, with our projections for $R_K^\nu$ slightly surpassing $K_L\to ee$ ($K_L\to\mu e$ and $K_L\to\mu\mu$ will remain the dominant constraints).
On the other hand, if the neutrino is assumed to be a $\tau$ neutrino, the $SU(2)_L$ rotation changing charged into neutral currents gives an expected effect in $\tau\to K\ell$, for which $R_K^\nu$ is already now a dominant or competitive constraint (depending on the lepton flavour), and even more so with a future measurement.
Moreover, depending on the alignment of the quark doublet, there can be additional effects in $B_s\to \ell\ell^{(')}$, or $D$ decays of the form $D\to K \ell\nu$ or $D\to\ell\nu$.
These constraints are comparably weaker than the ones from kaon physics, which is to be expected given the additional CKM suppression, but they show an interesting interplay between $K$, $B$, and $D$ physics, which is unavoidable to some degree.
Below we will discuss these interplays in a more concrete setting.
Finally, it should be mentioned that for all flavour combinations high-$p_T$ Drell-Yan tails also offer complementary constraints~\cite{Allwicher:2022gkm,Allwicher:2022mcg}.

\subsubsection*{$\cC_{\ell equ}^{(1)}$}

\begin{figure}[t]
    \centering
    \includegraphics[width=\linewidth]{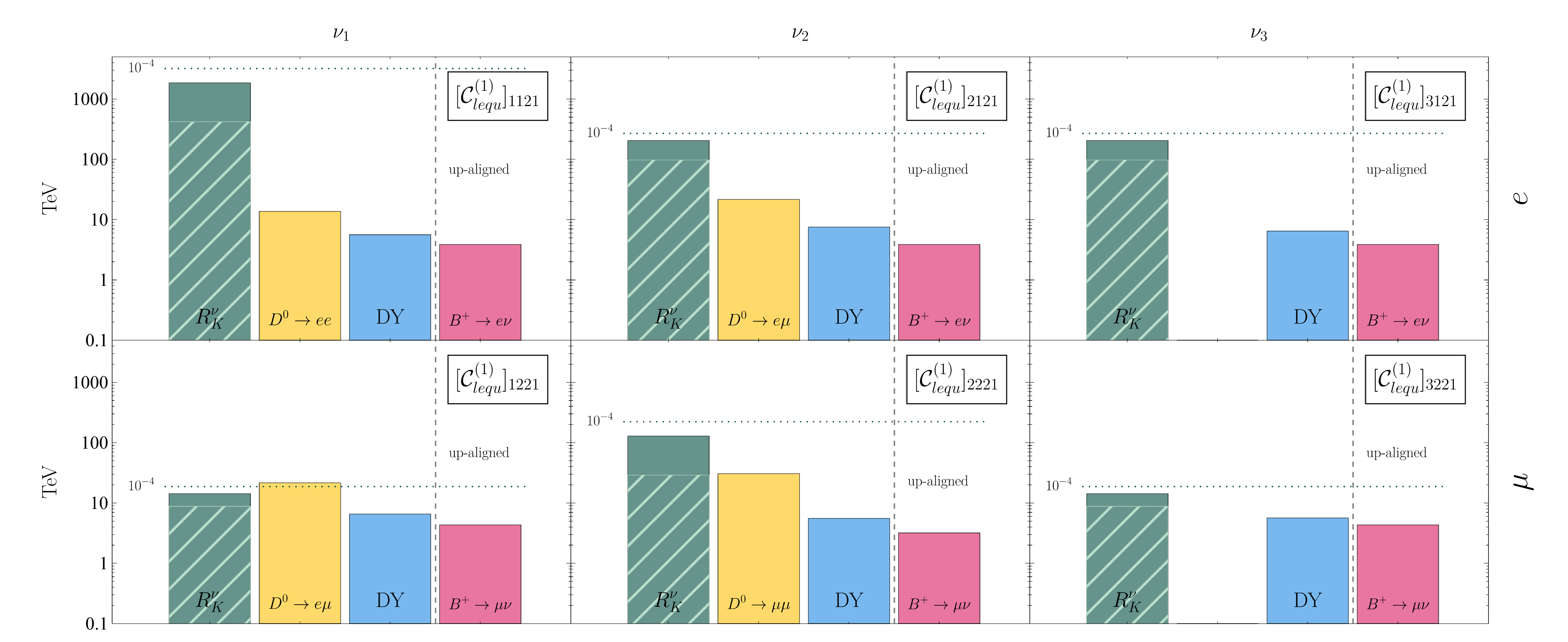}
    \caption{Bounds on the NP scale from the different flavour processes connected to $R_K^\nu$ under the assunmption of NP in $\cC_{\ell equ}^{(1)}$. For $R_K^\nu$ the hatched bars represent the bound from the current measurement, the solid bars the future projection with a precision of $3\times 10^{-4}$, and the dotted line the optimistic scenario of $1\times 10^{-4}$.}
    \label{fig:lequ1_bounds}
\end{figure}

Moving to the other scalar operator in the game, the picture is somewhat simpler in this case.
Similarly to before, we can expand out the quark and lepton currents to get
\begin{align}
    (\bar\ell e) \varepsilon (\bar q u) = (\bar \nu_L e_R) (\bar d_L u_R) - (\bar e_L e_R) (\bar u_L u_R) \,,
\end{align}
such that the charged-current kaon decays are now naturally linked to FCNCs in the charm sector, and through a flavour rotation in the case of up alignment, to charged current $B$ decays.
Given the scalar structure of the operator, the more significant constraints always come from purely leptonic decays, such as $D^0\to\ell\ell$, or $B^+\to \ell\nu$ (cf. Figure \ref{fig:lequ1_bounds}), but the kaon constraints tend to dominate the picture.
High-$p_T$ Drell--Yan measurements again offer complementary constraints, which are weaker than those from $R_K^\nu$.

\subsubsection*{$\cC_{\ell q}^{(3)}$}

The case of the semileptonic vector operator is the more involved one, since
\begin{align}
\begin{aligned}
    (\bar\ell\gamma_\mu \sigma^I \ell) (\bar q\gamma^\mu \sigma^I q) &= 2 (\bar \nu_L \gamma_\mu e_L)(\bar d_L \gamma^\mu u_L) + \text{ h.c.} \\
    &+ (\bar\nu_L \gamma_\mu \nu_L)\left[(\bar u_L \gamma^\mu u_L) - (\bar d_L \gamma^\mu d_L)\right] \\
    &+ (\bar e_L \gamma_\mu e_L)\left[(\bar d_L \gamma^\mu d_L) - (\bar u_L \gamma^\mu u_L)\right] \,.
\end{aligned}
\end{align}
Regarding connections in the neutral current kaon decays, our ratio $R_K^\nu$ is now linked to both dilepton and dineutrino final states.
For muon and electron neutrinos in $K\to\ell\nu$, the comparison with $K_L\to\ell\ell^{(')}$ follows the same pattern as for the $\cC_{\ell edq}$ case.
In the case of $K_L\to ee$, however, $K^+\to\pi^+\nu_e\nu_e$ provides an even stronger constraint, confirming the naive expectation that when both charged and neutral currents are present, the latter are usually the most sensitive to new physics contributions (cf. Figure \ref{fig:lq3_bounds}).
This also applies to the lepton-flavour violating case with tau leptons, where $K\to\pi\nu\nu$ is stronger than $\tau\to K_S\ell$. 
Another interesting observation is that there is not a large difference in sensitivity in $K\to\pi\nu\bar\nu$ between the lepton-flavour conserving and the violating case.
This is due to the fact that, even in the flavour-conserving (i.e. interfering with the SM) case, the fit is dominated by the new physics contribution, which tends to be larger than the SM and have opposite sign, thus resulting again in a SM-like central value.
This sort of double-solution behaviour is due to the current experimental and theoretical precisions, and has been already observed in previous works \cite{Allwicher:2026loe,Allwicher:2024ncl}.
The chiral enhancement also falling short in this case, Drell--Yan now plays a more important role as well, competitive with kaon decays in many cases.
We also show the related constraints in the charm sector, led by $D\to\ell\nu$.
These are however comparably weaker.
Similarly to the cases above, contributions from other flavour observables due to left-handed quark alignment lead to subleading contributions, and we choose to not show them here for brevity.
The corresponding plots are however reported in Appendix \ref{app:plots}.

\subsubsection*{$\cC_{H\ell}^{(3)}$}

Since $\cC_{H\ell}^{(3)}$ doesn't affect any quark current, its effect is essentially a lepton flavour (universality) breaking one.
As we can see in Figure \ref{fig:Hl3_bounds}, in the case of flavour-conserving currents, $R_K^\nu$ provides a better probe than current electroweak precision measurements.
However it is important to remark that this is the only case where the single ratio $R_K^\nu$ differs from the double ratio $R_{K\pi}^{\mu/e}$ as a probe of new physics.
Indeed, the double ratio (which is the measurement proposed in this work) would be insensitive to $\cC_{H\ell}^{(3)}$, as the lepton couplings systematically cancel out.
To recover the sensitivity to the lepton couplings, an improved measurement of the ratio $R_{\pi}^\nu$ is necessary.
We postpone a more detailed discussion of the issue to Section \ref{sec:lfu} in the context of lepton flavour universality tests.
Lepton-flavour violating indices, on the other hand, are quite effectively probed by $\mu\to 3e$ or $\tau\to 3\ell$ decays occurring through a virtual $Z$ exchange.

\subsubsection*{$\cC_{Hq}^{(3)}$}

Finally, the case of $\cC_{Hq}^{(3)}$ is conceptually different.
Since only the $W$ couplings to quarks are affected in this case, the ratio $R_K^\nu$ (or, equivalently, the double ratio $R_{K\pi}^{\mu/e}$) is not affected, and we need to compare our prospects for $R_{K\pi}^\ell$ with complementary flavour observables.
Drell-Yan in this case is not a very sensitive probe, as the energy-enhancement typical of four-fermion operators falls short in this case, as the SM Drell-Yan distribution is only shifted by a small constant term.
In the case of flavour observables, there is a direct connection with neutral-current kaon decays such as $K_L\to\mu\mu$ and $K\to\pi\nu\nu$, which are both slightly stronger than $R_{K\pi}^\mu$ (see Figure \ref{fig:Hq3_bounds}).

\subsubsection*{Global kaon fit}

\begin{figure}
    \centering
    \includegraphics[width=0.85\linewidth]{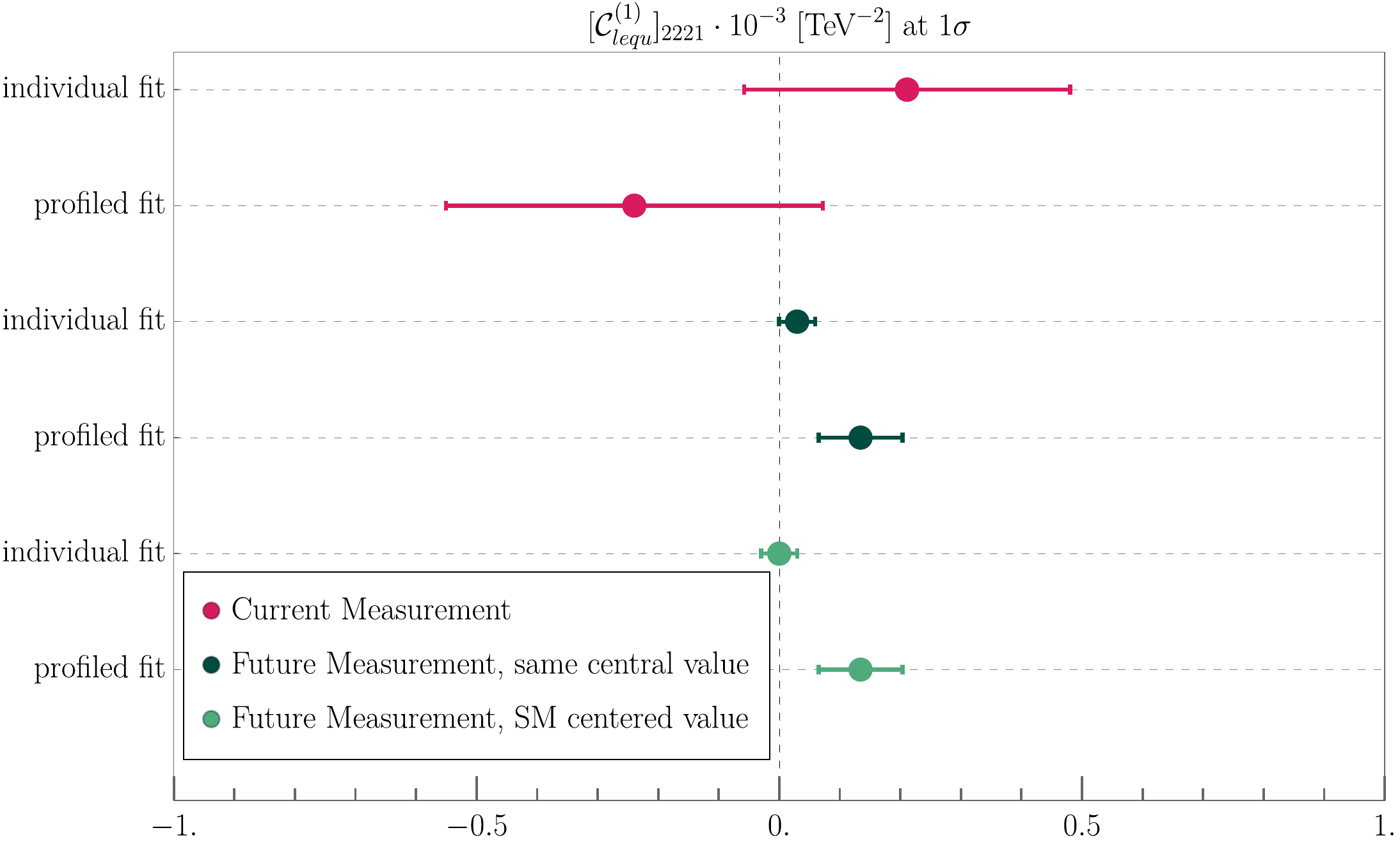}
    \caption{Best-fit interval for the SMEFT coefficient $[C_{\ell equ}^{(1)}]_{2221}$. Profiled fits are obtained by floating all SMEFT coefficients in Table \ref{tab:smeftops}.}
    \label{fig:lequ1Prec}
\end{figure}

\begin{figure}
    \centering
    \includegraphics[width=0.8\linewidth]{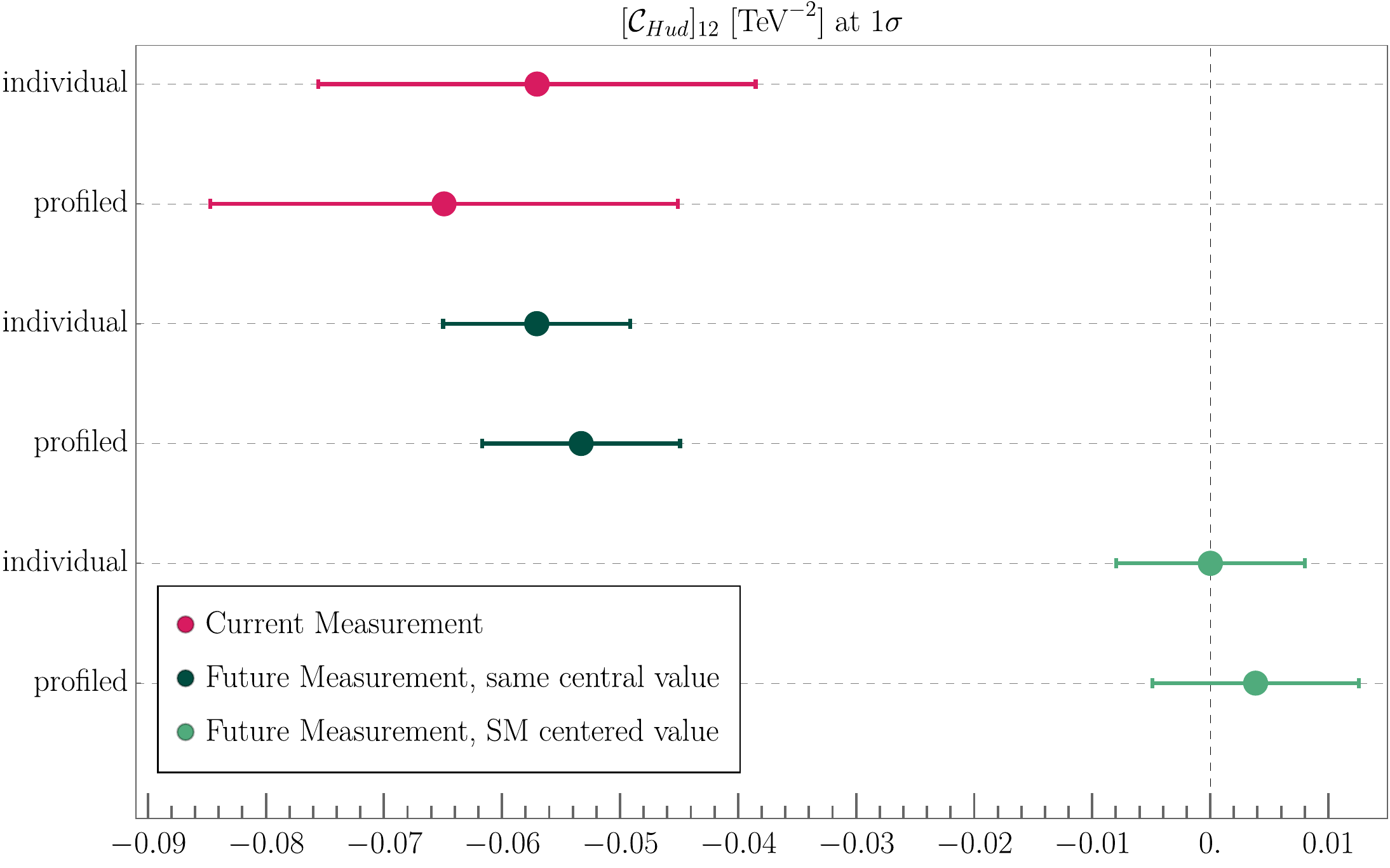}
    \caption{Best-fit interval for the SMEFT coefficient $[C_{Hud}]_{12}$. Profiled fits are obtained by floating all SMEFT coefficients in Table \ref{tab:smeftops}.}
    \label{fig:HudPrec}
\end{figure}

Having studied what happens in the case of individual coefficient fits, we now show how the constraints change assuming all of the five above coefficients are switched on at a time.
This will allow us to establish to which degree the combination of kaon data, together with their SMEFT-related processes, can give us insights about generic new physics affecting the kaon sector.
We do so assuming the down-aligned basis, although given the CKM suppression in heavier quark decays we find that the choice of alignment has no impact on the fit.
Regarding leptonic indices, we assume new physics in muons, but very similar results can be obtained in the electron case.
We show the fit results for two relevant coefficients in Figures \ref{fig:lequ1Prec} and \ref{fig:HudPrec}, assuming two different central values for our future projections of $R_{K\pi}^\mu$: once keeping the current central value, and once assuming the future measurement to be SM-like. In both cases we assume an improved theory error by a factor of two (see also following section).
Results are shown as 1$\sigma$-allowed intervals by profiling over all other coefficients in Table \ref{tab:smeftops}, and compared to the individual fits.
Overall, the fit converges with no flat directions (cf. also the full result reported in Appendix \ref{app:plots}), and we see a significant improvement in the sensitivity, especially for the scalar operator.
The tension in $\cC_{Hud}$ is related to the current anomaly in first-row CKM unitarity tests, and is discussed in further detail in the next Section.
The associated effective new physics scales in the global fit range from $\cO(10)$ to about 50 TeV.

\subsubsection*{$\cC_{Hud}$ and first-row unitarity}

\begin{figure}
    \centering
    \includegraphics[width=0.9\linewidth]{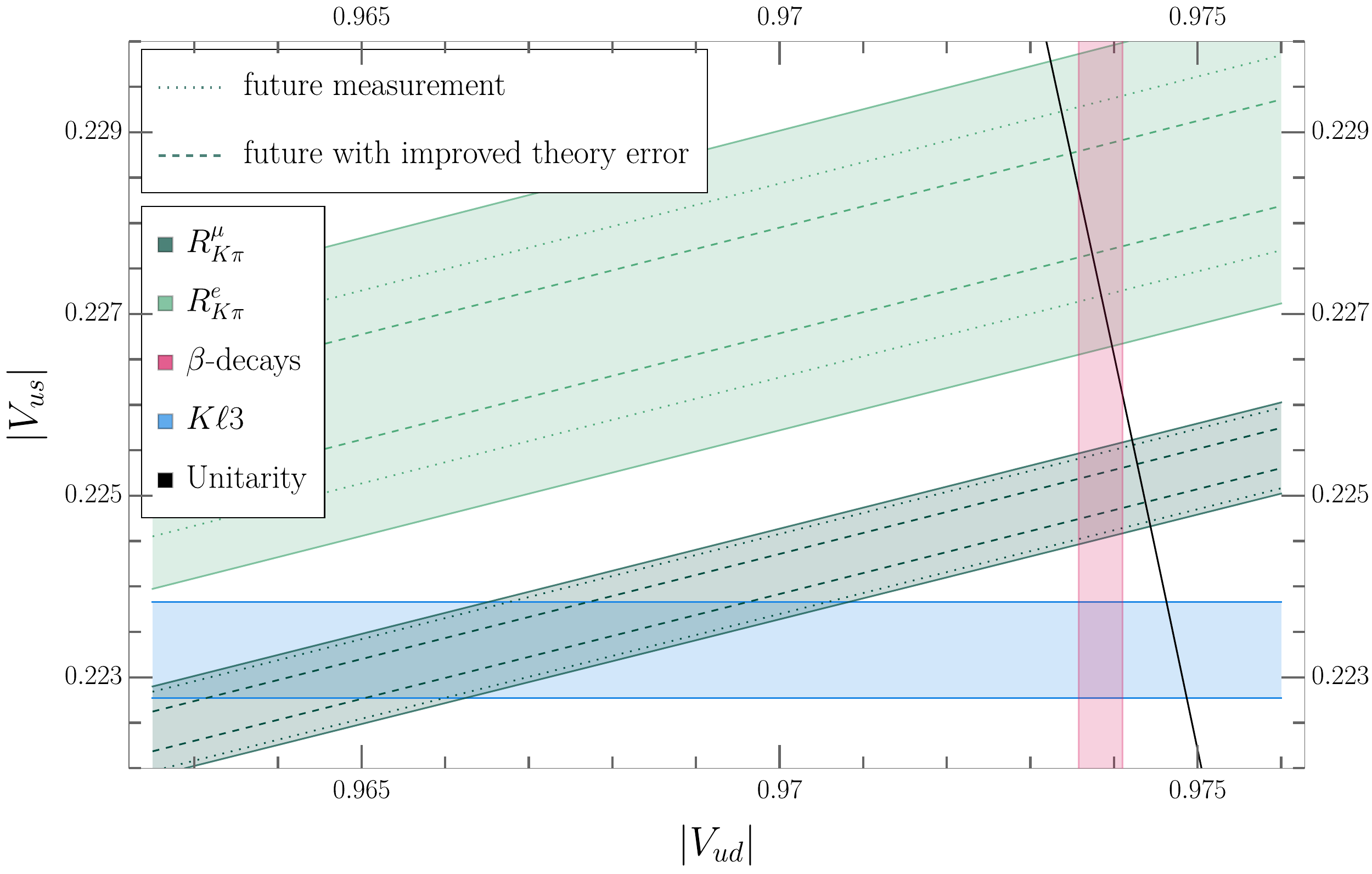}
    \caption{Current status of first-row CKM measurements, and future projection assuming current central value and our expected uncertainty for $R_{K\pi}^\mu$ and $R_{K\pi}^e$. The black line indicates the expectation from unitarity. The value for $K_{\ell 3}$ decays is taken from \cite{Cirigliano:2022yyo}, while the $\beta$-decay determination refers to a combination of neutron- and superallowed decays from \cite{Cirigliano:2022yyo}.}
    \label{fig:CKMTrianglePlot}
\end{figure}

An important precision measurement of the SM is the test of CKM unitarity \cite{10.1093/ptep/ptac097,Seng:2021nar,CKMMatrixReview2026, Eboli:2025vks, Crivellin:2021bkd, 3161023}. In the first row of the matrix this translates to 
\begin{align}
    |V_{ud}|^2+|V_{us}|^2+|V_{ub}|^2 = 1\,,
\end{align}
which is probed to very high precision. $V_{ud}$ and $V_{us}$ are determined by nuclear $\beta$ and kaon decays. 
As the current precision of those is one order of magnitude larger than the value of $|V_{ub}|^2 \approx 1.5 \times 10^{-5}$ \cite{10.1093/ptep/ptac097}, the latter can be safely neglected in our analysis, i.e. we will consider the unitarity constraint to be $|V_{ud}|^2 + |V_{us}|^2 = 1$.
The current measurements of kaon and $\beta$-decays show a deviation of $2-2.5 \sigma$ from 1 for this quantity. 

For $V_{ud}$ we will use the  combined value of super-allowed nuclear $\beta$ decays and neutron decays.
The value of $|V_{us}|$ instead can be estimated by kaon decays. One option is through the measurement of $K_{\ell 3} := K \rightarrow \pi \ell \nu_\ell$, using the form factors provided by lattice QCD \cite{FlavourLatticeAveragingGroupFLAG:2024oxs,Aoki:2017spo,FermilabLattice:2018zqv,Carrasco:2016kpy,Ishikawa:2022ulx}. In addition to that the decay $K\rightarrow \ell \nu$ through the ratio $R_{K\pi}^\ell$ allows the determination of $|V_{us}/V_{ud}|$. Using the current measurements of all the decays results in \cite{Cirigliano:2022yyo} (see also \cite{3161023,CKMMatrixReview2026})
\begin{align}
    |V_{ud}^\beta| &= 0.97384(26) \\
    |V_{us}^{K\ell 3}| &= 0.22330(35)_{\text{exp}}(39)_{f_+}(8)_{\text{IB}} = 0.22330(53)_{\text{total}} \\
    [|V_{us}|/|V_{ud}|]^{R_{K\pi}^\mu} &= 0.23108(23)_{\text{exp}}(42)_{F_K/F_\pi} (16)_{\text{IB}} = 0.23108(51)_{\text{total}},
\end{align}
where the ``IB'' component of the uncertainty refers to isospin breaking corrections. Improving the measurement of $R_{K\pi}^\mu$ to a precision of $10^{-4}$ will drastically lower the uncertainty coming from the experimental side from $23\times 10^{-5}$ to $1.2\times 10^{-5}$. Nevertheless, the total uncertainty is dominated by lattice calculations and it would only reduce marginally, to $45\times10^{-5}$.
Still, in the case of an improvement in the theoretical errors in the future, the reduction of the uncertainty on the experimental side would prove crucial. Assuming uncertainties reduced by a factor of two for both the isospin-breaking correction and for the decay constants, one would achieve (same central values considered):
\begin{align}
    [|V_{us}|/|V_{ud}|]^{R_{K\pi}^\mu, \text{ future}} &= 0.23108(1.2)_{\text{exp}}(21)_{F_K/F_\pi} (8)_{\text{IB}}=0.23108(23)_{\text{total}}.
\end{align}
This improvement can also be seen in Figure \ref{fig:CKMTrianglePlot}, where we show the current estimations of $|V_{us}|$ and $|V_{ud}|$ from different processes.

The same calculation can be done for the electron ratio, which is currently less precise and would improve a lot with the future measurement
\begin{align}
    [|V_{us}|/|V_{ud}|]^{R_{K\pi}^e, \text{ current}} &=0.2344(16)_{\text{total}}\\
    [|V_{us}|/|V_{ud}|]^{R_{K\pi}^e, \text{ future}} &=0.2344(6)_{\text{total}}\,.
\end{align}
As seen in the Figure \ref{fig:CKMTrianglePlot} a new measurement could provide an interesting input on the unitarity test. It is interesting to note that the present electron ratio input agrees with the unitarity constraint combined with the $V_{ud}$ determination from $\beta$-decays, while suggesting a tension with the kaon semi-leptonic determination of $V_{us}$. Nevertheless, the current estimation of the theoretical value of $R_{K\pi}^e$ is highly correlated to $R_{K\pi}^\mu$ due to $f_K/f_\pi$. Moreover, a separate estimation of the electromagnetic corrections for the electrons would be useful, in a similar manner as in the case of the muon \cite{DiCarlo:2019thl}.

As highlighted by our profiled fit to kaon (and related) data, new physics effects in $R_{K\pi}$ in our setup can be effectively described by one single operator, namely $[\mathcal{O}_{Hud}]_{12}$. Indeed, our global fit shows that the related coefficient can be constrained very well also in the profiled fit with the other lepton-flavour conserving coefficients turned on. In Figure \ref{fig:HudPrec}, we further compare the allowed ranges with two different assumptions for the future measurement: once with the same central value as the current measurement, and secondly with the central value being the SM value, both times assuming the improved theory uncertainty. Here one can observe two crucial things:
\begin{itemize}
    \item[(i)] In both cases the difference in precision between the individual and the profiled likelihood is nearly negligible. The ratio $R_{K\pi}^\mu$ is isolating the problem very well.
    \item[(ii)] With the same central value as the current measurement one could find a value of $C_{Hud}$ $6\sigma$ away from zero, showing again the great potential of having a precise determination of $R_{K\pi}^\mu$.
\end{itemize}

\subsection{Universality tests}\label{sec:lfu}

The ratio $R_K^\nu$ is effectively a test of lepton flavour universality in the kaon sector, measuring a possible deviation from universality in muons vs electrons.
In this sense, $R_K^\nu$ is crucially different from the double ratio $R_{K\pi}^{\mu/e}$ we present here.
Under the assumption of leptophilic new physics (i.e. SM-like quark couplings and modified lepton couplings), the double ratio can by definition not be used as a test of lepton flavour universality, and our simplifying substitution of $R_{K\pi}^{\mu/e}$ with the single ratio $R_K^\nu$ cannot be made.
In order to reinterpret the double ratio as an LFU test, an improved measurement on the pion side is needed, in particular of the ratio $R_\pi^\nu = \cB(\pi\to e \nu)/\cB(\pi\to\mu\nu)$.
The measurement of this ratio is precisely the target of the PIONEER experiment at PSI, which is expected to measure it at the $10^{-4}$ level~\cite{PIONEER:2022alm,PIONEER:2022yag}.
Nevertheless, a measurement of the single ratio $R_K^\nu$ is in fact possible with the setup proposed here, and would yield a precision of $10^{-3}$, still improving upon the current precision.
In the following we will assume this single-ratio measurement, which has the advantage of being an independent result.

\begin{figure}
    \centering
    \includegraphics[width=0.6\linewidth]{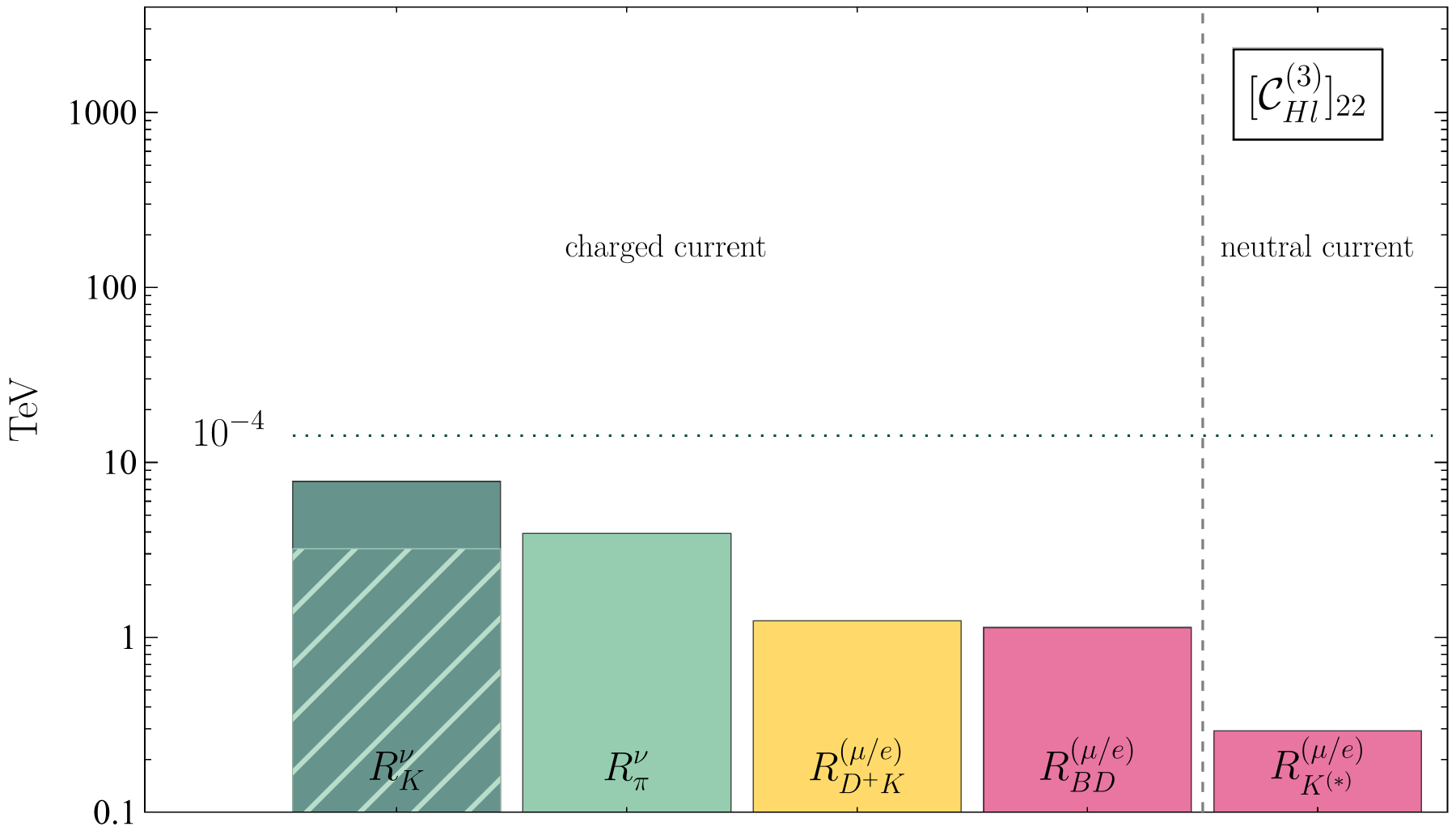}
    \caption{Comparison of LFU tests in different meson decays, assuming new physics in the $Z$- and $W$ couplings to muons. For $R_K^\nu$ the hatched bars represent the bound from the current measurement, the solid bars the future projection with a precision of $10^{-3}$, and the dotted line an optimistic scenario of $1\times 10^{-4}$.}
    \label{fig:lfu}
\end{figure}

In recent years, LFU tests have played a crucial role in our exploration of BSM, the more prominent examples given by the $B$-physics ratios $R_{D^{(*)}}$ and $R_{K^{(*)}}$ \cite{Bryman:2021sos,Hiller:2014ula,Guadagnoli:2022jdr,Bordone:2025elp,Bordone:2021olx}.
Here we want to compare the precision of different LFU tests in $B$, $D$, and $K$ decays.
The quark-level transitions are all different and potentially unrelated (see next section for a common approach), but we will consider for each sector observables measuring $\mu/e$ ratios.
An effective comparison can then be done by assuming that new physics modifies the lepton coupling of the $Z$ and/or $W$ bosons.
In practice, we will work again within SMEFT, and assume that, at the electroweak scale, NP effects can be encoded in the coefficient $[\cC_{H\ell}^{(3)}]_{22}$, i.e. we modify the $Z\bar\mu\mu (\bar\nu_\mu\nu_\mu)$ and the $W\bar\mu\nu_\mu$ couplings.
This choice also allows us to compare charged-current and neutral-current ratios, which tend to have very different sensitivity to NP scales due to the smallness of the neutral currents in the SM.
For illustration, we choose for each sector the most relevant observables, namely:
\begin{itemize}
    \item \textbf{kaons:} for charged currents, we use the estimate for $R_K^\nu$ presented here. On the neutral current side, the experimental upper bound on $K\to\pi\ell\ell$ modes is still one order of magnitude above the SM expectation, and hence no meaningful LFU test can be derived.
    \item \textbf{$D$ mesons:} We consider the ratio \\
    \begin{align}
        R_{DK}^{\mu/e} = \frac{\mathcal{B}(D^+\to \bar{K}^0 \mu\nu)}{\mathcal{B}(D^+\to \bar{K}^0 e\nu)} \,,
    \end{align}
    measured by BES-III to be \cite{10.1093/ptep/ptaa104}
    \begin{align}
        R_{DK}^{\mu/e,\, {\rm exp}} = 1.0003(25) \,,
    \end{align}
    with the SM expectation being $R_{DK}^{\mu/e,\, {\rm SM}} = 0.97510(10)$ \cite{Becirevic2021}. Neutral-current transitions involving charm quarks, however, do not have a comparable experimental precision as of today.
    \item \textbf{$B$ mesons:} the most precise charged-current $B$-decay ratio is $R_{D^{(*)}}^{\mu/e}$, defined as
    \begin{align}
        R_{D^{(*)}}^{\mu/e} = \frac{\mathcal{B}(B\to D^{(*)} \mu\nu)}{\mathcal{B}(B\to D^{(*)} e\nu)} \,,
    \end{align}
    for which the experimental measurement and SM expectation are \cite{Becirevic2021, Belle:2015pkj}
    \begin{align}
        R_{D^{(*)}}^{\mu/e,\,{\rm exp}} = 0.995(22)(39) 
        \qquad R_{D^{(*)}}^{\mu/e,\,{\rm SM}} = 0.9960(2)\,.
    \end{align}
    For neutral currents, we consider the well-known ratios
    \begin{align}
        R_{K^{(*)}} = \frac{\mathcal{B}(B\to K^{(*)} \mu\mu)}{\mathcal{B}(B\to K^{(*)} ee)} \,,
    \end{align}
    which are expected to be unity in the SM (with percent-level error \cite{Bordone2016,Isidori:2022bzw}), and whose experimental determinations are \cite{LHCb:2022vje}
    \begin{align}
        R_K^{\rm exp} = 0.95(7) \qquad R_{K^*}^{\rm exp} = 1.03(10) \,.
    \end{align}
\end{itemize}
We show again our results in form of a bar chart indicating the NP scale associated with each LFU test in Figure \ref{fig:lfu}.
One can clearly see the constraining power of $R_K^\nu$, as our future projection will provide the strongest test so far in this context, with effective scales reaching around 10 TeV.
It should also be added that including the prospected PIONEER result would yield a much more precise universality test in the pion sector ($R_\pi^\nu$), reaching the $10^{-4}$ line in Figure \ref{fig:lfu}.

\subsection{Third-generation new physics}

\begin{figure}
    \centering
    \includegraphics[width=0.40\linewidth]{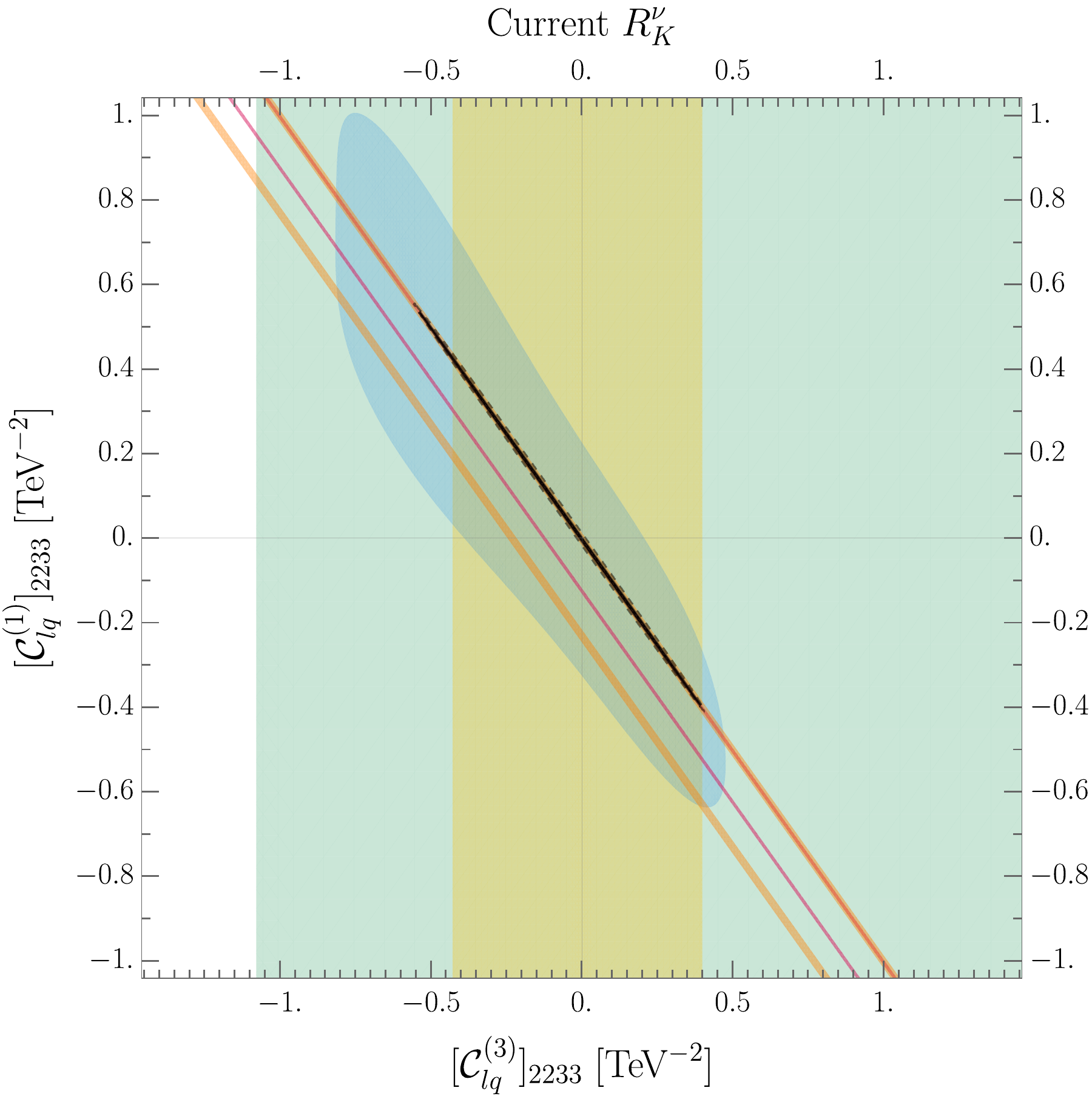}
    \includegraphics[width=0.58\linewidth]{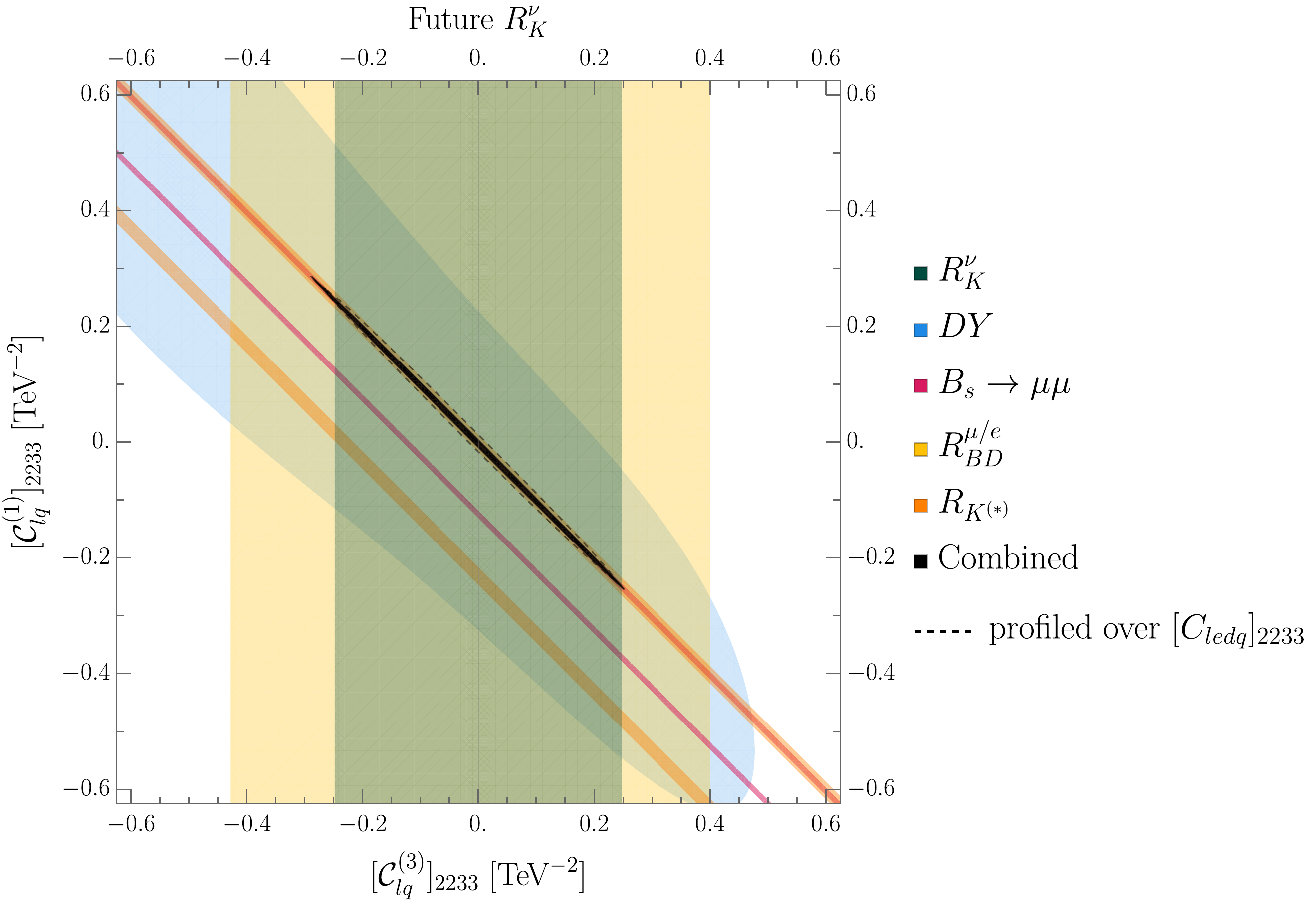}
    \caption{2$\sigma$ allowed ranges for Wilson coeficients with third-generation quark indices and muon indices.}
    \label{fig:U2}
\end{figure}

A common assumption which has risen in popularity in recent years is the idea that new physics should couple in a non-universal way to the three fermion families, with a preference for the third generation \cite{Allwicher:2023shc,Davighi:2025cqx,Demetriou:2025ewa,Allwicher:2025bub,Allwicher:2024ncl,Fuentes-Martin:2020zaz}.
This assumption allows on one hand to lower the NP scales involved (direct searches are less stringent without new physics in valence quarks), and on the other hand is a typical feature in models addressing both the Higgs hierarchy problem and the SM flavour puzzle.
In our SMEFT language, this implies that, on the quark side, we need to consider only coefficients with third-generation quark fields.
In particular, our SMEFT Lagrangian (at the electroweak scale) is
\begin{align}
\begin{aligned}
    \cL_{\rm SMEFT} &= \cL_{\rm SM} + [\cC_{\ell q}^{(1)}]_{ij33} (\bar\ell_i \gamma_\mu \ell_j)(\bar q_3 \gamma^\mu q_3) + [\cC_{\ell q}^{(3)}]_{ij33} (\bar\ell_i \gamma_\mu \sigma^I \ell_j)(\bar q_3 \gamma^\mu \sigma^I q_3) \\
    &+ [\cC_{\ell e d q}]_{ij33} (\bar\ell_i e_j)(\bar d_3 q_3) \,,
    \end{aligned}
\end{align}
where we have ignored operators involving right-handed top quarks.
Two important issues need to be addressed in this context.
The first is again the question of the alignment, this time relevant only for the third-generation doublet $q_3$, while the second is how we parametrise flavour-violating effects.
We chose to work in the down-aligned basis, i.e. with
\begin{align}
    q_3 = 
    \begin{pmatrix}
        V_{ub}^* u_L + V_{cb}^* c_L + V_{tb}^* t_L \\ b_L    
    \end{pmatrix} \,. 
\end{align}
On the flavour-violating side, a $U(2)_q$-scaling of the effective coefficients is compatible with the assumption of third-generation couplings \cite{Barbieri:2011ci,Barbieri:2012uh,Isidori:2012ts}.
This is characterised by assuming a $U(2)_q$ symmetry rotating the first- and second generation quark doublets into each other, at leading order.
The largest effects in the breaking of this symmetry can be parametrised by the $U(2)_q$ breaking spurion
\begin{align}
    \Tilde{V} = -\epsilon 
    \begin{pmatrix}
        \kappa V_{td} \\ V_{ts} 
    \end{pmatrix} \,,
\end{align}
with $\epsilon$ and $\kappa$ of $\cO(1)$.
In other words, flavour-violation is expected to follow a CKM-like structure, with $\cO(1)$ deviations possible due do the $U(2)$ assumption.
Most importantly, this assumption now allows us to connect different quark-level transitions, as for example those in the LFU tests discussed in the previous section.
Operationally, the effect of the spurion can be taken into account by replacing
\begin{align}
    q_3 \to q_3 + \Tilde{V}^i q_i \,,\quad i=1,2
\end{align}
in the operators listed above.
After doing so, it is easy to see that our charged-current kaon decays will only be affected by the coefficient $\cC_{\ell q}^{(3)}$ with two insertions of the spurion $\Tilde{V}$.
$\cC_{\ell q}^{(1)}$, instead, enters only neutral-current processes, and the scalar operator $\cC_{\ell edq}$ cannot be rotated into a $s\to d,u$ current, and hence will only affect $b$-physics and high-$p_T$ searches.
On the lepton side, we consider again muonic indices, focusing on five key observables: $pp\to \mu\mu (\mu\nu)$, $R_{K}^\nu$, $R_{BD}^{\mu/e}$, $B_s\to\mu\mu$, and $R_{K^{(*)}}$.
Di-neutrino modes, such as $B\to K \nu\bar\nu$ and $K\to\pi\nu\bar\nu$, are affected as well. However, not knowing the neutrino flavour, one can make the assumption of new physics effects involving $\tau$ neutrinos, coherently with the third-generation dominance picture. This has been shown to lead to a good compatibility between flavour, high-$p_T$, and electroweak precision tests with a $U(2)$ scaling \cite{Allwicher:2024ncl,Allwicher:2026loe}. Moreover, we can then use the result from said analysis to fix the values of $\epsilon$ and $\kappa$ in our fit, effectively leaving the Wilson coefficients with light leptons as only free parameters.
We thus fix $\epsilon=2.7$ and $\kappa=1$, compatibly with the results in Ref. \cite{Allwicher:2026loe}, and show the constraints in the $\cC_{\ell q}^{(1)}$-$\cC_{\ell q}^{(3)}$ plane in Figure \ref{fig:U2}.
$B_s\to\mu\mu$ strikes as a very strong constraint, but it is only sensitive to one combination of coefficients. Our future expectation for $R_{K}^\nu$
probes a complementary, independent direction, effectively making charged current kaon decays into a clean precision test of third-generation NP with additional muon couplings, and surpassing the sensitivity of Drell--Yan and $R_{BD}^{\mu/e}$ searches.
The dashed contours in Figure \ref{fig:U2} indicate the combined constraint, once the third relevant coefficient ($\cC_{\ell edq}$) has been profiled over.\footnote{The region is only slightly larger than in the two-parameter fit, as $\cC_{\ell edq}$ leads to a chirally-enhanced contribution in $B_s\to\mu\mu$.
The main constraint on the vector coefficients is then given by the LFU tests $R_{K^{(*)}}$, which have a similar sensitivity.}
The overall associated NP scale we extract from the 2$\sigma$ combined allowed range is
\begin{align}
    \Lambda \gtrsim 2 \text{ TeV} \,.
\end{align}
The results just described can also be interpreted in a $U(2)$-like picture on the lepton side.
The leading $U(2)_\ell \times U(2)_e$-symmetric term only contains vector operators, with $[\cC_{\ell q}^{(1,3)}]_{1133}=[\cC_{\ell q}^{(1,3)}]_{2233}$.
This combination is sensitive to Drell--Yan searches with electrons and muons in the final state, and to $B_s\to\mu\mu$.
Note that the scalar contribution in the latter falls short, as $\cC_{\ell edq}$ with light lepton indices is not allowed by the symmetry.
All other observables, being LFU tests, probe a breaking of $U(2)$.
An important exception is the ratio $R_{K\pi}^\mu$, which is indeed sensitive to lepton flavour universal effects.
The sensitivity however turns out to be significantly lower than the one from the double ratio (or $R_K^\nu$), as theoretical uncertainties limit the precision (cf. the discussion above).
The leading term of this breaking comes from rotating third-generation indices into light ones via a spurion of the form
\begin{align}
    \tilde{V}_\ell = \begin{pmatrix}
        0 \\ \delta
    \end{pmatrix} \,,
\end{align}
in analogy to the quark case.
LFU tests, along with LFV searches, can then be used to constrain the size of $\delta$ \cite{Covone:2025lee}.
A rough estimate of $\delta$ from $R_K^\nu$ can be made by considering an effective scale of approximately 1.5 TeV associated with $[\cC_{\ell q}^{(3)}]_{3333}$, and the scaling $[\cC_{\ell q}^{(3)}]_{2233} = \delta^2 [\cC_{\ell q}^{(3)}]_{3333}$.
One finds
\begin{align}
    \delta \lesssim 0.75 \,,
\end{align}
which is a looser constraint compared to the results in Ref. \cite{Covone:2025lee}.
As such, $R_K^\nu$ should be considered as a probe of additional (non-minimal) $U(2)_\ell$ breaking.

\subsection{A simplified leptoquark model}

Semileptonic interactions like the ones discussed in the SMEFT context in the previous sections can be generated in the UV by leptoquark states, which come with the additional advantage of suppressing four-quark operators and therefore meson mixing effects.
We consider here the example of a scalar $S_1$ leptoquark, with quantum numbers $S_1 \sim (\mathbf{3},\mathbf{1},-1/3)$ under the SM gauge group.
This leptoquark state has been studied extensively in the context of the $B$-anomalies \cite{Gherardi:2020qhc,Bauer:2016hbm,Hiller:2016kry,Angelescu:2021lln,Crivellin:2021egp}, but we now want to focus on the kaon sector.
The interaction Lagrangian with the SM fermion fields features two independent Yukawa coupling structures:
\begin{equation}
\mathcal{L}_{S_1} \supset y_{ij}^L \, \overline{(q^i)^c} \, i\sigma_2 \, \ell^j S_1 + y_{ij}^R \, \overline{(u^i)^c} \, e^j S_1 + \text{h.c.} \,.
\end{equation}
Assuming the mass of the leptoquark to be above the TeV scale, we can integrate it out and match its effects onto SMEFT, giving rise to the following contributions:
\begin{align}
    \begin{aligned}
        [\cC_{\ell q}^{(1)}]_{ijkl} = -[\cC_{\ell q}^{(3)}]_{ijkl} = \frac{y^{L^*}_{ki} y^{L}_{lj}}{4 m_{S_1}^2} \,, \\
        [\cC_{\ell e q u}^{(1)}]_{ijkl} = - 4 [\cC_{\ell e q u}^{(3)}]_{ijkl} = \frac{y^{L^*}_{ki} y^{R}_{lj}}{2 m_{S_1}^2} \,,\\
        [\cC_{eu}^{}]_{ijkl} = \frac{y^{R^*}_{ki} y^{R}_{lj}}{2 m_{S_1}^2} \,.
    \end{aligned}
\end{align}
Since we are looking for a chiral enhancement in $K\to \mu\nu$ decays, we choose to switch on the couplings
\begin{equation}
y_{22}^L \equiv \lambda_{s\mu}^L \neq 0 \quad (\text{coupling } q^2 \text{ to } \ell^2), \qquad 
y_{12}^R \equiv \lambda_{u\mu}^R \neq 0 \quad (\text{coupling } u_R \text{ to } \mu_R) \,,
\end{equation}
which in terms of activated SMEFT coefficients at the leptoquark mass scale yields non-zero $[\cC_{\ell q}^{(1,3)}]_{2222}$, $[\cC_{\ell e qu}^{(1,3)}]_{2221}$, and $[\cC_{eu}]_{2211}$.
Thus, in addition to the constraints from $\cC_{\ell e qu}^{(1)}$ discussed in the SMEFT analysis above, there will be additional tree-level constraints from Drell-Yan production, especially in the case of $u\bar u \to \mu\mu$, mediated by $\cC_{eu}$.
Moreover, working in a UV-complete setup allows us to precisely study one-loop phenomenology, the most important effect being $D^0$-$\bar D^0$ mixing.
Working in the down-aligned basis, the main contribution to the mixing comes from the operators $[\cC_{qq}^{(1,3)}]_{2222}$, generated at one loop.
The $(\bar c_L \gamma_\mu u_L)^2$ structure mediating the oscillation is then obtained through the misalignment of the mass eigenstates, giving \cite{Gherardi:2020qhc}
\begin{align}
    C_D^1 = - (V_{cs} V_{us}^*)^2 ([\cC_{qq}^{(1)}]_{2222} + [\cC_{qq}^{(3)}]_{2222}) = \frac{(V_{cs} V_{us}^*)^2}{128\pi^2} \frac{|y^L_{22}|^4}{m_{S_1}^2} \,.
\end{align}
The most recent result from the UTfit collaboration \cite{UTfit:2006onp} quotes
\begin{align}
    {\rm Re} (C_D^1) < 3.57 \times 10^{-7} \text{ TeV}^{-2} \,,
\end{align}
giving
\begin{align}
    \frac{|y^L_{22}|^4}{m_{S_1}^2} \lesssim 9.5 \times 10^{-3} \text{ TeV}^{-2} \,.
\end{align}

\begin{figure}
    \centering
    \includegraphics[width=0.49\linewidth]{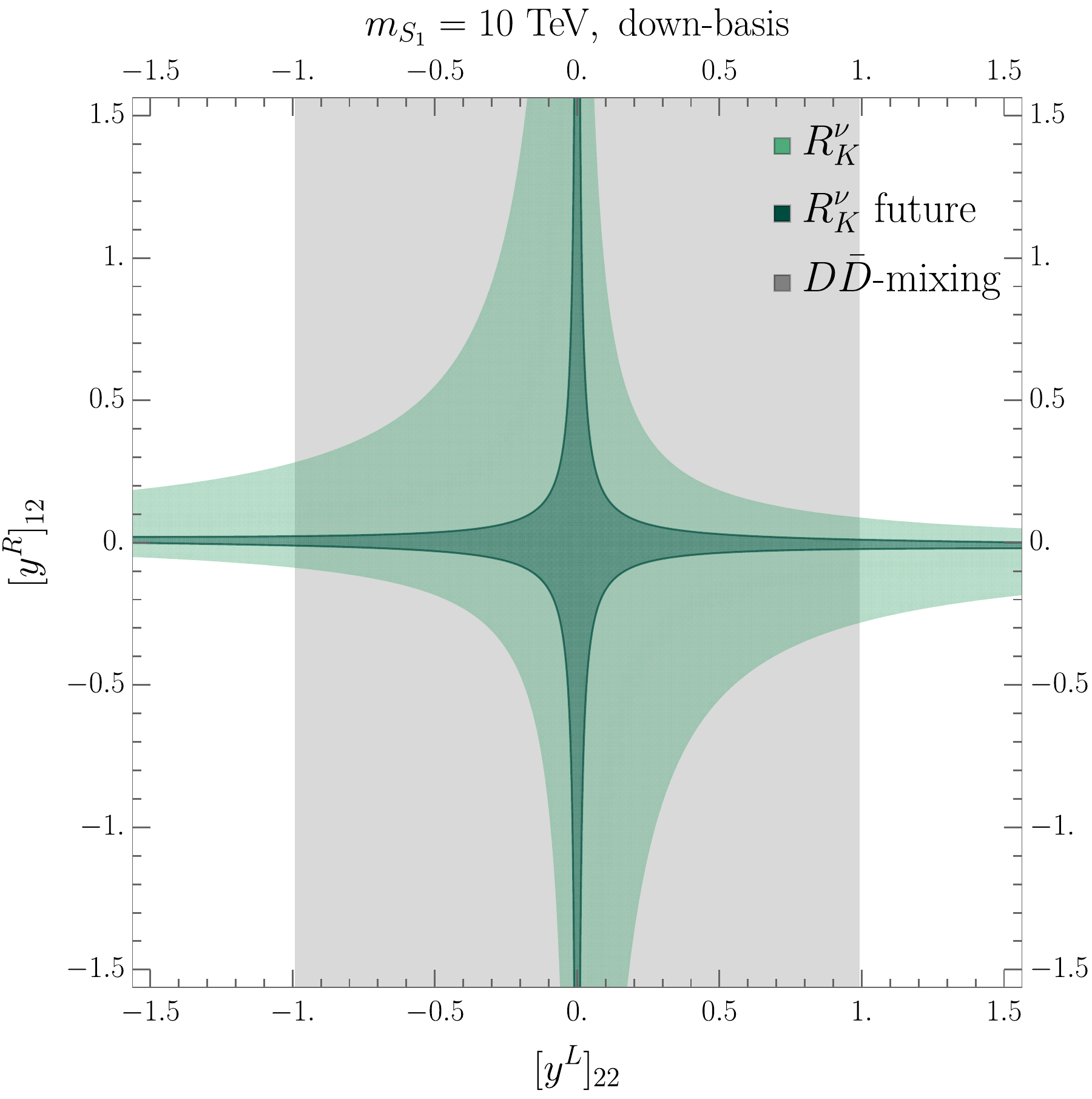}
    \includegraphics[width=0.49\linewidth]{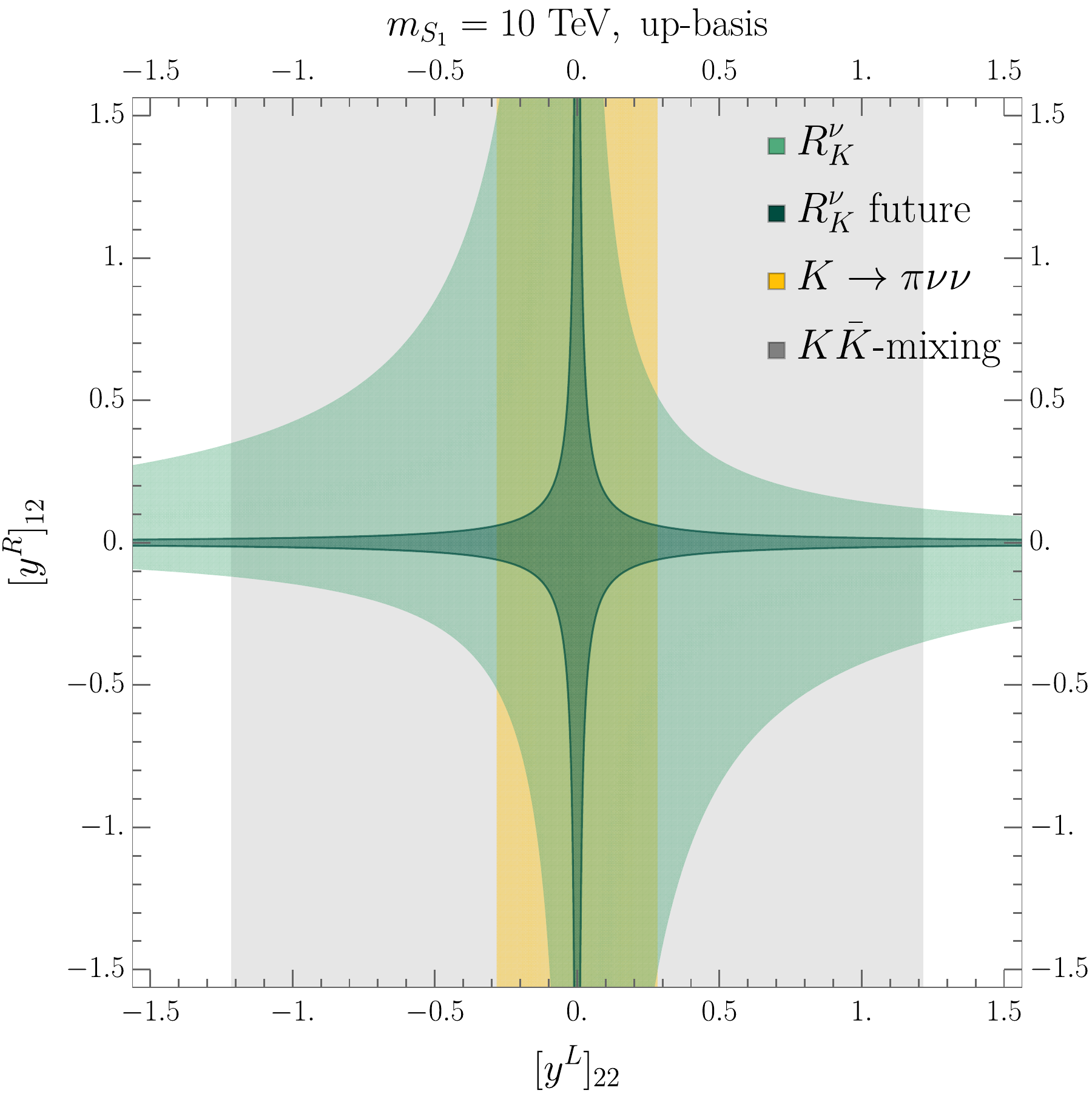}
    \caption{Allowed parameter space for the $S_1$ leptoquark couplings considered in our analysis, for $m_{S_1} = 10$ TeV. All contours are shown at 2$\sigma$ confidence level. \textit{Left}: down-alignment in the left-handed quark is assumed, i.e. $q_2 = (V_{is}^* u_L^i \quad s_L)^T$. \textit{Right}: up-alignment, i.e. $q_2= (c_L \quad V_{ci} d_L^i)^T$.}
    \label{fig:LQ}
\end{figure}

Should one assume up alignment instead, the $D$ meson mixing constraint is substituted by $K^0$-$\bar K^0$ mixing, and on top of that a contribution to $K^+\to\pi^+\nu\bar\nu$ arises from the down-quark content of the doublet $q_L^2$.
The kaon mixing constraint is given by \cite{Gherardi:2020qhc,UTfit:2006onp}
\begin{align}
    {\rm Re} (C_K^1) = \frac{(V_{cd} V_{cs}^*)^2}{128\pi^2} \frac{|y^L_{22}|^4}{m_{S_1}^2} < 8.04\times 10^{-7} \text{TeV}^{-2} \,.
\end{align}
The relevant constraints are shown in the $y^L$-$y^R$ plane in Figure \ref{fig:LQ}, where the impact of the ratio $R_K^\nu$ is apparent. 
With current data it is already competitive with $D^0\to\mu\mu$ or $K\to\pi\nu\nu$, and the future projection will make it the most relevant constraint in this sector.
A comment is due regarding the Drell--Yan constraints.
Given our choice of $m_{S_1}=10$ TeV, we still work in the EFT approximation and don't consider the effect of the leptoquark $t$-channel propagation, as this has been shown to have a small impact in this regime \cite{Allwicher:2024mzw}.
However, on the $y^L$ side, the constraints are comparably weak, since at tree-level there is no effect in neutral-current Drell--Yan, and the main effect comes from $s\bar c \to \mu\nu$.
On the $y^R$ side, typical scales lie in the 10 TeV range, corresponding to $\cO(1)$ couplings for our choice for the mass.

\section{Conclusions}\label{sec:conclusions}

A dedicated next-generation experiment measuring the kaon leptonic width ratio $R_K^\nu$, the kaon-to-pion leptonic decay width ratios, $R_{K\pi}^{e}$ for electron modes and $R_{K\pi}^{\mu}$ for muon modes and the double ratio $R_{K\pi}^e/R_{K\pi}^\mu$ represents a timely opportunity for precision flavour physics. Making use of slow-extracted proton beams such as those from the CERN SPS used for a number of experiments at the CERN North Area, approximately $4.5\times10^{13}$ tagged pions and $2.2\times10^{12}$ tagged kaons can be obtained per year.

Within just a few years of data taking, this setup will achieve unprecedented statistical and systematic precision: $10^{-3}$ for $R_K^\nu$, $3\times10^{-4}$ for $R_{K\pi}^{e}$ and $10^{-4}$ for $R_{K\pi}^{\mu}$. While the proposed $R_{K\pi}^\ell$ ratios allow cancellation of experimental systematic uncertainties, their SM expectations derive from estimates of the ratio of decay constants $f_K/f_\pi$ and of the parameters of the CKM matrix $V_{us}/V_{ud}$, presently known to within a few $10^{-3}$. 

The double ratio $R_{K\pi}^{e}/R_{K\pi}^{\mu}$ evades from the above limitations and, while allowing a thorough cancellation of the experimental systematic effects, it is also theoretically extremely clean and an important probe of new physics. In the experiment setup here discussed, the double ratio can be measured with total uncertainties of a few $10^{-4}$ after few years, thus improving on the state of the art by a factor of 10 or more. Moreover, the experiment will provide state-of-the-art sensitivities to the production of heavy neutral leptons in a variety of flavour coupling patterns, improving the sensitivity on the related squared couplings. 

We thoroughly reviewed the connection of $R_{K}^\nu$, $R_{K\pi}^{e}$, $R_{K\pi}^{\mu}$, and $R_{K\pi}^{e}/R_{K\pi}^{\mu}$ to other observables, both in the flavour sector and in high-$p_T$ and electroweak precision tests, under the assumption of heavy new physics above the electroweak scale.
Starting from an EFT approach (SMEFT), we showed the constraints and relevant connections for NP affecting kaon decays.
Moreover, we analysed the impact of our ratios in the context of tests of lepton flavour universality, and physics coupled mainly to the third generation. 
As a UV application, we showed a simplified leptoquark model, and the additional correlations arising in this case.

Ultimately, we have demonstrated that in many cases a next-generation measurement of kaon and pion leptonic decay widths with the quoted uncertainties will lead to dominant or competitive sensitivity to heavy BSM physics. In particular, the best sensitivities are obtained for chirally-enhanced scalar operators, for which scales up to 10$^3$ TeV can be probed. 

It is well known that the unitarity test for the first row of the CKM matrix presents two to three sigma tensions. Here for the first time we used $R_{K\pi}^e$ and noticed that the related $V_{us}/V_{ud}$ determination agrees with unitarity and with $V_{ud}$, while being in tension with $V_{us}$. Future measurements of $R_{K\pi}^e$ and $R_{K\pi}^\mu$ would greatly improve the sensitivity of such test.

In the context of LFU tests the measurement of $R_K^\nu$ will provide the leading constraint, which would be only surpassed by a possible future result from PIONEER \cite{PIONEER:2022yag}. For third-generation new physics, kaon LFU ratios will provide a useful complementary probe to current measurements. Finally, the allowed parameter space for an example leptoquark model can be probed with largely improved sensitivity.

\acknowledgments

We would like to thank Matheus Martines and Olcyr Sumensari for helpful discussions.
This project has received support from the Deutsche Forschungsgemeinschaft under Germany’s Excellence Strategy EXC 2121 “Quantum Universe” – 390833306, as well as from the grant 491245950.
LT’s research is funded by the Deutsche Forschungsgemeinschaft (DFG, German Research Foundation) - Projektnummer 417533893/GRK2575 “Rethinking Quantum Field Theory”. TS would like to thank Giuseppe Ruggiero for the useful discussions. SG acknowledges the support of the Institute of Atomic Physics through the CERN-RO Project No. 10/25.11.2024. TS and SG would like to thank Johannes Bernhard and Lau Gatignon for the enlightening discussions. We would like to thank Nazario Tantalo, Giuseppe Gagliardi and Agostino Patella for the input on the lattice calculation of $R_{K\pi}^e$.
We acknowledge the hospitality and support of the \href{https://theory.lnf.infn.it/events/lnf-theory-institute/}{Theoretical Phenomenology Visiting Institute @ LNF}, where this paper was initiated.

\appendix

\section{Details on flavour observables}\label{app:obs}

In this section we report further details on the flavour observables used in our fits.
A list of all observables, with their current experimental determinations and SM predictions, can be found in Table \ref{tab:obs}.

\begin{table}[]
    \centering
    \renewcommand{\arraystretch}{1.2}
    \begin{tabular}{|c|cc|cc|}
        \hline
        Observable & Exp. value & Ref. & SM prediction & Ref. \\ \hline
        $R_{K}^\nu$ & $2.488(9)\times10^{-5}$ & \cite{ParticleDataGroup:2026mpi} & $2.47653(34) \times 10^{-5}$ & \cite{Boyle:2026lrz}\\
        $R_{K\pi}^\mu$ & 1.3367(40) & \cite{ParticleDataGroup:2026mpi} & 1.316(6) & \ref{app:RKpiSM} \\
        \hline
        $\cB(K_L\to ee)$ & $(9^{+6}_{-4}) \times 10^{-12}$ & \cite{ParticleDataGroup:2026mpi} & $8.46(37) \times 10^{-12}$& \cite{Hoferichter:2023wiy} \\
        $\cB(K_L\to \mu e)$ & $<4.7 \times 10^{-12}$ & \cite{ParticleDataGroup:2026mpi} & -- & \\
        $\cB(K_L\to \mu\mu)$ & $6.84(11)\times 10^{-9}$ & \cite{ParticleDataGroup:2026mpi} & $(7.44^{+0.44}_{-34})\times 10^{-9}$& \cite{Hoferichter:2023wiy} \\
        $\cB(K^+\to \pi^+\nu\bar\nu)$ & $(1.14^{+0.40}_{-0.33})\times 10^{-10}$& \cite{ParticleDataGroup:2026mpi} & $8.09(63)\times10^{-11}$ & \cite{Allwicher:2026loe}\\ \hline
        $\cB(D^0\to ee)$ & $<7.9 \times 10^{-8}$ & \cite{ParticleDataGroup:2026mpi} & $\sim 10^{-23}$ & \cite{Burdman:2001tf} \\
        $\cB(D^0\to \mu e)$ & $<1.3 \times 10^{-8}$ & \cite{ParticleDataGroup:2026mpi} & -- & \\
        $\cB(D^0\to \mu\mu)$ & $<3.1 \times 10^{-9}$ & \cite{ParticleDataGroup:2026mpi} & $\sim 3 \times 10^{-13}$ & \cite{Burdman:2001tf}\\
        $\cB(D^+\to \mu^+\nu)$ & $4.02(09)\times 10^{-4}$ & \cite{ParticleDataGroup:2026mpi} & $3.51(5)\times 10^{-4}$ & \cite{FlavourLatticeAveragingGroupFLAG:2024oxs} \\
        $\cB(D^+\to \tau^+\nu)$ & $9.9(1.2)\times 10^{-4}$ & \cite{ParticleDataGroup:2026mpi} & $0.935(13)\times 10^{-3}$ & \cite{FlavourLatticeAveragingGroupFLAG:2024oxs} \\
        $\cB(D_s\to \mu^+\nu)$ &$0.537(11)\times 10^{-2}$ & \cite{ParticleDataGroup:2026mpi} & 
        $5.35(7)\times10^{-3}$ & \cite{FlavourLatticeAveragingGroupFLAG:2024oxs} \\
        $\cB(D^+\to K^0 e^+\nu)$ & $8.76 (9)\times 10^-2$ & \cite{ParticleDataGroup:2026mpi} & $8.85(10)$ & \cite{Becirevic:2026tle}\\ 
        $R_{DK}^{\mu/e} = \frac{\cB(D^+\to K^0\mu\nu)}{\cB(D^+\to K^0e\nu)}$ & $1.0003(25)$ & \cite{10.1093/ptep/ptaa104} & $0.97510(10)$ & \cite{Becirevic2021}  \\ \hline
        $\cB(B^+\to e^+\nu)$ & $<9.8 \times 10^{-7}$& \cite{ParticleDataGroup:2026mpi} & $(8.9 \pm 1.2)\times 10^{-12}$& \cite{FlavourLatticeAveragingGroupFLAG:2024oxs} \\
        $\cB(B^+\to \mu^+\nu)$ &$<9.8 \times 10^{-7}$ & \cite{ParticleDataGroup:2026mpi} & $(3.8 \pm 0.5)\times 10^{-7}$& \cite{FlavourLatticeAveragingGroupFLAG:2024oxs} \\
        $\cB(B_s\to ee)$ & $<9.4\times 10^{-9}$& \cite{ParticleDataGroup:2026mpi}& $8.6(4)\times 10^{-14}$& \cite{Bobeth:2013uxa}\\
        $\cB(B_s\to \mu e)$ & $<5.4\times 10^{-9}$& \cite{ParticleDataGroup:2026mpi} & -- & \\
        $\cB(B_s\to \mu\mu)$ & $3.34(27)\times 10^{-9}$& \cite{ParticleDataGroup:2026mpi}& $3.64(12)\times 10^{-9}$& \cite{Beneke:2019slt} \\
        $\cB(B_s\to \tau e)$ & $<1.4\times 10^{-3}$ & \cite{ParticleDataGroup:2026mpi} & -- & \\
        $\cB(B_s\to \tau\mu)$ & $<4.2\times 10^{-5}$ & & -- & \\
        $\cB(B^+ \to K^+\nu\bar\nu)$ & $2.3(7)\times 10^{-5}$ & \cite{ParticleDataGroup:2026mpi} & $4.72(27) \times 10^{-6}$ & \cite{Allwicher:2026loe} \\
        $R_{D}^{\mu/e} = \frac{\cB(B\to D\mu\nu)}{\cB(B\to De \nu)}$ & 0.995(45) & \cite{PhysRevD.93.032006} & 0.9960(2) & \cite{Becirevic2021} \\ 
        $R_{K} = \frac{\cB(B\to K\mu\mu)}{\cB(B\to Kee)}$ & 0.95(7)& \cite{LHCb:2022vje} & 1.00(1) & \cite{Bordone2016} \\ 
        $R_{K^*} = \frac{\cB(B\to K^*\mu\mu)}{\cB(B\to K^*ee)}$& 1.03(10)& \cite{LHCb:2022vje} & 1.00(1) & \cite{Bordone2016}\\ \hline
        $\cB(\mu\to 3e)$ & $<1.3\times 10^{-12}$ & \cite{ParticleDataGroup:2026mpi} & --  & \\ \hline
        $\cB(\tau\to 3e)$ & $<2.7 \times 10^{-8}$ & \cite{Hayasaka:2010np} & -- & \\
        $\cB(\tau\to 3\mu)$ & $<1.9\times 10^{-8}$ & \cite{Belle-II:2024sce} & -- & \\
        $\cB(\tau\to K_S e)$ & $<0.8\times 10^{-8}$ & \cite{Belle:2025iff} & -- & \\
        $\cB(\tau\to K_S \mu)$ & $<1.8\times 10^{-8}$ & \cite{Belle:2025iff} & -- & \\ \hline
    \end{tabular}
    \caption{Numerical values used for the observables in our analyses. All upper bounds are given at 90\% C.L.. Entries with ``--" denote processes forbidden by the SM.
    }
    \label{tab:obs}
\end{table}

\begin{table}[]
    \centering
    \begin{tabular}{c|cc|c}
        Parameter & Value & Ref. & \\ \hline
        $V_{us}$ & 0.22330 & \cite{Cirigliano:2022yyo} & From $K_{\ell 3}$ decays\\
        $V_{ud}$ & 0.97384 & \cite{Cirigliano:2022yyo} & From $\beta$-decays\\
        $f_{K^{\pm}}/ f_{\pi^{\pm}}$ & 1.1932(19) & \cite{FlavourLatticeAveragingGroupFLAG:2024oxs} & $N_f = N+1+1$\\
        $f_{K^{\pm}}$ & 155.7(3) MeV & \cite{FlavourLatticeAveragingGroupFLAG:2024oxs} & $N_f = N+1+1$ \\
        $f_{\pi^{\pm}}$ & 130.2(8) MeV & \cite{FlavourLatticeAveragingGroupFLAG:2024oxs} & $N_f = N+1+1$ \\
        $f_{B}$ & 190.0(1.3) MeV & \cite{FlavourLatticeAveragingGroupFLAG:2024oxs} & $N_f = N+1+1$ \\
        $f_{B_s}$ & 230.1(1.5) MeV & \cite{FlavourLatticeAveragingGroupFLAG:2024oxs} & $N_f = N+1+1$ \\
        $f_{D}$ & 212.0(0.7) MeV & \cite{FlavourLatticeAveragingGroupFLAG:2024oxs} & $N_f = N+1+1$ \\
        $f_{D_s}$ & 249.9(0.5) MeV & \cite{FlavourLatticeAveragingGroupFLAG:2024oxs} &  $N_f = N+1+1$ \\
    \end{tabular}
    \caption{Numerical values for theory parameters relevant in our analyses.}
    \label{tab:inputs}
\end{table}

\subsection{Leptonic meson decays in LEFT}

Using the Lagrangian \ref{eq:LagrLEFT} with the operators defined in \ref{eq:leftcoeffs} one can describe all scalar meson decays to $\ell \nu$ by
\begin{align}
    \begin{aligned}
    \mathcal{B}(P \rightarrow \ell \nu)=&\frac{\tau_P}{64\pi}\,
    f_P^2\,m_P\,m_\ell^2
    \left(1-\frac{m_\ell^2}{m_P^2}\right)^2 S_{\rm ew}\left(1+\delta_{\rm ew}^{P\ell}\right)
    \sum_{j=1}^{3}
    \left[
    [L_{\mathrm{\nu edu}}^{V,LL}]_{jixr}
    -[L_{\mathrm{\nu edu}}^{V,LR}]_{jixr} \right. \\
    & \left.-\frac{m_P^2}{m_\ell(m_{q_x}+m_{q_r})}
    [L_{\mathrm{\nu edu}}^{S,RR}]_{jixr}
    +\frac{m_P^2}{m_\ell(m_{q_x}+m_{q_r})}[L_{\mathrm{\nu edu}}^{S,RL}]_{jixr}
    \right]^2 \,,
    \end{aligned}
    \label{eq:bPlnu}
\end{align}
see in Appendix~\ref{app:radcorr} the discussion about the radiative corrections $S_{\rm ew}\left(1+\delta_{\rm ew}^{P\ell}\right)$.
The decay of neutral mesons to two leptons in LEFT is mediated by the following LEFT operators,
\begin{align}
\label{eq:coeffLEFTeq}
\begin{aligned}
    [O_{eq}^{V,LL}]_{ijxr} &= (\bar e_{L,i} \gamma_\mu e_{L,j})(\bar q_{L,x} \gamma^\mu q_{L,r}) \,, \qquad [O_{eq}^{V,LR}]_{ijxr} = (\bar e_{L,i} \gamma_\mu e_{L,j})(\bar q_{R,x} \gamma^\mu q_{R,r}) \,,  \\
    [O_{eq}^{V,RR}]_{ijxr} &= (\bar e_{R,i} \gamma_\mu e_{R,j})(\bar q_{R,x} \gamma^\mu q_{R,r}) \,, \qquad [O_{qe}^{V,LR}]_{ijxr} = (\bar q_{L,x} \gamma^\mu q_{L,r})(\bar e_{R,i} \gamma_\mu e_{R,j}) \,,  \\
    [O_{eq}^{S,RR}]_{ijxr} &= (\bar e_{L,i} e_{R,j})(\bar q_{L,x} q_{R,r}) \,, \qquad\qquad [O_{eq}^{S,RL}]_{ijxr} = (\bar e_{L,i} e_{R,j})(\bar q_{R,x} q_{L,r}) \,,
    \end{aligned}
\end{align}
where the $q$ can be replaced by either $u$ or $d$ depending on the quark content of the neutral meson.
The branching ratio to two general leptons $\ell_i \ell_j$ can then be written as 
\begin{align}
    &\mathcal{B}(P \rightarrow \ell_i \ell_j) =\frac{\tau_P f_P^2}{40\pi m_P^2}
    \cdot p_f  \\ 
    &\cdot 
    \left[
    \left(-m_jm_i + \sqrt{m_j^2+p_f^2}\,\sqrt{m_i^2+p_f^2}+p_f^2\right)
    \left(
    (m_j-m_i)C_{VA}
    -\frac{m_P^2}{m_{d_i}+m_{d_j}}C_{SP}
    \right)^2 \right. \\
    & \left. +\left(m_jm_i+\sqrt{m_j^2+p_f^2}\,\sqrt{m_i^2+p_f^2}+p_f^2\right)
    \left(
    (m_j+m_i)C_{AA}
    +\frac{m_P^2}{m_{q_x}+m_{q_r}}C_{PP}
    \right)^2
    \right].
\end{align}
where
\begin{align}
    p_f = \frac{\sqrt{
    m_P^4 - 2m_P^2m_i^2 + m_i^4
    - 2m_P^2m_j^2 - 2m_i^2m_j^2 + m_j^4
    }}{2m_P} \,,
\end{align}
and 
\begin{align}
C_{SP} &=
-[L_{eq}^{S,RL}]_{xrij}
+[L_{eq}^{S,RR}]_{xrij}
+[L_{eq}^{S,RL}]_{rxji}^*
-[L_{eq}^{S,RR}]_{rxji}^* \,,
\\[4pt]
C_{PP} &=
-[L_{eq}^{S,RL}]_{xrij}
+[L_{eq}^{S,RR}]_{xrij}
-[L_{eq}^{S,RL}]_{rxji}^*
+[L_{eq}^{S,RR}]_{rxji}^* \,,
\\[4pt]
C_{VA} &=
-[L_{eq}^{V,LR}]_{xrij}
+[L_{eq}^{V,LL}]_{xrij}
+[L_{qe}^{V,LR}]_{xrij}
-[L_{eq}^{V,RR}]_{xrij} \,,
\\[4pt]
C_{AA} &=
+[L_{eq}^{V,LR}]_{xrij}
-[L_{eq}^{V,LL}]_{xrij}
+[L_{qe}^{V,LR}]_{xrij}
-[L_{eq}^{V,RR}]_{xrij} \,,
\end{align}
where depending on the quark content as in \ref{eq:coeffLEFTeq} of the meson the $eq$ is replaced by $ed$ or $eu$.

\subsubsection{Radiative corrections for leptonic decays}\label{app:radcorr}

Here, we comment about the radiative corrections $S_{\rm ew}\left(1+\delta_{\rm ew}^{P\ell}\right)$ in \ref{eq:bPlnu}.
The universal electroweak short-distance enhancement factor for semileptonic decays in the SM~\cite{Marciano:1993sh} is $S_{\rm ew} = 1.0232(3)$ at the scale $\mu = m_\rho$ for $\pi$ and $K$~, and $S_{\rm ew} = 1.0066(3)$ at the scale $\mu = m_B$ for $B$ decays.
The long-distance electromagnetic corrections are denoted by $\delta^{P\ell}_{\rm ew}$ are channel-dependent corrections~\cite{Marciano:1993sh,Finkemeier:1995gi,Cirigliano:2007xi,Cirigliano:2007ga,Cirigliano:2011tm,Giusti:2017dwk,Desiderio:2020oej,DiCarlo:2019thl,Boyle:2026lrz,Cornella:2026lkp,Cornella:2026ask,Cornella:2022ubo,Beneke:2020vnb,Becirevic:2009aq,Boer:2023vsg}. 
In the case of $P=\pi, K$, they are given~\cite{Marciano:1993sh,Finkemeier:1995gi,Cirigliano:2007xi,Cirigliano:2007ga,Cirigliano:2011tm} by
\begin{equation}
\delta^{P\ell}_{\rm ew} = \frac{\alpha}{\pi}\left[F(m_\ell^2/m_P^2)+\frac{3}{2}\log\frac{m_P}{m_\rho}-c_1^P\right], \label{eq:Kl2rad}
\end{equation}
where $\alpha$ is the fine-structure constant and $F(z)$ describes the leading universal long-distance radiative corrections for a point-like meson~\cite{Marciano:1993sh}
\begin{equation}
    \begin{aligned}
F(z) &= \frac{3}{2} \log z + \frac{13 - 19z}{8(1 - z)} - \frac{8 - 5z}{4(1 - z)^2} z \log z  \\
&\quad - \left( 2 + \frac{1 + z}{1 - z} \log z \right) \log(1 - z)- 2 \frac{1 + z}{1 - z} \text{Li}_2(1 - z) ~.
\end{aligned}
\end{equation}
The constant $c_1^P$ encodes hadronic structure effects calculated in Chiral Perturbation Theory (ChPT) \cite{Cirigliano:2007xi,Cirigliano:2007ga,Cirigliano:2011tm} and drops out in the double ratios. In addition to ChPT, $\delta^{P\mu}_{\rm ew}$ for $P=\pi, K$ has recently been evaluated using first-principles lattice QCD calculations~\cite{Giusti:2017dwk,Desiderio:2020oej,DiCarlo:2019thl,Boyle:2026lrz}, yielding compatible results. Since lattice determination is not yet available for $\delta^{Pe}_{\rm ew}$, we use the ChPT estimate for the electronic mode. 
For leptonic $B$-meson decays, long-distance QED corrections cannot be evaluated via ChPT and are instead calculated using SCET and $B$-meson LCDAs~\cite{Cornella:2026lkp,Cornella:2026ask,Cornella:2022ubo,Beneke:2020vnb,Becirevic:2009aq,Boer:2023vsg}, while for $D$-meson decays first attempts are in~\cite{Kitahara:2025jqk}. 

\subsection{\texorpdfstring{$R_{K\pi}^\ell$ in the Standard Model}{RKpimu in the Standard Model}}
\label{app:RKpiSM}

When using $R_{K\pi}^\mu$ for our fits to new physics, we are determining the SM value with
\begin{align}
    R_{K\pi}^{\mu,\,{\rm SM}} = \left|\frac{V_{us}}{V_{ud}}\right|^2\frac{f_K^2}{f_\pi^2}\,\frac{ m_{K^+} \left(1 - \frac{m_{\mu}^2}{m_{K^+}^2}\right)^2}{m_{\pi^+} \left(1 - \frac{m_{\mu}^2}{m_{\pi^+}^2}\right)^2}   \left(1+\delta_{\rm ew}^{K\pi\mu}\right) \, ,
\end{align}
where $\delta_{\rm ew}^{K\pi\mu}=-0.0126(14)$ \cite{DiCarlo:2019thl} are the radiative and isospin-breaking correction. To not take the the isospin-breaking twice, the decay constants are taken uncorrected as $f_K/f_\pi= 1.1978(22)$ \cite{FlavourLatticeAveragingGroupFLAG:2024oxs}. Furthermore, we take into account the CKM elements determined from $K\ell 3$ and $\beta$-decays as in Tab. \ref{tab:inputs}. This results in a SM value of
\begin{align}
    R_{K\pi}^{\mu,\,{\rm SM}} = 1.316(6) \,.
\end{align}
The ratio $R_{K\pi}^e$ is estimated through the results of the other ratios, therefore using as input the same values for $V_{us}/V_{ud}$ as in the muon case. The SM value we will use is therefore 
\begin{align}
    [R_{K\pi}^e]_{\SM}= \frac{[R_K^\nu]_\SM}{[R_\pi^\nu]_\SM}\; [R_{K\pi}^\mu]_\SM = 0.2639(12)\,.
\end{align}
An estimation of the radiative correction for the electron case would be worth to be performed in future.

\subsection{Semileptonic decays and LFU ratios}

\subsubsection{Di-neutrino modes}

At the $B$ and $K$ mass scale, the new physics contributions to the $d_i\to d_j \nu\bar\nu$ modes can be expressed in the LEFT as
\begin{align}
    \frac{\cB(B^+\to K^+ \nu\bar\nu)}{\cB(B^+\to K^+\nu\bar\nu)_{\rm SM}} = \frac{1}{3 \left|[L_{\nu d}^{V,LL}]_{kk23}^{\rm SM}\right|^2} 
    \sum_{i,j} \left|[L_{\nu d}^{V,LL}]_{ij23}+[L_{\nu d}^{V,LR}]_{ij23}\right|^2 \,,
\end{align}
\begin{align}
    \cB(K^+\to \pi^+ \nu\bar\nu) = 
    \sum_{i,j} A_{ij} \frac{\left|[L_{\nu d}^{V,LL}]_{ij21}+[L_{\nu d}^{V,LR}]_{ij21}\right|^2}{\left|[L_{\nu d}^{V,LL}]_{ii12}^{\rm SM}\right|^2} \,,
\end{align}
where
\begin{align}
    A_{ij} = \begin{cases}
        \frac{\cB(K^+\to\pi^+\nu_e\bar\nu_e)_{\rm SM}}{\left|[L_{\nu d}^{V,LL}]_{ii12}^{\rm SM}\right|^2} \qquad i,j = 1,2 \\ 
    \frac{\cB(K^+\to\pi^+\nu_\tau\bar\nu_\tau)_{\rm SM}}{\left|[L_{\nu d}^{V,LL}]_{3312}^{\rm SM}\right|^2} \qquad i \text{ and/or } j = 3
    \end{cases}
    \,,
\end{align}
and
\begin{align}
\begin{aligned}
    \cB(K^+\to\pi^+\nu_e\bar\nu_e)_{\rm SM} &= 2.87\times 10^{-11} \,,\\
    \cB(K^+\to\pi^+\nu_\tau\bar\nu_\tau)_{\rm SM} &= 2.35 \times 10^{-11} \,.
    \end{aligned}
\end{align}
The SM values for the LEFT coefficients are
\begin{align}
\begin{aligned}
    [L_{\nu d}^{V,LL}]_{ij23}^{\rm SM} &= - \frac{4 G_F}{\sqrt{2}} V_{tb} V_{ts}^* \frac{e^2}{16\pi^2} \frac{X_t}{s_W^2} \delta_{ij} \,,\\
    [L_{\nu d}^{V,LL}]_{ij12}^{\rm SM} &= - \frac{4 G_F}{\sqrt{2}} V_{ts} V_{td}^* \frac{e^2}{16\pi^2}\frac{1}{s_W^2} \left(X_t + \frac{V_{cs}V_{cd}^*}{V_{ts}V_{td}^*} X_c^i \right) \delta_{ij} \,,
    \end{aligned}
\end{align}
with \cite{Buchalla:1998ba,Brod:2008ss,Buras:2006gb,Brod:2010hi,Brod:2021hsj,Isidori:2005xm,Lunghi:2024sjy,Mescia:2007kn}
\begin{align}
\begin{aligned}
    X_t &= 1.462 \,, \\
    X_c^i &= (1.053,1.053,0.711) \times 10^{-3} \,,
    \end{aligned}
\end{align}
and the ones with right-handed quark currents are zero.
In all expressions above, the Wilson coefficients contain both the SM and the new physics contribution.

\subsubsection{\texorpdfstring{$D^+\rightarrow K^0 \ell \nu$}{D+ -> K0 l nu}}

In the charm sector we use the semi-leptonic decays \cite{Becirevic:2026tle}
\begin{align}
\cB(D^+\rightarrow K^0 \mu \nu)=& \, \tau_{D^+} \,
161.318\left(
\left|[g_S]_{2221}\right|^2
+\left|[g_S]_{2222}\right|^2
+\left|[g_S]_{2223}\right|^2
\right)
+64.623\left|[g_T]_{2222}\right|^2
\nonumber\\
&+83.6437\left(
\left|[g_V]_{2221}\right|^2
+\left|1+[g_V]_{2222}\right|^2
+\left|[g_V]_{2223}\right|^2
\right)
\nonumber\\
&+19.3545\,\operatorname{Re}\left\{
[g_T]_{2222}^{*}
\left(1+[g_V]_{2222}\right)
\right\}
\nonumber\\
&-20.2402\left(
\operatorname{Re}\left\{
[g_S]_{2221}^{*}[g_V]_{2221}
\right\}
+\operatorname{Re}\left\{
[g_S]_{2222}^{*}\left(1+[g_V]_{2222}\right)
\right\}
\right.
\nonumber\\
&\left.
+\operatorname{Re}\left\{
[g_S]_{2223}^{*}[g_V]_{2223}
\right\}
\right),
\end{align}
and 
\begin{align}
\cB(D^+\rightarrow K^0 \mu \nu)=& \,\tau_{D^+} \,
165.789\left(
\left|[g_S]_{2211}\right|^2
+\left|[g_S]_{2212}\right|^2
+\left|[g_S]_{2213}\right|^2
\right)
+64.7835\left|[g_T]_{2211}\right|^2
\nonumber\\
&+85.7041\left(
\left|1+[g_V]_{2211}\right|^2
+\left|[g_V]_{2212}\right|^2
+\left|[g_V]_{2213}\right|^2
\right)
\nonumber\\
&+0.104132\,\operatorname{Re}\left\{
[g_T]_{2211}^{*}
\left(1+[g_V]_{2211}\right)
\right\}
\nonumber\\
&-0.105541\left(
\operatorname{Re}\left\{
[g_S]_{2211}^{*}
\left(1+[g_V]_{2211}\right)
\right\}
+\operatorname{Re}\left\{
[g_S]_{2212}^{*}[g_V]_{2212}
\right\}
\right.
\nonumber\\
&\left.
+\operatorname{Re}\left\{
[g_S]_{2213}^{*}[g_V]_{2213}
\right\}
\right) \,.
\end{align}
The coefficients $g_i$ can be replaced by the LEFT coefficients as
\begin{align}
[g_V]_{ij\ell x}
&=
\frac{-1}{2\sqrt{2}\,G_F V_{ij}}
\left(
[L_{\nu d}^{V,LR}]_{x\ell ji}
+
[L_{\nu d}^{V,LL}]_{x\ell ji}
\right),
\\[4pt]
[g_S]_{ij\ell x}
&=
\frac{-1}{2\sqrt{2}\,G_F V_{ij}}
\left(
[L_{\nu d}^{S,RL}]_{x\ell ji}
+
[L_{\nu d}^{S,RR}]_{x\ell ji}
\right),
\\[4pt]
[g_A]_{ij\ell x}
&=
\frac{-1}{2\sqrt{2}\,G_F V_{ij}}
\left(
[L_{\nu d}^{V,LR}]_{x\ell ji}
-
[L_{\nu d}^{V,LL}]_{x\ell ji}
\right),
\\[4pt]
[g_P]_{ij\ell x}
&=
\frac{-1}{2\sqrt{2}\,G_F V_{ij}}
\left(
[L_{\nu d}^{S,RL}]_{x\ell ji}
-
[L_{\nu d}^{S,RR}]_{x\ell ji}
\right),
\\[4pt]
[g_T]_{ij\ell x}
&=
\frac{-1}{2\sqrt{2}\,G_F V_{ij}}
[L_{\nu d}^{T}]_{x\ell ji}.
\end{align}
The two branching ratios can then be combined to the LFU ratio
\begin{align}
    R_{DK}^{\mu/e} = \frac{\cB(D^+\to K^0\mu\nu)}{\cB(D^+\to K^0e\nu)}.
\end{align}

\subsubsection{\texorpdfstring{$R_{BD}^{\mu/e}$}{R BD {mu/e}}}
Another LFU ratio we consider is
\begin{align}
    R_{BD}^{\mu/e}= \frac{\mathcal{B}(B\rightarrow D\mu \bar{\nu})}{\mathcal{B}(B\rightarrow De \bar{\nu})}\, ,
\end{align}
which can be described by \cite{Becirevic2021}
\begin{align}
    R_{BD}^{\mu/e}= R_{BD, \text{SM}}^{\mu/e}\ &\left(
    |1+g_V|^2 + 1.13(3) |g_S|^2 + 0.68(6)|g_T|^2 \right. \\
    &\left. + 0.154(2) \text{Re}[(1+g_V)g_S^*]+ 0.188(9)\text{Re}[(1+g_V)g_T^*]
    \right) \, .
\end{align}
The coefficients describe the vectorial, scalar and tonsorial coupling, which in SMEFT can be written as
\begin{align}
    g_V &= -\frac{v^2}{\Lambda^2}\frac{1}{V_{23}}\sum_k\left(
    - \frac{[C_{Hud}]_{23}}{2} + [C^{(3)}_{lq}]_{22k3} \, + [C^{(3)}_{Hq}]_{k3} \, - \delta_{k3} [C^{(3)}_{Hl}]_{22} 
    \right) \,, \\
    g_S &= -\frac{v^2}{2 \Lambda^2}\frac{1}{V_{23}}\sum_k\left(
    [C^{(1)}_{lequ}]_{2232}^* + V_{2k}^*[C_{ledq}]_{223k}^* \right) \,, \\
    g_T &= -\frac{v^2}{2 \Lambda^2}\frac{1}{V_{23}}
    [C^{(3)}_{lequ}]_{2232}^*.
\end{align}

\subsubsection{\texorpdfstring{$R_K/R_{K^\ast}$}{RK/RK*}}

Lastly we also considered the ratio $R_K/R_{K^*}$, where
\begin{align}
    R_{K^{(*)}}^{[q_{\text{min}}^2,q_{\text{max}}^2]}=
    \int^{q_{\text{max}}^2}_{q_{\text{min}}^2} dq^2
    \frac{\frac{d\mathcal{B}}{dq^2}(B \rightarrow K^{(*)}\mu^+\mu^-)}{\frac{d\mathcal{B}}{dq^2}(B \rightarrow K^{(*)}e^+e^-)}. 
\end{align}

Assuming NP couplings only to the $\mu$ in the lepton sector, we can write the ratio including NP as \cite{Cornella2021}
\begin{align}
    [R_K/R_{K^*}]^{[1.1,6]} =& [R_K/R_{K^*}]_{\text{SM}}^{[1.1,6]}\frac{1+0.24 \, \text{Re}[C_{9, \text{NP}}^\mu]-0.26 \, \text{Re}[C_{10,\text{NP}}^\mu]}{1 + 0.18 \, \text{Re}[C_{9, \text{NP}}^\mu]-0.29 \, \text{Re}[C_{10,\text{NP}}^\mu]} \\
    \approx& [R_K/R_{K^*}]_{\text{SM}}^{[1.1,6]} (1+0.06 \, \text{Re}[C_{9, \text{NP}}^\mu]+0.03 \, \text{Re}[C_{10,\text{NP}}^\mu]) \,,
\end{align}
up to linear order in the Wilson coefficient, where the values of $q^2_\mathrm{min,max}$ are in GeV$^2/c^2$, and
\begin{align}
    O_9^\mu = (\bar b_L \gamma_\alpha s_L)(\bar \mu \gamma^\alpha \mu) \,,
    \qquad
    O_{10}^\mu = (\bar b_L \gamma_\alpha s_L)(\bar \mu \gamma^\alpha\gamma_5 \mu) \,.
\end{align}

\subsection{\texorpdfstring{LFV $\mu$ and $\tau$ decays}{LFV mu and tau decays}}

The $\ell_j \to 3\ell_i$ decays can be expressed as (keeping only the relevant contributions for our analysis) \cite{Plakias:2023esq}
\begin{align}
    \cB_{\ell_j \to 3\ell_i} = \frac{\tau_{\ell_j}m_{\ell_j}^5}{768 \pi^3} |[L_{ee}^{V,LL}]_{iiij}|^2 \,,
\end{align}
where $[O_{ee}^{V,LL}]_{ijkl} = (\bar e_{L,i} \gamma_\mu e_{L,j})(\bar e_{L,k} \gamma_\mu e_{L,l})$.
The semileptonic $\tau \to K_S \ell$ is given by \cite{Plakias:2023esq}
\begin{align}
\begin{split}
   \mathcal{B}(\tau \to \ell_i K_S)= \tau_{\tau} \, &\dfrac{f_{K}^2 m_{\tau}^3}{256 \pi v^4} \left(1-\dfrac{m_{K^0}^2}{m_{\tau}^2}\right)^2\\
   &\times\bigg{\lbrace} \bigg{|}C_{VA} +  \dfrac{m_{K^0}^2\,{C}_{SP}}{m_{\tau}(m_s+m_d)}\bigg{|}^2+ \bigg{|}C_{AA} - \dfrac{m_{K^0}^2\,{C}_{PP}}{m_{\tau}(m_s+m_d)}\bigg{|}^2\bigg{\rbrace} \,.
\end{split}
\end{align}

\section{Additional results}\label{app:plots}

In this section we report all the numerical results and additional plots which we omitted from the main text for the sake of brevity.

\subsection{Individual fits}

\begin{figure}[htbp]
    \centering
    \includegraphics[width=\linewidth]{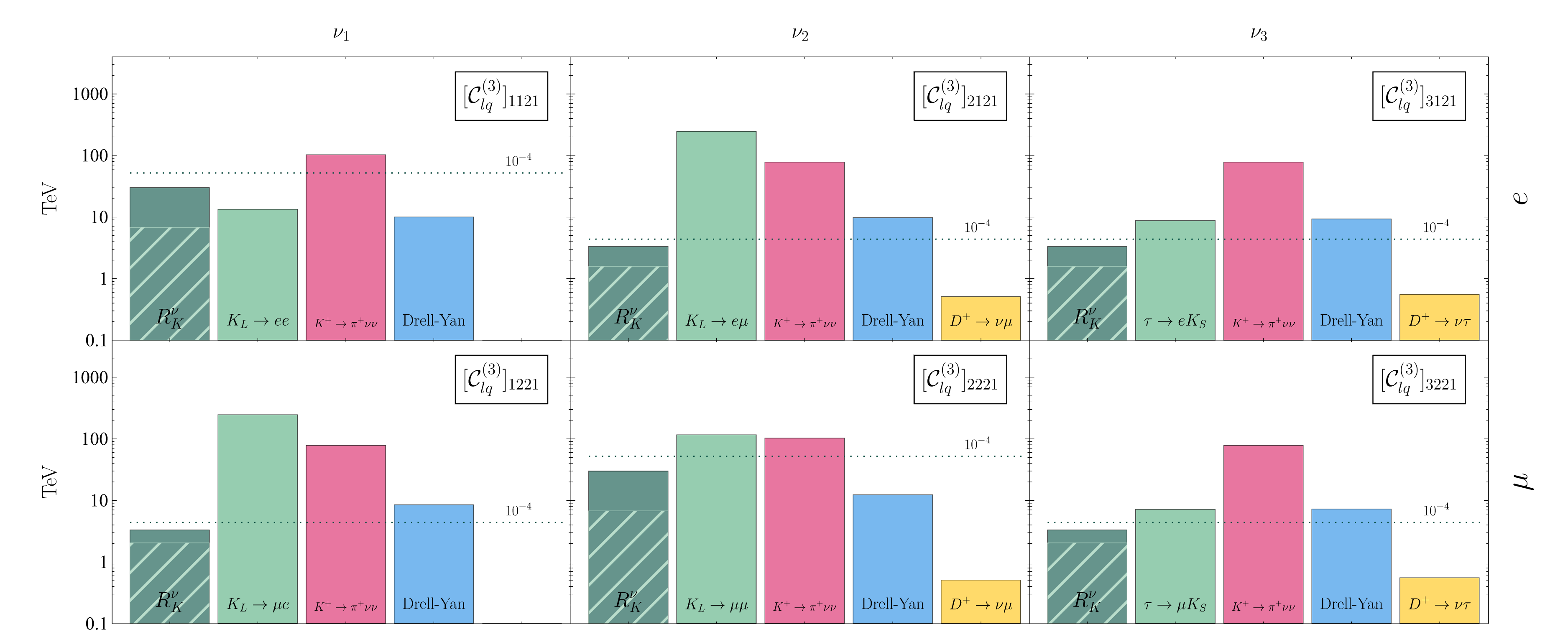}
    \caption{Bounds on the NP scale from the different flavour processes connected to $R_K^\nu$ under the assunmption of NP in $\cC_{\ell q}^{(3)}$. For $R_K^\nu$ the hatched bars represent the bound from the current measurement, the solid bars the future projection with a precision of $3\times 10^{-4}$, and the dotted line the optimistic scenario of $1\times 10^{-4}$.}
    \label{fig:lq3_bounds}
\end{figure}

\begin{figure}
    \centering
    \includegraphics[width=0.9\linewidth]{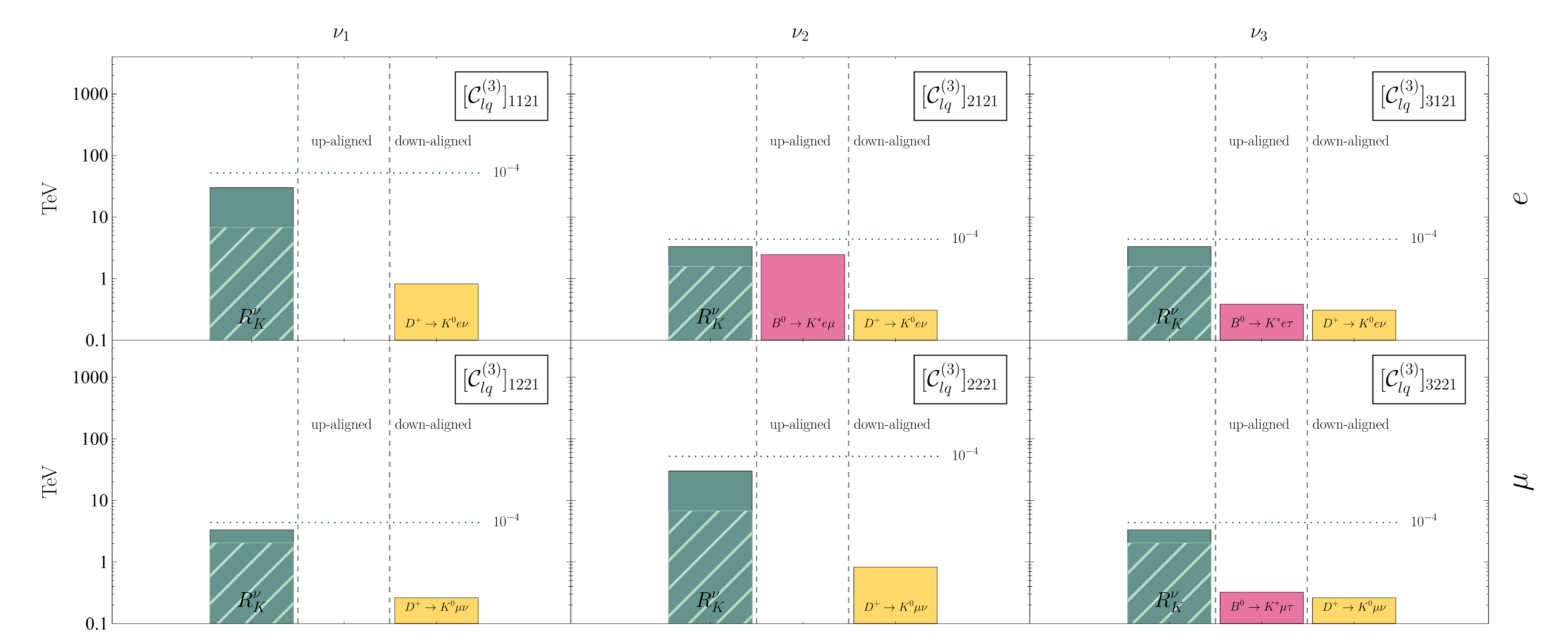}
    \caption{ Main effects due to left-handed quark misalignment under the assumption of NP in $\cC_{\ell q}^{(3)}$, compared to the $R_K^\nu$ sensitivity. For $R_K^\nu$ the hatched bars represent the bound from the current measurement, the solid bars the future projection with a precision of $3\times 10^{-4}$, and the dotted line the optimistic scenario of $1\times 10^{-4}$.}
    \label{fig:lq3basis}
\end{figure}

\begin{figure}[htbp]
    \centering
    \includegraphics[width=\linewidth]{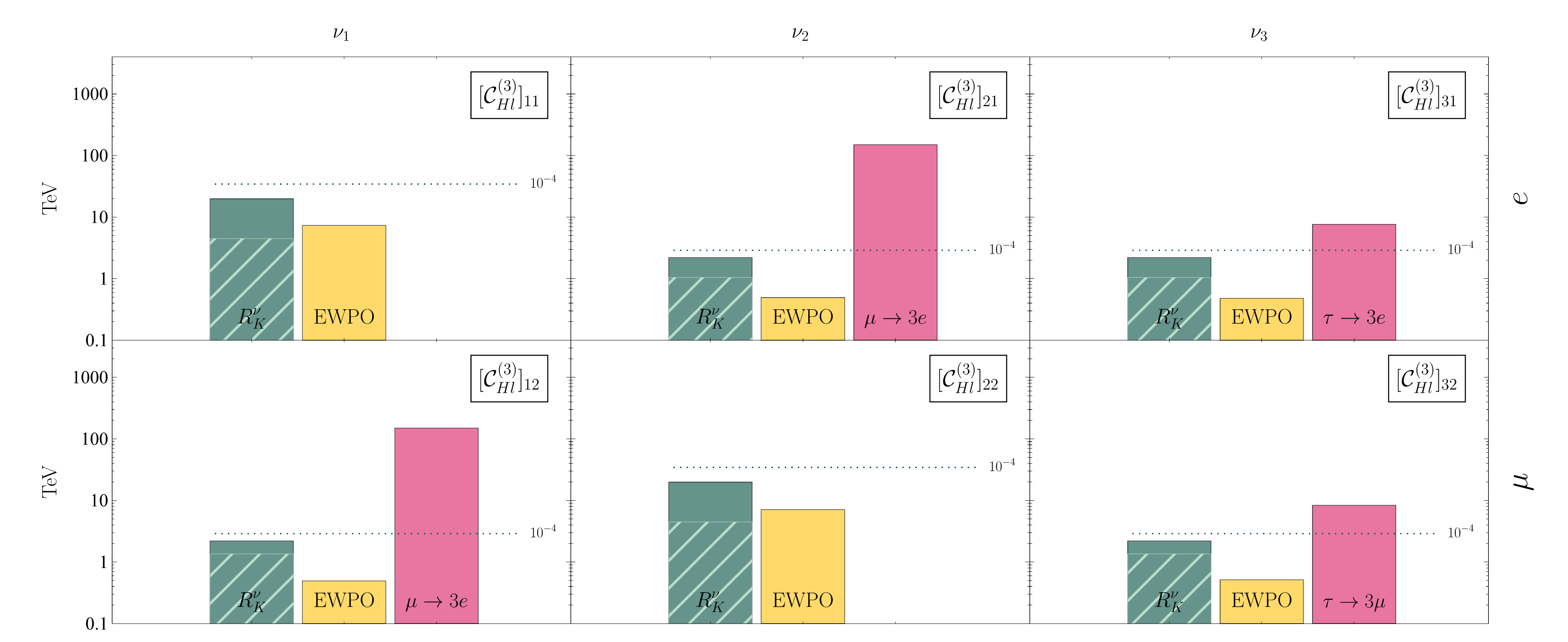}
    \caption{Bounds on the NP scale from the different flavour processes connected to $R_{K\pi}^\mu$ under the assunmption of NP in $[\cC_{H \ell}^{(3)}]_{ij}$. For $R_K^\nu$ the hatched bars represent the bound from the current measurement, the solid bars the future projection with a precision of $3\times 10^{-4}$, and the dotted line the optimistic scenario of $1\times 10^{-4}$.}
    \label{fig:Hl3_bounds}
\end{figure}

\begin{figure}[htbp]
    \centering
    \includegraphics[width=0.5\linewidth]{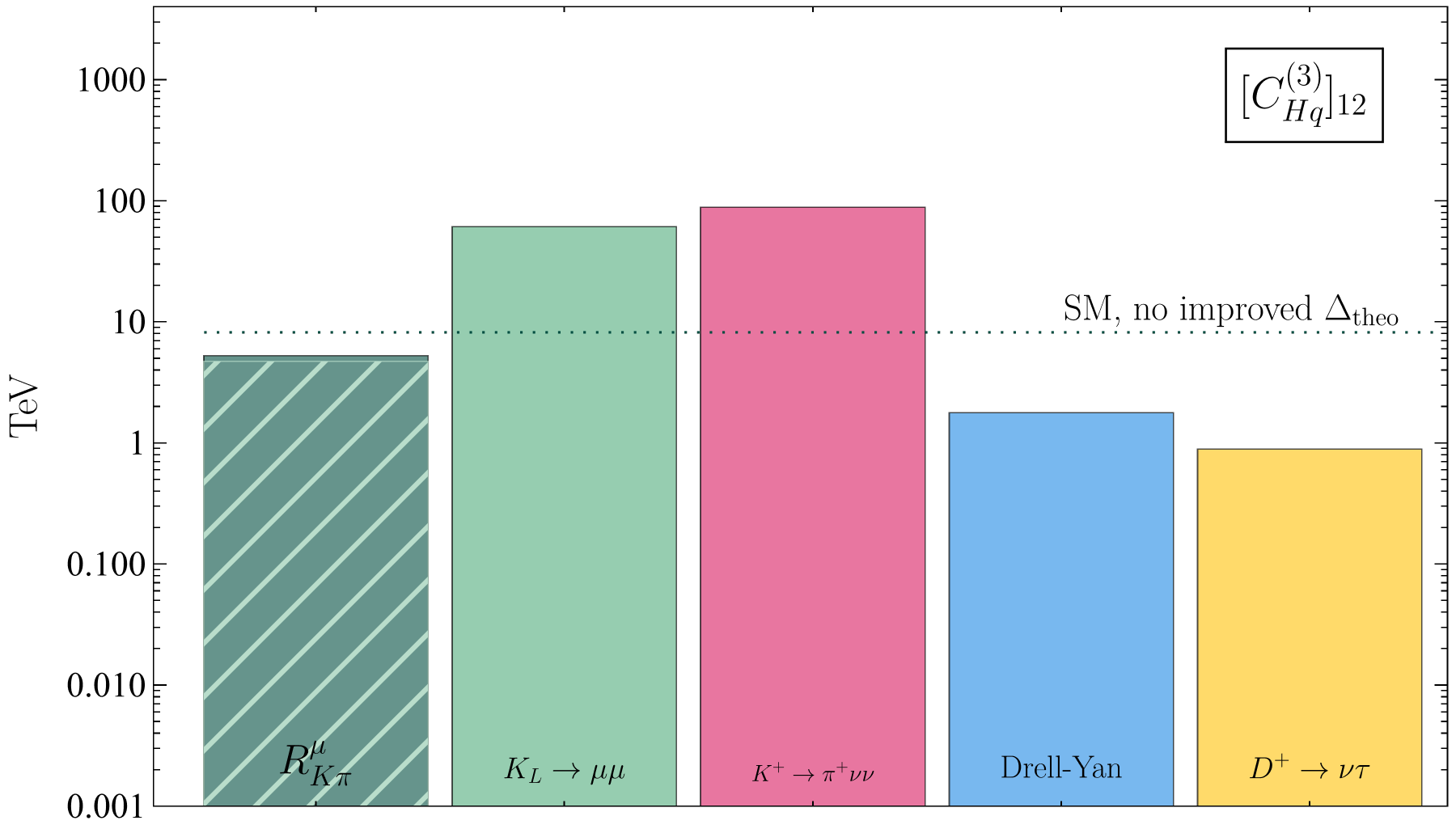}
    \caption{Bounds on the NP scale from the different flavour processes connected to $R_{K\pi}^\mu$ under the assunmption of NP in $[\cC_{H q}^{(3)}]_{12}$. For $R_K^\nu$ the hatched bars represent the bound from the current measurement, the solid bars the future projection with a precision of $3\times 10^{-4}$, and the dotted line the optimistic scenario of $1\times 10^{-4}$.}
    \label{fig:Hq3_bounds}
\end{figure}

Figures \ref{fig:lq3_bounds}-\ref{fig:Hq3_bounds} contain the single-coefficient bounds for the SMEFT operators in Table \ref{tab:smeftops}, except for the scalar ones.
Moreover, all bounds obtained from the individual fits to the $R_K^\nu$ ratio are collected in Table \ref{tab:ScalesRKnu}.
Finally, for the coefficient $[\cC_{Hq}^{(3)}]_{12}$, to which $R_K^\nu$ is insensitive, we find
\begin{align}
    \Lambda^{\rm current} \gtrsim 4.6\text{ TeV}\,, \qquad \Lambda^{\rm future (SM)} \gtrsim 8.2\text{ TeV} \,.
\end{align}


\begin{table}[]
\renewcommand{\arraystretch}{1.4}
\setlength{\tabcolsep}{8pt}
\centering
\begin{tabular}{|ll|lll|lll|}
\hline
\multicolumn{2}{|l|}{\multirow{2}{*}{Coefficient}} &
  \multicolumn{6}{c|}{lepton indices $ij$} \\
\multicolumn{2}{|l|}{} &
  11 &
  12 &
  13 &
  21 &
  22 &
  23 \\ \hline
\multirow{2}{*}{$[C_{ledq}]_{ij21}$} &
  current & 420 & 98 & 79 & 8.9 & 29 & 8.9 \\ 
  & future & 1860 & 206 & 206 & 14.3 & 129 & 14.3 \\ \hline
\multirow{2}{*}{$[C_{lequ}^{(1)}]_{ij21}$} &
  current & 418 & 97 & 97 & 8.8 & 29 & 8.8 \\ 
  & future & 1850 & 205 & 205 & 14.2 & 129 & 14.2 \\ \hline
\multirow{2}{*}{$[C_{lq}^{(3)}]_{ij21}$} &
  current & 6.8 & 1.6 & 1.6 & 2.1 & 6.8 & 2.1 \\
  & future & 30 & 3.3 & 3.3 & 3.3 & 30 & 3.3 \\ \hline
\multirow{2}{*}{$[C_{Hl}^{(3)}]_{ij21}$} &
  current & 4.5 & 1.0 & 1.0 & 1.4 & 4.5 & 1.4 \\
  & future & 19.9 & 2.2 & 2.2 & 2.2 & 19.9 & 2.2 \\ \hline
\end{tabular}
\caption{The $2\sigma$ bounds in TeV on the different coefficients from current and future measurements of $R_K^\nu$.}
\label{tab:ScalesRKnu}
\end{table}

\subsection{Global kaon fit}

\begin{figure}
    \centering
    \includegraphics[width=0.7\linewidth]{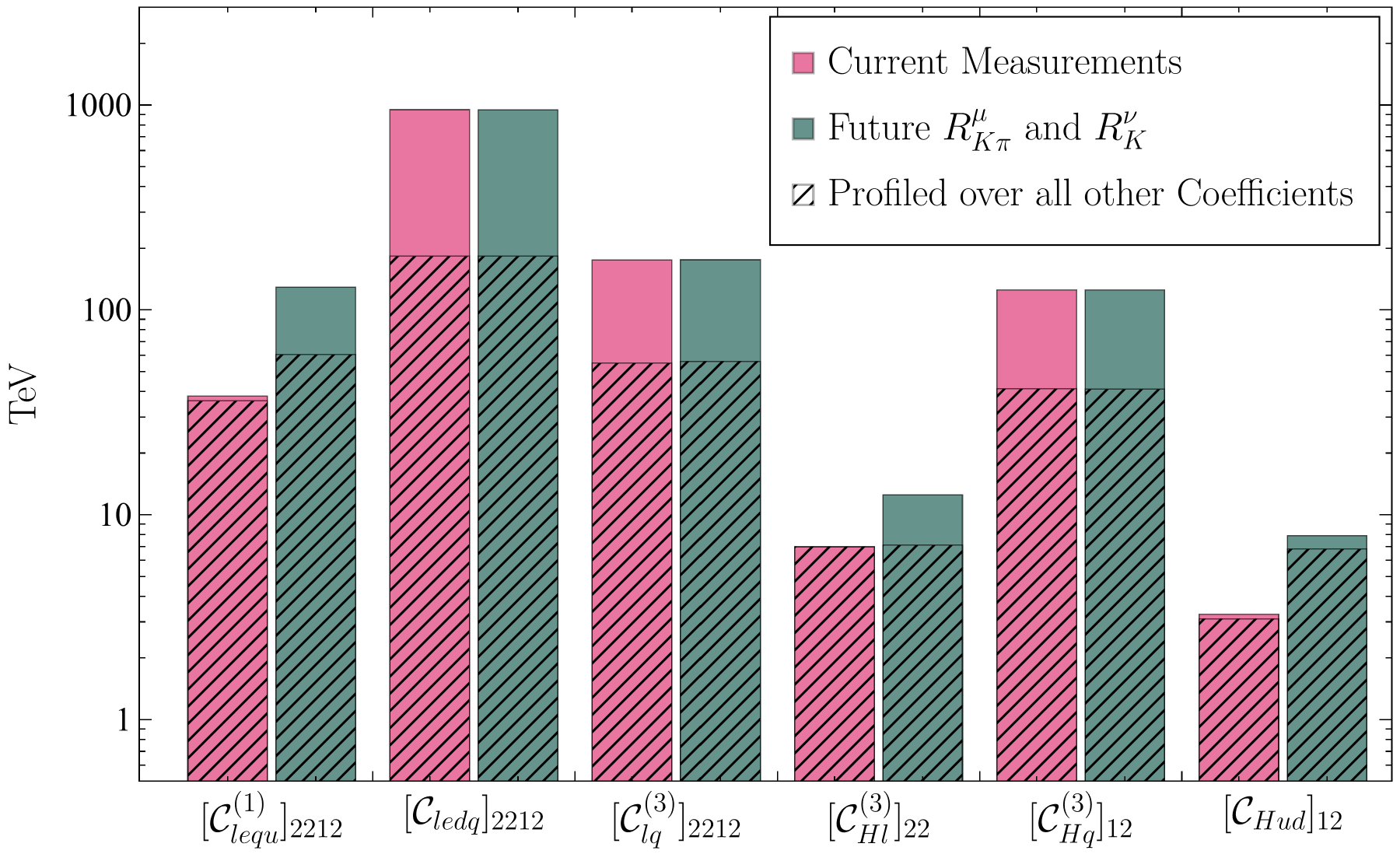}
    \caption{Individual and marginalised fits on the five SMEFT coefficients affecting the ratio $R_K^\nu$. For all future projections we assumed SM-like central values.}
    \label{fig:marginalised}
\end{figure}

Figure \ref{fig:marginalised} contains the individual and profiled fits for all six SMEFT coefficients in Table \ref{tab:smeftops}, with muon indices on the lepton side.

\bibliographystyle{JHEP}
\bibliography{refs}

\providecommand{\href}[2]{#2}\begingroup\raggedright\begin{thebibliography}{100}

\bibitem{Masiero:2005wr}
A.~Masiero, P.~Paradisi and R.~Petronzio, \emph{{Probing new physics through
  $\mu - e$ universality in $K\to l \nu$}},
  \href{https://doi.org/10.1103/PhysRevD.74.011701}{\emph{Phys. Rev. D}
  {\bfseries 74} (2006) 011701}
  [\href{https://arxiv.org/abs/hep-ph/0511289}{{\ttfamily hep-ph/0511289}}].

\bibitem{FlaviaNetWorkingGrouponkaonDecays:2010lot}
{\scshape FlaviaNet Working Group on Kaon Decays} collaboration, \emph{{An
  Evaluation of $|V_{us}|$ and precise tests of the Standard Model from world
  data on leptonic and semileptonic kaon decays}},
  \href{https://doi.org/10.1140/epjc/s10052-010-1406-3}{\emph{Eur. Phys. J. C}
  {\bfseries 69} (2010) 399} [\href{https://arxiv.org/abs/1005.2323}{{\ttfamily
  1005.2323}}].

\bibitem{FlaviaNetWorkingGrouponkaonDecays:2008hpm}
{\scshape FlaviaNet Working Group on Kaon Decays} collaboration,
  \emph{{Precision tests of the Standard Model with leptonic and semileptonic
  kaon decays}},  in \emph{{5th International Workshop on e+ e- Collisions from
  Phi to Psi}}, 1, 2008 [\href{https://arxiv.org/abs/0801.1817}{{\ttfamily
  0801.1817}}].

\bibitem{FlavourLatticeAveragingGroupFLAG:2024oxs}
{\scshape Flavour Lattice Averaging Group (FLAG)} collaboration, \emph{{FLAG
  review 2024}}, \href{https://doi.org/10.1103/nfzp-p5dn}{\emph{Phys. Rev. D}
  {\bfseries 113} (2026) 014508}
  [\href{https://arxiv.org/abs/2411.04268}{{\ttfamily 2411.04268}}].

\bibitem{Aebischer:2025mwl}
J.~Aebischer et~al., \emph{{Kaon physics: a cornerstone for future
  discoveries}}, \href{https://doi.org/10.1088/1361-6471/ae05b4}{\emph{J. Phys.
  G} {\bfseries 52} (2025) 100501}
  [\href{https://arxiv.org/abs/2503.22256}{{\ttfamily 2503.22256}}].

\bibitem{Masiero:2008cb}
A.~Masiero, P.~Paradisi and R.~Petronzio, \emph{{Anatomy and Phenomenology of
  the Lepton Flavor Universality in SUSY Theories}},
  \href{https://doi.org/10.1088/1126-6708/2008/11/042}{\emph{JHEP} {\bfseries
  11} (2008) 042} [\href{https://arxiv.org/abs/0807.4721}{{\ttfamily
  0807.4721}}].

\bibitem{Ellis:2008st}
J.~Ellis, S.~Lola and M.~Raidal, \emph{{Supersymmetric Grand Unification and
  Lepton Universality in $K\to l \nu$}},
  \href{https://doi.org/10.1016/j.nuclphysb.2008.12.006}{\emph{Nucl. Phys. B}
  {\bfseries 812} (2009) 128}
  [\href{https://arxiv.org/abs/0809.5211}{{\ttfamily 0809.5211}}].

\bibitem{Girrbach:2012km}
J.~Girrbach and U.~Nierste, \emph{{$\Gamma(K\to e \nu)/\Gamma(K\to \mu \nu)$ in
  the Minimal Supersymmetric Standard Model}},
  \href{https://arxiv.org/abs/1202.4906}{{\ttfamily 1202.4906}}.

\bibitem{Fonseca:2012kr}
R.M.~Fonseca, J.C.~Romao and A.M.~Teixeira, \emph{{Revisiting the $\Gamma(K \to
  e \nu)/\Gamma(K \to \mu \nu)$ Ratio in Supersymmetric Unified Models}},
  \href{https://doi.org/10.1140/epjc/s10052-012-2228-2}{\emph{Eur. Phys. J. C}
  {\bfseries 72} (2012) 2228}
  [\href{https://arxiv.org/abs/1205.1411}{{\ttfamily 1205.1411}}].

\bibitem{Abada:2012mc}
A.~Abada, D.~Das, A.M.~Teixeira, A.~Vicente and C.~Weiland, \emph{{Tree-level
  lepton universality violation in the presence of sterile neutrinos: impact
  for $R_K$ and $R_\pi$}},
  \href{https://doi.org/10.1007/JHEP02(2013)048}{\emph{JHEP} {\bfseries 02}
  (2013) 048} [\href{https://arxiv.org/abs/1211.3052}{{\ttfamily 1211.3052}}].

\bibitem{Abada:2013aba}
A.~Abada, A.M.~Teixeira, A.~Vicente and C.~Weiland, \emph{{Sterile neutrinos in
  leptonic and semileptonic decays}},
  \href{https://doi.org/10.1007/JHEP02(2014)091}{\emph{JHEP} {\bfseries 02}
  (2014) 091} [\href{https://arxiv.org/abs/1311.2830}{{\ttfamily 1311.2830}}].

\bibitem{Jung:2010ik}
M.~Jung, A.~Pich and P.~Tuzon, \emph{{Charged-Higgs phenomenology in the
  Aligned two-Higgs-doublet model}},
  \href{https://doi.org/10.1007/JHEP11(2010)003}{\emph{JHEP} {\bfseries 11}
  (2010) 003} [\href{https://arxiv.org/abs/1006.0470}{{\ttfamily 1006.0470}}].

\bibitem{Gonzalez-Alonso:2016etj}
M.~Gonz{\'a}lez-Alonso and J.~Martin~Camalich, \emph{{Global
  Effective-Field-Theory analysis of New-Physics effects in (semi)leptonic kaon
  decays}}, \href{https://doi.org/10.1007/JHEP12(2016)052}{\emph{JHEP}
  {\bfseries 12} (2016) 052}
  [\href{https://arxiv.org/abs/1605.07114}{{\ttfamily 1605.07114}}].

\bibitem{deBlas:2025gyz}
J.~de~Blas et~al., \emph{{Physics Briefing Book: Input for the 2026 update of
  the European Strategy for Particle Physics}},
  \href{https://arxiv.org/abs/2511.03883}{{\ttfamily 2511.03883}}.

\bibitem{Boyle:2026lrz}
P.~Boyle, N.H.~Christ, X.~Feng, T.~Izubuchi, L.~Jin, C.T.~Sachrajda et~al.,
  \emph{{First Lattice QCD Determination of Lepton-Flavor-Universality Ratios
  in Light-Meson Leptonic Decays}},
  \href{https://arxiv.org/abs/2607.22358}{{\ttfamily 2607.22358}}.

\bibitem{DiPalma:2025iud}
R.~Di~Palma, R.~Frezzotti, G.~Gagliardi, V.~Lubicz, G.~Martinelli,
  C.T.~Sachrajda et~al., \emph{{Kaon radiative leptonic decay rates from
  lattice QCD simulations at the physical point}},
  \href{https://doi.org/10.1103/r55k-cnqg}{\emph{Phys. Rev. D} {\bfseries 111}
  (2025) 114523} [\href{https://arxiv.org/abs/2504.08680}{{\ttfamily
  2504.08680}}].

\bibitem{DiCarlo:2019thl}
M.~Di~Carlo, D.~Giusti, V.~Lubicz, G.~Martinelli, C.T.~Sachrajda, F.~Sanfilippo
  et~al., \emph{{Light-meson leptonic decay rates in lattice QCD+QED}},
  \href{https://doi.org/10.1103/PhysRevD.100.034514}{\emph{Phys. Rev. D}
  {\bfseries 100} (2019) 034514}
  [\href{https://arxiv.org/abs/1904.08731}{{\ttfamily 1904.08731}}].

\bibitem{PIONEER:2022alm}
{\scshape PIONEER} collaboration, \emph{{Testing Lepton Flavor Universality and
  CKM Unitarity with Rare Pion Decays in the PIONEER experiment}},  in
  \emph{{Snowmass 2021}}, 3, 2022
  [\href{https://arxiv.org/abs/2203.05505}{{\ttfamily 2203.05505}}].

\bibitem{PIONEER:2022yag}
{\scshape PIONEER} collaboration, \emph{{PIONEER: Studies of Rare Pion
  Decays}},  \href{https://arxiv.org/abs/2203.01981}{{\ttfamily 2203.01981}}.

\bibitem{lubos:LatticeLab2026}
L.~Bician and M.~Koval, \emph{${R}_{K} = \gamma({K}_{e2})/\gamma({K}_{\mu2})$
  from {NA62}},  in \emph{From Lattice to the Lab: Illuminating Kaon Decays},
  2026,
  \href{https://agenda.infn.it/event/50498/contributions/288100/attachments/147616/225451/RK\_at\_NA62.pdf}{https://agenda.infn.it/event/50498/contributions/288100/attachments/147616/225451/RK\_at\_NA62.pdf}.

\bibitem{ParticleDataGroup:2026mpi}
{\scshape Particle Data Group} collaboration, \emph{{Review of Particle
  Physics}}, \href{https://doi.org/10.1142/s0217751x26300115}{\emph{Int. J.
  Mod. Phys. A} {\bfseries 41} (2026) 2630011}.

\bibitem{VanDijk:2641500}
M.~Van~Dijk and M.~Rosenthal, \emph{{Target studies for the proposed KLEVER
  experiment}}, .

\bibitem{NA62:2017rwk}
{\scshape NA62} collaboration, \emph{{The Beam and detector of the NA62
  experiment at CERN}},
  \href{https://doi.org/10.1088/1748-0221/12/05/P05025}{\emph{JINST} {\bfseries
  12} (2017) P05025} [\href{https://arxiv.org/abs/1703.08501}{{\ttfamily
  1703.08501}}].

\bibitem{Bovet:1975bx}
C.~Bovet, S.~Milner and A.~Placci, \emph{{The Cedar Project. Cherenkov
  Differential Counters with Achromatic Ring Focus}},
  \href{https://doi.org/10.1109/TNS.1978.4329375}{\emph{IEEE Trans. Nucl. Sci.}
  {\bfseries 25} (1978) 572}.

\bibitem{ANDRIEUX2022166069}
V.~Andrieux, A.~Berlin, N.~Doshita, M.~Finger, M.~Finger, F.~Gautheron et~al.,
  \emph{The large compass polarized solid ammonia target for drell–yan
  measurements with a pion beam},
  \href{https://doi.org/https://doi.org/10.1016/j.nima.2021.166069}{\emph{Nuclear
  Instruments and Methods in Physics Research Section A: Accelerators,
  Spectrometers, Detectors and Associated Equipment} {\bfseries 1025} (2022)
  166069}.

\bibitem{NA62:2024pjp}
{\scshape NA62} collaboration, \emph{{Observation of the $ {K}^{+}\to
  {\pi}^{+}\nu \overline{\nu} $ decay and measurement of its branching ratio}},
  \href{https://doi.org/10.1007/JHEP02(2025)191}{\emph{JHEP} {\bfseries 02}
  (2025) 191} [\href{https://arxiv.org/abs/2412.12015}{{\ttfamily
  2412.12015}}].

\bibitem{Seng:2021nar}
C.-Y.~Seng, D.~Galviz, W.J.~Marciano and U.-G.~Mei{\ss}ner, \emph{{Update on
  $|V_{us}|$ and $|_{Vus}/V_{ud}|$ from semileptonic kaon and pion decays}},
  \href{https://doi.org/10.1103/PhysRevD.105.013005}{\emph{Phys. Rev. D}
  {\bfseries 105} (2022) 013005}
  [\href{https://arxiv.org/abs/2107.14708}{{\ttfamily 2107.14708}}].

\bibitem{CKMMatrixReview2026}
A.~Ceccucci, Z.~Ligeti and Y.~Sakai, \emph{{CKM Quark-Mixing Matrix}},  in
  \emph{{Review of Particle Physics}}, vol.~41, p.~2630011 (2026).

\bibitem{3161023}
E.~Blucher, W.D.~Marciano and G.~D'Ambrosio, \emph{{Vud, Vus the Cabibbo Angle,
  and CKM Unitarity}}, {\emph{Particle Data Group} (2026) }.

\bibitem{Cirigliano:2023nol}
V.~Cirigliano, W.~Dekens, J.~de~Vries, E.~Mereghetti and T.~Tong,
  \emph{{Anomalies in global SMEFT analyses. A case study of first-row CKM
  unitarity}}, \href{https://doi.org/10.1007/JHEP03(2024)033}{\emph{JHEP}
  {\bfseries 03} (2024) 033}
  [\href{https://arxiv.org/abs/2311.00021}{{\ttfamily 2311.00021}}].

\bibitem{Jenkins:2017jig}
E.E.~Jenkins, A.V.~Manohar and P.~Stoffer, \emph{{Low-Energy Effective Field
  Theory below the Electroweak Scale: Operators and Matching}},
  \href{https://doi.org/10.1007/JHEP03(2018)016}{\emph{JHEP} {\bfseries 03}
  (2018) 016} [\href{https://arxiv.org/abs/1709.04486}{{\ttfamily
  1709.04486}}].

\bibitem{Grzadkowski:2010es}
B.~Grzadkowski, M.~Iskrzynski, M.~Misiak and J.~Rosiek, \emph{{Dimension-Six
  Terms in the Standard Model Lagrangian}},
  \href{https://doi.org/10.1007/JHEP10(2010)085}{\emph{JHEP} {\bfseries 10}
  (2010) 085} [\href{https://arxiv.org/abs/1008.4884}{{\ttfamily 1008.4884}}].

\bibitem{Jenkins2018}
E.E.~Jenkins, A.V.~Manohar and P.~Stoffer, \emph{Low-energy effective field
  theory below the electroweak scale: operators and matching},
  \href{https://doi.org/10.1007/jhep03(2018)016}{\emph{Journal of High Energy
  Physics} {\bfseries 2018} (2018) }.

\bibitem{Fuentes-Martin:2020zaz}
J.~Fuentes-Martin, P.~Ruiz-Femenia, A.~Vicente and J.~Virto, \emph{{DsixTools
  2.0: The Effective Field Theory Toolkit}},
  \href{https://doi.org/10.1140/epjc/s10052-020-08778-y}{\emph{Eur. Phys. J. C}
  {\bfseries 81} (2021) 167}
  [\href{https://arxiv.org/abs/2010.16341}{{\ttfamily 2010.16341}}].

\bibitem{Allwicher:2022gkm}
L.~Allwicher, D.A.~Faroughy, F.~Jaffredo, O.~Sumensari and F.~Wilsch,
  \emph{{Drell-Yan tails beyond the Standard Model}},
  \href{https://doi.org/10.1007/JHEP03(2023)064}{\emph{JHEP} {\bfseries 03}
  (2023) 064} [\href{https://arxiv.org/abs/2207.10714}{{\ttfamily
  2207.10714}}].

\bibitem{Allwicher:2022mcg}
L.~Allwicher, D.A.~Faroughy, F.~Jaffredo, O.~Sumensari and F.~Wilsch,
  \emph{{HighPT: A tool for~ high-$p_T$ Drell-Yan tails beyond the standard
  model}}, \href{https://doi.org/10.1016/j.cpc.2023.108749}{\emph{Comput. Phys.
  Commun.} {\bfseries 289} (2023) 108749}
  [\href{https://arxiv.org/abs/2207.10756}{{\ttfamily 2207.10756}}].

\bibitem{Allwicher:2026loe}
L.~Allwicher and M.~Bordone, \emph{{Implications of $K\to \pi \nu \bar \nu$ for
  new physics in $B$ decays}},
  \href{https://arxiv.org/abs/2607.24494}{{\ttfamily 2607.24494}}.

\bibitem{Allwicher:2024ncl}
L.~Allwicher, M.~Bordone, G.~Isidori, G.~Piazza and A.~Stanzione,
  \emph{{Probing third-generation New Physics with
  K{\textrightarrow}{\ensuremath{\pi}}{\ensuremath{\nu}}{\ensuremath{\nu}}{\textasciimacron}
  and
  B{\textrightarrow}K({\textasteriskcentered}){\ensuremath{\nu}}{\ensuremath{\nu}}{\textasciimacron}}},
  \href{https://doi.org/10.1016/j.physletb.2025.139295}{\emph{Phys. Lett. B}
  {\bfseries 861} (2025) 139295}
  [\href{https://arxiv.org/abs/2410.21444}{{\ttfamily 2410.21444}}].

\bibitem{Cirigliano:2022yyo}
V.~Cirigliano, A.~Crivellin, M.~Hoferichter and M.~Moulson, \emph{{Scrutinizing
  CKM unitarity with a new measurement of the
  K{\ensuremath{\mu}}3/K{\ensuremath{\mu}}2 branching fraction}},
  \href{https://doi.org/10.1016/j.physletb.2023.137748}{\emph{Phys. Lett. B}
  {\bfseries 838} (2023) 137748}
  [\href{https://arxiv.org/abs/2208.11707}{{\ttfamily 2208.11707}}].

\bibitem{10.1093/ptep/ptac097}
P.D.~Group, R.L.~Workman, V.D.~Burkert, V.~Crede, E.~Klempt, U.~Thoma et~al.,
  \emph{Review of particle physics},
  \href{https://doi.org/10.1093/ptep/ptac097}{\emph{Progress of Theoretical and
  Experimental Physics} {\bfseries 2022} (2022) 083C01}
  [\href{https://arxiv.org/abs/https://academic.oup.com/ptep/article-pdf/2022/8/083C01/49175539/ptac097.pdf}{{\ttfamily
  https://academic.oup.com/ptep/article-pdf/2022/8/083C01/49175539/ptac097.pdf}}].

\bibitem{Eboli:2025vks}
O.J.P.~{\'E}boli, L.P.S.~Leal, M.~Martines and O.~Sumensari, \emph{{Flavor
  constraints from $pp\to Vh$ and $pp\to VW$ at the LHC}},
  \href{https://doi.org/10.1007/JHEP03(2026)054}{\emph{JHEP} {\bfseries 03}
  (2026) 054} [\href{https://arxiv.org/abs/2509.08437}{{\ttfamily
  2509.08437}}].

\bibitem{Crivellin:2021bkd}
A.~Crivellin, M.~Hoferichter, M.~Kirk, C.A.~Manzari and L.~Schnell,
  \emph{{First-generation new physics in simplified models: from low-energy
  parity violation to the LHC}},
  \href{https://doi.org/10.1007/JHEP10(2021)221}{\emph{JHEP} {\bfseries 10}
  (2021) 221} [\href{https://arxiv.org/abs/2107.13569}{{\ttfamily
  2107.13569}}].

\bibitem{Aoki:2017spo}
{\scshape JLQCD} collaboration, \emph{{Chiral behavior of $K \to \pi l \nu$
  decay form factors in lattice QCD with exact chiral symmetry}},
  \href{https://doi.org/10.1103/PhysRevD.96.034501}{\emph{Phys. Rev. D}
  {\bfseries 96} (2017) 034501}
  [\href{https://arxiv.org/abs/1705.00884}{{\ttfamily 1705.00884}}].

\bibitem{FermilabLattice:2018zqv}
{\scshape Fermilab Lattice, MILC} collaboration, \emph{{$|V_{us}|$ from
  $K_{\ell 3}$ decay and four-flavor lattice QCD}},
  \href{https://doi.org/10.1103/PhysRevD.99.114509}{\emph{Phys. Rev. D}
  {\bfseries 99} (2019) 114509}
  [\href{https://arxiv.org/abs/1809.02827}{{\ttfamily 1809.02827}}].

\bibitem{Carrasco:2016kpy}
N.~Carrasco, P.~Lami, V.~Lubicz, L.~Riggio, S.~Simula and C.~Tarantino,
  \emph{{$K \to \pi$ semileptonic form factors with $N_f=2+1+1$ twisted mass
  fermions}}, \href{https://doi.org/10.1103/PhysRevD.93.114512}{\emph{Phys.
  Rev. D} {\bfseries 93} (2016) 114512}
  [\href{https://arxiv.org/abs/1602.04113}{{\ttfamily 1602.04113}}].

\bibitem{Ishikawa:2022ulx}
{\scshape PACS} collaboration, \emph{{K{\ensuremath{\ell}}3 form factors at the
  physical point: Toward the continuum limit}},
  \href{https://doi.org/10.1103/PhysRevD.106.094501}{\emph{Phys. Rev. D}
  {\bfseries 106} (2022) 094501}
  [\href{https://arxiv.org/abs/2206.08654}{{\ttfamily 2206.08654}}].

\bibitem{Bryman:2021sos}
D.~Bryman, V.~Cirigliano, A.~Crivellin and G.~Inguglia, \emph{{Testing Lepton
  Flavor Universality with Pion, Kaon, Tau, and Beta Decays}},
  \href{https://doi.org/10.1146/annurev-nucl-110121-051223}{\emph{Ann. Rev.
  Nucl. Part. Sci.} {\bfseries 72} (2022) 69}
  [\href{https://arxiv.org/abs/2111.05338}{{\ttfamily 2111.05338}}].

\bibitem{Hiller:2014ula}
G.~Hiller and M.~Schmaltz, \emph{{Diagnosing lepton-nonuniversality in $b \to s
  \ell \ell$}}, \href{https://doi.org/10.1007/JHEP02(2015)055}{\emph{JHEP}
  {\bfseries 02} (2015) 055} [\href{https://arxiv.org/abs/1411.4773}{{\ttfamily
  1411.4773}}].

\bibitem{Guadagnoli:2022jdr}
D.~Guadagnoli and P.~Koppenburg, \emph{{Lepton Flavor Violation and Lepton
  Flavor Universality Violation in $b$ and $c$ Decays}},
  \href{https://doi.org/10.1146/annurev-nucl-101122-045442}{\emph{Ann. Rev.
  Nucl. Part. Sci.} {\bfseries 73} (2023) 1}
  [\href{https://arxiv.org/abs/2207.01851}{{\ttfamily 2207.01851}}].

\bibitem{Bordone:2025elp}
M.~Bordone, G.~Isidori, C.~Mayer and J.-N.~Toelstede, \emph{{Probing lepton
  flavour universality with $\Lambda _b$ decays to $\tau ^+\tau ^-$ final
  states}}, \href{https://doi.org/10.1140/epjc/s10052-026-15464-y}{\emph{Eur.
  Phys. J. C} {\bfseries 86} (2026) 398}
  [\href{https://arxiv.org/abs/2511.05654}{{\ttfamily 2511.05654}}].

\bibitem{Bordone:2021olx}
M.~Bordone, C.~Cornella, G.~Isidori and M.~K{\"o}nig, \emph{{The LFU ratio
  $R_\pi $ in the Standard Model and beyond}},
  \href{https://doi.org/10.1140/epjc/s10052-021-09618-3}{\emph{Eur. Phys. J. C}
  {\bfseries 81} (2021) 850}
  [\href{https://arxiv.org/abs/2101.11626}{{\ttfamily 2101.11626}}].

\bibitem{10.1093/ptep/ptaa104}
P.D.~Group, P.A.~Zyla, R.M.~Barnett, J.~Beringer, O.~Dahl, D.A.~Dwyer et~al.,
  \emph{Review of particle physics},
  \href{https://doi.org/10.1093/ptep/ptaa104}{\emph{Progress of Theoretical and
  Experimental Physics} {\bfseries 2020} (2020) 083C01}
  [\href{https://arxiv.org/abs/https://academic.oup.com/ptep/article-pdf/2020/8/083C01/34673722/ptaa104.pdf}{{\ttfamily
  https://academic.oup.com/ptep/article-pdf/2020/8/083C01/34673722/ptaa104.pdf}}].

\bibitem{Becirevic2021}
D.~Bečirević, F.~Jaffredo, A.~Peñuelas and O.~Sumensari, \emph{New physics
  effects in leptonic and semileptonic decays},
  \href{https://doi.org/10.1007/jhep05(2021)175}{\emph{Journal of High Energy
  Physics} {\bfseries 2021} (2021) }.

\bibitem{Belle:2015pkj}
{\scshape Belle} collaboration, \emph{{Measurement of the decay $B\to
  D\ell\nu_\ell$ in fully reconstructed events and determination of the
  Cabibbo-Kobayashi-Maskawa matrix element $|V_{cb}|$}},
  \href{https://doi.org/10.1103/PhysRevD.93.032006}{\emph{Phys. Rev. D}
  {\bfseries 93} (2016) 032006}
  [\href{https://arxiv.org/abs/1510.03657}{{\ttfamily 1510.03657}}].

\bibitem{Bordone2016}
M.~Bordone, G.~Isidori and A.~Pattori, \emph{On the standard model predictions
  for $r_k$ and $r_{K^*}$},
  \href{https://doi.org/10.1140/epjc/s10052-016-4274-7}{\emph{The European
  Physical Journal C} {\bfseries 76} (2016) }.

\bibitem{Isidori:2022bzw}
G.~Isidori, D.~Lancierini, S.~Nabeebaccus and R.~Zwicky, \emph{{QED in $
  \overline{B}\to\overline{K}\ell^{+}\ell^{-}$ LFU ratios: theory versus
  experiment, a Monte Carlo study}},
  \href{https://doi.org/10.1007/JHEP10(2022)146}{\emph{JHEP} {\bfseries 10}
  (2022) 146} [\href{https://arxiv.org/abs/2205.08635}{{\ttfamily
  2205.08635}}].

\bibitem{LHCb:2022vje}
{\scshape LHCb} collaboration, \emph{{Measurement of lepton universality
  parameters in $B^+\to K^+\ell^+\ell^-$ and $B^0\to K^{*0}\ell^+\ell^-$
  decays}}, \href{https://doi.org/10.1103/PhysRevD.108.032002}{\emph{Phys. Rev.
  D} {\bfseries 108} (2023) 032002}
  [\href{https://arxiv.org/abs/2212.09153}{{\ttfamily 2212.09153}}].

\bibitem{Allwicher:2023shc}
L.~Allwicher, C.~Cornella, G.~Isidori and B.A.~Stefanek, \emph{{New physics in
  the third generation. A comprehensive SMEFT analysis and future prospects}},
  \href{https://doi.org/10.1007/JHEP03(2024)049}{\emph{JHEP} {\bfseries 03}
  (2024) 049} [\href{https://arxiv.org/abs/2311.00020}{{\ttfamily
  2311.00020}}].

\bibitem{Davighi:2025cqx}
J.~Davighi and G.~Isidori, \emph{{A Composite Theory of Higgs and Flavour}},
  \href{https://arxiv.org/abs/2512.19650}{{\ttfamily 2512.19650}}.

\bibitem{Demetriou:2025ewa}
G.~Demetriou, G.~Isidori, G.~Piazza and E.~Pinsard, \emph{{The
  third-generation-philic WIMP: an EFT analysis}},
  \href{https://doi.org/10.1140/epjc/s10052-025-14580-5}{\emph{Eur. Phys. J. C}
  {\bfseries 85} (2025) 865}
  [\href{https://arxiv.org/abs/2505.04708}{{\ttfamily 2505.04708}}].

\bibitem{Allwicher:2025bub}
L.~Allwicher, G.~Isidori and M.~Pesut, \emph{{Flavored circular collider:
  cornering New Physics at FCC-ee via flavor-changing processes}},
  \href{https://doi.org/10.1140/epjc/s10052-025-14359-8}{\emph{Eur. Phys. J. C}
  {\bfseries 85} (2025) 631}
  [\href{https://arxiv.org/abs/2503.17019}{{\ttfamily 2503.17019}}].

\bibitem{Barbieri:2011ci}
R.~Barbieri, G.~Isidori, J.~Jones-Perez, P.~Lodone and D.M.~Straub,
  \emph{{$U(2)$ and Minimal Flavour Violation in Supersymmetry}},
  \href{https://doi.org/10.1140/epjc/s10052-011-1725-z}{\emph{Eur. Phys. J. C}
  {\bfseries 71} (2011) 1725}
  [\href{https://arxiv.org/abs/1105.2296}{{\ttfamily 1105.2296}}].

\bibitem{Barbieri:2012uh}
R.~Barbieri, D.~Buttazzo, F.~Sala and D.M.~Straub, \emph{{Flavour physics from
  an approximate $U(2)^3$ symmetry}},
  \href{https://doi.org/10.1007/JHEP07(2012)181}{\emph{JHEP} {\bfseries 07}
  (2012) 181} [\href{https://arxiv.org/abs/1203.4218}{{\ttfamily 1203.4218}}].

\bibitem{Isidori:2012ts}
G.~Isidori and D.M.~Straub, \emph{{Minimal Flavour Violation and Beyond}},
  \href{https://doi.org/10.1140/epjc/s10052-012-2103-1}{\emph{Eur. Phys. J. C}
  {\bfseries 72} (2012) 2103}
  [\href{https://arxiv.org/abs/1202.0464}{{\ttfamily 1202.0464}}].

\bibitem{Covone:2025lee}
S.~Covone, P.~Morell and A.~Tinari, \emph{{Constraints on lepton-flavor mixing
  with third-generation new physics}},
  \href{https://doi.org/10.1016/j.physletb.2026.140582}{\emph{Phys. Lett. B}
  {\bfseries 879} (2026) 140582}
  [\href{https://arxiv.org/abs/2511.15800}{{\ttfamily 2511.15800}}].

\bibitem{Gherardi:2020qhc}
V.~Gherardi, D.~Marzocca and E.~Venturini, \emph{{Low-energy phenomenology of
  scalar leptoquarks at one-loop accuracy}},
  \href{https://doi.org/10.1007/JHEP01(2021)138}{\emph{JHEP} {\bfseries 01}
  (2021) 138} [\href{https://arxiv.org/abs/2008.09548}{{\ttfamily
  2008.09548}}].

\bibitem{Bauer:2016hbm}
M.~Bauer and M.~Neubert, \emph{{Minimal Leptoquark Explanation for the
  $R_{D^{(*)}}$, $R_K$, and $(g-2)_\mu$ Anomalies}},
  \href{https://doi.org/10.1103/PhysRevLett.116.141802}{\emph{Phys. Rev. Lett.}
  {\bfseries 116} (2016) 141802}
  [\href{https://arxiv.org/abs/1511.01900}{{\ttfamily 1511.01900}}].

\bibitem{Hiller:2016kry}
G.~Hiller, D.~Loose and K.~Schönwald, \emph{{Leptoquark Flavor Patterns \& B
  Decay Anomalies}}, \href{https://doi.org/10.1007/JHEP12(2016)027}{\emph{JHEP}
  {\bfseries 12} (2016) 027}
  [\href{https://arxiv.org/abs/1609.08895}{{\ttfamily 1609.08895}}].

\bibitem{Angelescu:2021lln}
A.~Angelescu, D.~Bečirević, D.A.~Faroughy, F.~Jaffredo and O.~Sumensari,
  \emph{{Single leptoquark solutions to the $B$-physics anomalies}},
  \href{https://doi.org/10.1103/PhysRevD.104.055017}{\emph{Phys. Rev. D}
  {\bfseries 104} (2021) 055017}
  [\href{https://arxiv.org/abs/2103.12504}{{\ttfamily 2103.12504}}].

\bibitem{Crivellin:2021egp}
A.~Crivellin, D.~M{\"u}ller and L.~Schnell, \emph{{Combined constraints on
  first generation leptoquarks}},
  \href{https://doi.org/10.1103/PhysRevD.103.115023}{\emph{Phys. Rev. D}
  {\bfseries 103} (2021) 115023}
  [\href{https://arxiv.org/abs/2104.06417}{{\ttfamily 2104.06417}}].

\bibitem{UTfit:2006onp}
{\scshape UTfit} collaboration, \emph{{Constraints on new physics from the
  quark mixing unitarity triangle}},
  \href{https://doi.org/10.1103/PhysRevLett.97.151803}{\emph{Phys. Rev. Lett.}
  {\bfseries 97} (2006) 151803}
  [\href{https://arxiv.org/abs/hep-ph/0605213}{{\ttfamily hep-ph/0605213}}].

\bibitem{Allwicher:2024mzw}
L.~Allwicher, D.A.~Faroughy, M.~Martines, O.~Sumensari and F.~Wilsch, \emph{{On
  the EFT validity for Drell{\textendash}Yan tails at the LHC}},
  \href{https://doi.org/10.1140/epjc/s10052-025-14171-4}{\emph{Eur. Phys. J. C}
  {\bfseries 85} (2025) 463}
  [\href{https://arxiv.org/abs/2412.14162}{{\ttfamily 2412.14162}}].

\bibitem{Hoferichter:2023wiy}
M.~Hoferichter, B.-L.~Hoid and J.~Ruiz~de Elvira, \emph{{Improved
  Standard-Model prediction for $K_{L}\to\ell^{+}\ell^{-}$}},
  \href{https://doi.org/10.1007/JHEP04(2024)071}{\emph{JHEP} {\bfseries 04}
  (2024) 071} [\href{https://arxiv.org/abs/2310.17689}{{\ttfamily
  2310.17689}}].

\bibitem{Burdman:2001tf}
G.~Burdman, E.~Golowich, J.L.~Hewett and S.~Pakvasa, \emph{{Rare charm decays
  in the standard model and beyond}},
  \href{https://doi.org/10.1103/PhysRevD.66.014009}{\emph{Phys. Rev. D}
  {\bfseries 66} (2002) 014009}
  [\href{https://arxiv.org/abs/hep-ph/0112235}{{\ttfamily hep-ph/0112235}}].

\bibitem{Becirevic:2026tle}
D.~Be{\v{c}}irevi{\'c}, M.~Martines, S.~Rosauro-Alcaraz and O.~Sumensari,
  \emph{{Interpreting the results on exclusive $c\rightarrow s\mu\nu$ modes}},
  \href{https://doi.org/10.1016/j.physletb.2026.140713}{\emph{Phys. Lett. B}
  {\bfseries 880} (2026) 140713}
  [\href{https://arxiv.org/abs/2603.25837}{{\ttfamily 2603.25837}}].

\bibitem{Bobeth:2013uxa}
C.~Bobeth, M.~Gorbahn, T.~Hermann, M.~Misiak, E.~Stamou and M.~Steinhauser,
  \emph{{$B_{s,d} \to l^+ l^-$ in the Standard Model with Reduced Theoretical
  Uncertainty}},
  \href{https://doi.org/10.1103/PhysRevLett.112.101801}{\emph{Phys. Rev. Lett.}
  {\bfseries 112} (2014) 101801}
  [\href{https://arxiv.org/abs/1311.0903}{{\ttfamily 1311.0903}}].

\bibitem{Beneke:2019slt}
M.~Beneke, C.~Bobeth and R.~Szafron, \emph{{Power-enhanced leading-logarithmic
  QED corrections to $B_q \to \mu^+\mu^-$}},
  \href{https://doi.org/10.1007/JHEP10(2019)232}{\emph{JHEP} {\bfseries 10}
  (2019) 232} [\href{https://arxiv.org/abs/1908.07011}{{\ttfamily
  1908.07011}}].

\bibitem{PhysRevD.93.032006}
{\scshape Belle Collaboration} collaboration, \emph{Measurement of the decay
  $b\ensuremath{\rightarrow}d\ensuremath{\ell}{\ensuremath{\nu}}_{\ensuremath{\ell}}$
  in fully reconstructed events and determination of the
  cabibbo-kobayashi-maskawa matrix element $|{V}_{cb}|$},
  \href{https://doi.org/10.1103/PhysRevD.93.032006}{\emph{Phys. Rev. D}
  {\bfseries 93} (2016) 032006}.

\bibitem{Hayasaka:2010np}
K.~Hayasaka et~al., \emph{{Search for Lepton Flavor Violating Tau Decays into
  Three Leptons with 719 Million Produced Tau+Tau- Pairs}},
  \href{https://doi.org/10.1016/j.physletb.2010.03.037}{\emph{Phys. Lett. B}
  {\bfseries 687} (2010) 139}
  [\href{https://arxiv.org/abs/1001.3221}{{\ttfamily 1001.3221}}].

\bibitem{Belle-II:2024sce}
{\scshape Belle-II} collaboration, \emph{{Search for lepton-flavor-violating
  {\ensuremath{\tau}}$^{-}${\textrightarrow}
  {\ensuremath{\mu}}$^{-}${\ensuremath{\mu}}$^{+}${\ensuremath{\mu}}$^{-}$
  decays at Belle II}},
  \href{https://doi.org/10.1007/JHEP09(2024)062}{\emph{JHEP} {\bfseries 09}
  (2024) 062} [\href{https://arxiv.org/abs/2405.07386}{{\ttfamily
  2405.07386}}].

\bibitem{Belle:2025iff}
{\scshape Belle, Belle-II} collaboration, \emph{{Search for
  lepton-flavor-violating ${\tau }^{-}\to {{\ell}}^{-}{K}_{s}^{0}$ decays at
  Belle and Belle II}},
  \href{https://doi.org/10.1007/JHEP08(2025)092}{\emph{JHEP} {\bfseries 08}
  (2025) 092} [\href{https://arxiv.org/abs/2504.15745}{{\ttfamily
  2504.15745}}].

\bibitem{Marciano:1993sh}
W.J.~Marciano and A.~Sirlin, \emph{{Radiative corrections to pi(lepton 2)
  decays}}, \href{https://doi.org/10.1103/PhysRevLett.71.3629}{\emph{Phys. Rev.
  Lett.} {\bfseries 71} (1993) 3629}.

\bibitem{Finkemeier:1995gi}
M.~Finkemeier, \emph{{Radiative corrections to pi(l2) and K(l2) decays}},
  \href{https://doi.org/10.1016/0370-2693(96)01030-1}{\emph{Phys. Lett. B}
  {\bfseries 387} (1996) 391}
  [\href{https://arxiv.org/abs/hep-ph/9505434}{{\ttfamily hep-ph/9505434}}].

\bibitem{Cirigliano:2007xi}
V.~Cirigliano and I.~Rosell, \emph{{Two-loop effective theory analysis of pi
  (K) ---{\ensuremath{>}} e anti-nu/e [gamma] branching ratios}},
  \href{https://doi.org/10.1103/PhysRevLett.99.231801}{\emph{Phys. Rev. Lett.}
  {\bfseries 99} (2007) 231801}
  [\href{https://arxiv.org/abs/0707.3439}{{\ttfamily 0707.3439}}].

\bibitem{Cirigliano:2007ga}
V.~Cirigliano and I.~Rosell, \emph{{pi/K ---{\ensuremath{>}} e anti-nu(e)
  branching ratios to O(e**2 p**4) in Chiral Perturbation Theory}},
  \href{https://doi.org/10.1088/1126-6708/2007/10/005}{\emph{JHEP} {\bfseries
  10} (2007) 005} [\href{https://arxiv.org/abs/0707.4464}{{\ttfamily
  0707.4464}}].

\bibitem{Cirigliano:2011tm}
V.~Cirigliano and H.~Neufeld, \emph{{A note on isospin violation in Pl2(gamma)
  decays}}, \href{https://doi.org/10.1016/j.physletb.2011.04.038}{\emph{Phys.
  Lett. B} {\bfseries 700} (2011) 7}
  [\href{https://arxiv.org/abs/1102.0563}{{\ttfamily 1102.0563}}].

\bibitem{Giusti:2017dwk}
D.~Giusti, V.~Lubicz, G.~Martinelli, C.T.~Sachrajda, F.~Sanfilippo, S.~Simula
  et~al., \emph{{First lattice calculation of the QED corrections to leptonic
  decay rates}},
  \href{https://doi.org/10.1103/PhysRevLett.120.072001}{\emph{Phys. Rev. Lett.}
  {\bfseries 120} (2018) 072001}
  [\href{https://arxiv.org/abs/1711.06537}{{\ttfamily 1711.06537}}].

\bibitem{Desiderio:2020oej}
A.~Desiderio et~al., \emph{{First lattice calculation of radiative leptonic
  decay rates of pseudoscalar mesons}},
  \href{https://doi.org/10.1103/PhysRevD.103.014502}{\emph{Phys. Rev. D}
  {\bfseries 103} (2021) 014502}
  [\href{https://arxiv.org/abs/2006.05358}{{\ttfamily 2006.05358}}].

\bibitem{Cornella:2026lkp}
C.~Cornella, M.~Ferr{\'e}, M.~K{\"o}nig and M.~Neubert, \emph{{The simplest B
  decay, precisely}},
  \href{https://doi.org/10.1007/JHEP06(2026)027}{\emph{JHEP} {\bfseries 06}
  (2026) 027} [\href{https://arxiv.org/abs/2601.14361}{{\ttfamily
  2601.14361}}].

\bibitem{Cornella:2026ask}
C.~Cornella, M.~Ferr{\'e}, M.~K{\"o}nig and M.~Neubert, \emph{{QED Corrections
  to $B^-\to\tau^-\bar\nu_\tau$}},
  \href{https://arxiv.org/abs/2607.13144}{{\ttfamily 2607.13144}}.

\bibitem{Cornella:2022ubo}
C.~Cornella, M.~K{\"o}nig and M.~Neubert, \emph{{Structure-dependent QED
  effects in exclusive B decays at subleading power}},
  \href{https://doi.org/10.1103/PhysRevD.108.L031502}{\emph{Phys. Rev. D}
  {\bfseries 108} (2023) L031502}
  [\href{https://arxiv.org/abs/2212.14430}{{\ttfamily 2212.14430}}].

\bibitem{Beneke:2020vnb}
M.~Beneke, P.~B{\"o}er, J.-N.~Toelstede and K.K.~Vos, \emph{{QED factorization
  of non-leptonic $B$ decays}},
  \href{https://doi.org/10.1007/JHEP11(2020)081}{\emph{JHEP} {\bfseries 11}
  (2020) 081} [\href{https://arxiv.org/abs/2008.10615}{{\ttfamily
  2008.10615}}].

\bibitem{Becirevic:2009aq}
D.~Becirevic, B.~Haas and E.~Kou, \emph{{Soft Photon Problem in Leptonic
  B-decays}}, \href{https://doi.org/10.1016/j.physletb.2009.10.017}{\emph{Phys.
  Lett. B} {\bfseries 681} (2009) 257}
  [\href{https://arxiv.org/abs/0907.1845}{{\ttfamily 0907.1845}}].

\bibitem{Boer:2023vsg}
P.~B{\"o}er and T.~Feldmann, \emph{{Structure-dependent QED effects in
  exclusive $B$-meson decays}},
  \href{https://doi.org/10.1140/epjs/s11734-024-01091-9}{\emph{Eur. Phys. J.
  ST} {\bfseries 233} (2024) 299}
  [\href{https://arxiv.org/abs/2312.12885}{{\ttfamily 2312.12885}}].

\bibitem{Kitahara:2025jqk}
T.~Kitahara, J.~Miyamoto and K.~Sasaki, \emph{{Complete one-loop QED
  corrections to $ {D}_s^{+} $ leptonic decays and impact on the CKM unitarity
  test}}, \href{https://doi.org/10.1007/JHEP05(2026)041}{\emph{JHEP} {\bfseries
  05} (2026) 041} [\href{https://arxiv.org/abs/2511.22383}{{\ttfamily
  2511.22383}}].

\bibitem{Buchalla:1998ba}
G.~Buchalla and A.J.~Buras, \emph{{The rare decays $K\to \pi \nu\bar\nu$, $B
  \to X \nu\bar\nu$ and $B \to l^+ l^-$: An Update}},
  \href{https://doi.org/10.1016/S0550-3213(99)00149-2}{\emph{Nucl. Phys. B}
  {\bfseries 548} (1999) 309}
  [\href{https://arxiv.org/abs/hep-ph/9901288}{{\ttfamily hep-ph/9901288}}].

\bibitem{Brod:2008ss}
J.~Brod and M.~Gorbahn, \emph{{Electroweak Corrections to the Charm Quark
  Contribution to K+ ---{\ensuremath{>}} pi+ nu anti-nu}},
  \href{https://doi.org/10.1103/PhysRevD.78.034006}{\emph{Phys. Rev. D}
  {\bfseries 78} (2008) 034006}
  [\href{https://arxiv.org/abs/0805.4119}{{\ttfamily 0805.4119}}].

\bibitem{Buras:2006gb}
A.J.~Buras, M.~Gorbahn, U.~Haisch and U.~Nierste, \emph{{Charm Quark
  Contribution to $K^+ \to \pi^+ \nu \bar \nu$ at Next-to-Next-to-Leading
  Order}}, \href{https://doi.org/10.1007/JHEP11(2012)167}{\emph{JHEP}
  {\bfseries 11} (2006) 002}
  [\href{https://arxiv.org/abs/hep-ph/0603079}{{\ttfamily hep-ph/0603079}}].

\bibitem{Brod:2010hi}
J.~Brod, M.~Gorbahn and E.~Stamou, \emph{{Two-Loop Electroweak Corrections for
  the $K \to \pi \nu \bar{\nu}$ Decays}},
  \href{https://doi.org/10.1103/PhysRevD.83.034030}{\emph{Phys. Rev. D}
  {\bfseries 83} (2011) 034030}
  [\href{https://arxiv.org/abs/1009.0947}{{\ttfamily 1009.0947}}].

\bibitem{Brod:2021hsj}
J.~Brod, M.~Gorbahn and E.~Stamou, \emph{{Updated Standard Model Prediction for
  $K \to \pi \nu \bar{\nu}$ and $\epsilon_K$}},
  \href{https://doi.org/10.22323/1.391.0056}{\emph{PoS} {\bfseries BEAUTY2020}
  (2021) 056} [\href{https://arxiv.org/abs/2105.02868}{{\ttfamily
  2105.02868}}].

\bibitem{Isidori:2005xm}
G.~Isidori, F.~Mescia and C.~Smith, \emph{{Light-quark loops in K
  ---{\ensuremath{>}} pi nu anti-nu}},
  \href{https://doi.org/10.1016/j.nuclphysb.2005.04.008}{\emph{Nucl. Phys. B}
  {\bfseries 718} (2005) 319}
  [\href{https://arxiv.org/abs/hep-ph/0503107}{{\ttfamily hep-ph/0503107}}].

\bibitem{Lunghi:2024sjy}
E.~Lunghi and A.~Soni, \emph{{Light quark loops in $ {K}^{\pm}\to
  {\pi}^{\pm}\nu \overline{\nu} $ from vector meson dominance and update on the
  Kaon Unitarity Triangle}},
  \href{https://doi.org/10.1007/JHEP12(2024)097}{\emph{JHEP} {\bfseries 12}
  (2024) 097} [\href{https://arxiv.org/abs/2408.11190}{{\ttfamily
  2408.11190}}].

\bibitem{Mescia:2007kn}
F.~Mescia and C.~Smith, \emph{{Improved estimates of rare K decay
  matrix-elements from Kl3 decays}},
  \href{https://doi.org/10.1103/PhysRevD.76.034017}{\emph{Phys. Rev. D}
  {\bfseries 76} (2007) 034017}
  [\href{https://arxiv.org/abs/0705.2025}{{\ttfamily 0705.2025}}].

\bibitem{Cornella2021}
C.~Cornella, D.A.~Faroughy, J.~Fuentes-Martín, G.~Isidori and M.~Neubert,
  \emph{Reading the footprints of the b-meson flavor anomalies},
  \href{https://doi.org/10.1007/jhep08(2021)050}{\emph{Journal of High Energy
  Physics} {\bfseries 2021} (2021) }.

\bibitem{Plakias:2023esq}
I.~Plakias and O.~Sumensari, \emph{{Lepton flavor violation in semileptonic
  observables}}, \href{https://doi.org/10.1103/PhysRevD.110.035016}{\emph{Phys.
  Rev. D} {\bfseries 110} (2024) 035016}
  [\href{https://arxiv.org/abs/2312.14070}{{\ttfamily 2312.14070}}].

\end{thebibliography}\endgroup

\end{document}